\documentclass[aps,prr,twocolumn,showpacs,preprintnumbers,amsmath,amssymb]{revtex4-1}
\usepackage[dvipdfmx]{graphicx}
\usepackage{amsmath,braket}
\usepackage{graphicx,float,color}
\usepackage{hyperref}

\renewcommand{\theequation}{\thesection.\arabic{equation}}

\newcommand{\de}{\partial}
\newcommand{\be}{\begin{equation}}
\newcommand{\ba}{\begin{eqnarray}}
\newcommand{\ea}{\end{eqnarray}}
\newcommand{\ee}{\end{equation}}

\newcommand{\f}{\frac}
\newcommand{\s}{\sqrt}

\newcommand{\ti}{\tilde}

\newcommand{\ddd}{\cdot\cdot\cdot}
\newcommand{\no}{\nonumber \\}
\newcommand{\la}{\langle}
\newcommand{\lb}{\rangle}
\newcommand{\bea}{\begin{eqnarray}}
\newcommand{\eea}{\end{eqnarray}}
\newcommand{\bes}{\begin{equation*}}
\newcommand{\beas}{\begin{eqnarray*}}
\newcommand{\eeas}{\end{eqnarray*}}
\newcommand{\bas}{\begin{array*}}
\newcommand{\eas}{\end{array*}}
\newcommand{\ees}{\end{equation*}}
\newcommand{\nn}{\nonumber}

\newcommand{\ep}{\epsilon}

\newcommand{\bpm}{\begin{pmatrix}}
\newcommand{\epm}{\end{pmatrix}}
\newcommand{\bbm}{\begin{bmatrix}}
\newcommand{\ebm}{\end{bmatrix}}

\newcommand{\CO}{{\cal{O}}}

\def\CO{{\mathcal{O}}}

\def\CM{{\mathcal{M}}}

\usepackage{bm}

\begin{document}

\title{Aspects of Pseudo Entropy in Field Theories}
\preprint{YITP-21-52;\ IPMU21-0033}

\author{Ali Mollabashi$^{a,b}$, Noburo Shiba$^a$, Tadashi Takayanagi$^{a,c,d}$, 
Kotaro Tamaoka$^{a,e}$, and Zixia Wei$^{a}$}

\affiliation{$^a$Center for Gravitational Physics,
Yukawa Institute for Theoretical Physics,
Kyoto University,
Kitashirakawa Oiwakecho, Sakyo-ku, Kyoto 606-8502, Japan}

\affiliation{$^b$Max-Planck-Institut for Physics
Werner-Heisenberg-Institut 80805 Munich, Germany}

\affiliation{$^c$Inamori Research Institute for Science,
620 Suiginya-cho, Shimogyo-ku,
Kyoto 600-8411 Japan}

\affiliation{$^{d}$Kavli Institute for the Physics and Mathematics
 of the Universe (WPI),
University of Tokyo, Kashiwa, Chiba 277-8582, Japan}

\affiliation{$^e$Department of Physics, College of Humanities and Sciences, Nihon University, Sakura-josui, Tokyo 156-8550, Japan}

\begin{abstract}
In this article, we explore properties of pseudo entropy \cite{Nakata:2021ubr} 
in quantum field theories and spin systems from several approaches.
Pseudo entropy is a generalization of entanglement entropy such that it depends on both an initial and 
final state and has a clear gravity dual via the AdS/CFT. We numerically analyze 
a class of free scalar field theories and the XY spin model. This reveals basic properties of 
pseudo entropy in quantum many-body systems, namely, the area law behavior,  the saturation behavior, and the non-positivity of difference between the pseudo entropy and averaged entanglement entropy in the same quantum phase. In addition, our numerical analysis finds an example where the strong subadditivity of pseudo entropy 
gets violated. Interestingly, we find that  the non-positivity of the difference can be violated only if the initial and final state belong to different quantum phases. We also present analytical arguments which support these properties by both conformal field theoretic and holographic calculations. When the initial and final state belong to different topological phases, we expect a gapless mode localized along an interface, which enhances the pseudo entropy, leading to the violation of the non-positivity of the difference. Moreover, we also compute the time evolution of pseudo entropy after a global quantum quench, 
where we observe that the imaginary part of pseudo entropy shows an interesting characteristic behavior.

\end{abstract}

\maketitle

\tableofcontents

\section{Introduction}

As one of the most fundamental quantum resources, entanglement plays key roles in almost all areas of quantum physics, not only practically but also theoretically. One branch is its application to quantum many body systems, which has revealed many important properties of correlation structures \cite{Peschel02, Vidal:2002rm, CC04, Casini:2009sr}, quantum phase transitions \cite{Vidal:2002rm, Latorre:2003kg}, thermalization process \cite{Calabrese:2007rg, Calabrese:2016xau} and emergence of spacetime \cite{RT, Swingle:2009bg, VanRaamsdonk:2010pw, Maldacena:2013xja}. Entanglement entropy is the most often used quantity to measure the amount of entanglement between two parts of a quantum system. 

Recently, a generalization of entanglement entropy called pseudo entropy has been proposed in \cite{Nakata:2021ubr}. Instead of a quantum state which is described by a density matrix, let us consider a matrix defined from two pure quantum states $\ket{\psi_1}$ and $\ket{\psi_2}$: 
\begin{align}
	\tau^{1|2} = \frac{|\psi_1\rangle\langle\psi_2|}{\langle\psi_2|\psi_1\rangle}. 
\end{align}
This is called a transition matrix. A transition matrix describes an experimental process called post-selection, which is realized by setting the initial state as $\ket{\psi_1}$ and the final state as $\ket{\psi_2}$. The expectation value of an observable $\CO$, ${\rm Tr}\left(\CO\tau^{1|2}\right)$ is known as a weak value \cite{Aharonov:1988xu, Dressel2013} in the post-selection experiment. 
Dividing the whole system into $A$ and its complementary $B$, the pseudo entropy of $A$ is defined as  
\begin{align}
    S(\tau_A^{1|2}) = -{\rm Tr} \left[\tau_A^{1|2}\log\tau_A^{1|2} \right],
\end{align}
where $\tau_A^{1|2} = {\rm Tr}_B \left(\tau^{1|2}\right)$ is called the reduced transition matrix. Note that transition matrices and pseudo entropy are reduced to conventional density matrices and entanglement entropy when $\ket{\psi_1} = \ket{\psi_2}$. Practically, it is convenient to introduce R\'{e}nyi pseudo entropy
\begin{align}
    S^{(n)}(\tau_A^{1|2}) = \frac{1}{1-n} \log {\rm Tr} \left(\tau_A^{1|2} \right)^n,
\end{align}
whose $n\rightarrow1$ limit gives the pseudo entropy $S(\tau_A^{1|2})$. 

Pseudo entropy is originally proposed with a motivation from AdS/CFT: it serves as a CFT quantity corresponding to the area of a generic codimension-2 surface in Euclidean AdS. In spite of this, pseudo entropy is clearly an important fundamental quantity in general quantum systems. For example, it is shown in \cite{Nakata:2021ubr} that pseudo entropy can be regarded as the number of Bell pairs that one can distill from a post-selection process for a specific class of transition matrices. On the other hand, although pseudo entropy has been studied in simple qubit systems, random systems, and holographic systems in \cite{Nakata:2021ubr}, less is known in ordinary many body systems. All of these motivates us to study pseudo entropy in more familiar quantum many-body systems, such as free field theories and spin systems. 

The letter \cite{Mollabashi:2020yie} initializes the study of pseudo entropy in free scalar theories and the transverse Ising model. This paper is an extended version of \cite{Mollabashi:2020yie} with all the technical details presented. Besides, it includes many new results on different types of factorization, fermionic systems, quantum phase transitions, dynamical setups and holographic setups. 

Although a wide variety of analysis has been performed, three properties are found to be universal. 
\begin{enumerate}
    \item {\bf Saturation of pseudo entropy} \\For states near the ground state of QFT and holographic states, we observe the saturation property
    \begin{align}
        S(\tau^{1|2}_A) \sim \min[S(\rho_A^1), S(\rho_A^2)]
    \end{align}
    where $\rho^i = |\psi_i\rangle\langle\psi_i|$. 
    \item {\bf Nonpositivity of difference and its violation}\\ For states near the ground state of QFT and holographic states, we observe the inequality
    \begin{align}
      \Delta S_{12} = S(\tau^{1|2}_A) - \frac{1}{2}\left(S(\rho_A^1)+S(\rho_A^2)\right) \leq 0  \label{difpea}
    \end{align}
    as long as $\ket{\psi_1}$ and $\ket{\psi_2}$ lie in the same quantum phase. If there is a quantum phase transition that brings $\ket{\psi_1}$ to $\ket{\psi_2}$, then the above inequality can be violated. This implies that the pseudo entropy can distinguish between different quantum phases.
    \item {\bf Area law} \\ For states near the ground state of QFT and for holographic states, $S(\tau^{1|2}_A)$ satisfies an area law. 
\end{enumerate}
These properties will be verified for multiple times in different systems in this paper. 

In the following, we will summarize the main results contained in this paper and then show a road map which is useful to read this paper.

\subsection{Summary of This Paper}

Here we summarize the results for each section. Note that section \ref{sec:scacorr} and section \ref{sec:scalarsingle} contain old results which has been already presented in the letter \cite{Mollabashi:2020yie} but with more technical details, while all other parts of this paper are new results which do not appear in \cite{Mollabashi:2020yie}. 

In section \ref{sec:correlator}, methods for computing pseudo entropy in both bosonic and fermionic free theories are presented by generalizing the standard correlator method \cite{Casini:2009sr,Peschel09} and operator method \cite{Shiba2014, Shiba2020} for entanglement entropy computation. 

In section \ref{sec:freescalar}, we numerically study universal properties 1 - 3 in free Lifshitz scalar theories. We start with results which have been already mentioned in the original letter \cite{Mollabashi:2020yie}. We further investigate several important inequalities which the ordinary entanglement entropy satisfies. In particular, we find that the strong subadditivity of the pseudo entropy can break in general (section \ref{subsec:ssa}). We also give an analytic result based on periodic subsystems which supports the saturation of pseudo entropy (section \ref{subsec:periodic_sub}). 

In section \ref{sec:spinex}, we study the pseudo entropy in the quantum XY model \cite{Vidal:2002rm, XY1, XY2}, which largely extends the Ising model computations in  \cite{Mollabashi:2020yie}.
Quantum XY model is a spin model which contains the transverse Ising model as a special case, and shows colorful quantum phase structures. It can be mapped to free fermions by performing Jordan-Wigner transformation. By numerical computation with the correlator method for free fermions, as well as direct computation in the spin system, we confirm that the non-positivity of difference always holds when $\ket{\psi_1}$ and $\ket{\psi_2}$ lie in the same quantum phase, and it can be violated when they belong to different phases. 

In section \ref{sec:globalquench}, we study the pseudo entropy between two different states after the same global quench, in CFTs (section \ref{sec:quenchCFT}) and in free scalar theories (section \ref{sec:quenchscalar}). The real time evolution introduces an imaginary part to the pseudo entropy. While the real part shows a linear behavior similar to the entanglement entropy after a global quench, the imaginary part has a plateau behavior and measures a sort of negativity of the reduced transition matrix. We also present a gravity dual for a global quench in section \ref{sec:quenchGraDual}, which turns out to be a black hole geometry with an end-of-the-world brane, whose location is given by complexified coordinates. 

In section \ref{sec:pertpe}, we analyze the pseudo entropy for perturbed CFTs.
In general, we find that the pseudo entropy after a perturbation is smaller than that of an unperturbed CFT. For exactly marginal perturbations, we show that the coefficient of logarithmic divergence decreases, which implies that the difference (\ref{difpea}) is negative. 

In section \ref{sec:hol}, we study the behavior of holographic pseudo entropy in gravity setups which are dual to two vacuum states in two different quantum field theories.
When we consider two field theories related by an exactly marginal perturbation, whose gravity dual is given by Janus solutions, we are able to show that the  coefficient of logarithmic divergence decreases under rather general assumptions. This is consistent with our field theory perturbation result in section  \ref{sec:pertpe}. Note that our holographic results here go beyond the perturbation regime. Moreover, we also examine the holographic pseudo entropy when two massive field theory ground states can be deformed to each other only through that of a gapless theory (CFT). In this case we find that the difference (\ref{difpea}) can be positive for a range of parameters, which control relevant perturbations. This is consistent with our spin system results that the difference between pseudo entropy and averaged entanglement entropy tends to be positive when the two states are in two different quantum phases.

In the appendix A, we present some details of path integral calculations of Gaussian transition matrices.

\subsection{How to Read This Paper: A Road Map}
Except for section \ref{sec:correlator} which contains basic computing methods, all other sections are independent from each other so that readers can either go through the paper in order or pick up certain sections to read. However, since several different methods are used throughout paper, and certain methods maybe associated to certain motivations, we provide a road map here for readers to refer. 

Analytic methods used in this paper can be classified into three types: numerical methods, CFT computations and holographic analysis. Results based on each method and their motivations are summarized as below. 
\begin{itemize}
\item Numerical methods are presented in section \ref{sec:correlator}, and results obtained from them are in section \ref{sec:freescalar}, \ref{sec:spinex} and \ref{sec:quenchscalar}. These results and discussions are mostly condensed matter oriented. 
\item Results from CFT computations are in \ref{sec:quenchCFT} and \ref{sec:pertpe}. Readers who are mostly interested in analytical results may firstly check these sections. 
\item Holographic analysis are presented in \ref{sec:quenchGraDual} and \ref{sec:hol}. These results and discussions are mostly high energy oriented. 
\end{itemize}
Accordingly, besides picking up reading contents by topics, readers can also do it by the methods applied.


\section{Calculation of Pseudo Entropy in Free Theories}\label{sec:correlator}
From a long time ago it has been known how to calculate the spectrum of the reduced density matrix and as a result the entanglement and R\'{e}nyi entropies for Gaussian states of quadratic Hamiltonians. To our knowledge, this goes back to \cite{Peschel02} for generic subregions in fermionic systems, to \cite{Latorre:2003kg, Vidal:2002rm} for a single block of spins, and to \cite{Audenaert:2002xfl, Cramer:2005mx} for generic subregions in bosonic systems. We encourage interested readers to look at very nice reviews on this topic in \cite{Peschel09, Casini:2009sr}.

Since the Hamiltonian of interest is quadratic, the basic idea is to use Wick's theorem to reduce all correlation functions to two point functions and find a clever way to read off the spectrum of the reduced density matrix from the two point functions restricted to the subregion of interest. An important advantage of this method, which we will often refer to as the correlator method, is that one can utilize it to study finite subregions on infinite systems. As a result, we can get rid of the effects due to finiteness of the whole system, but in principle we should be still careful about lattice effects, since for practical reasons there is no way around considering discretized subregions.

In this section, we are going to generalize these methods for calculation of the spectrum of the \textit{transition matrix}. As a result, pseudo entropy and its R\'{e}nyi generalizations can also be computed accordingly. We generalize the methods for bosonic and fermionic systems.

\subsection{Scalar Theories}\label{sec:scacorr}
In this part we first present our generalization for the correlator method for bosonic systems as well as introducing an operator method to calculate pseudo entropy.  
As briefly mentioned above, we will not directly calculate the transition matrix, but extract its spectrum in some sort of indirect way. For a direct path integral calculation of the reduced transition matrix please see appendix \ref{sec:app1}. 

The methodology introduced in this  section is valid for arbitrary quadratic scalar theories. More precisely, we consider real free theories in $(d+1)$ dimensional spacetime. We regularize the theory by putting it on a lattice with $N$ sites in each spatial dimensions. We will stick to this finite size notation with periodic boundary condition in this section, though it is straightforward to take the continuum limit on infinite system or find the corresponding relations for different boundary conditions. We denote the scalar field on the $r$-th site with $\phi_{r}$ and its conjugate momentum with $\pi_{r}$. These two obey the canonical commutation relations 
\begin{equation}\label{eq:comutation}
[\phi_{r} , \pi_{s}]=i \delta_{rs}. 
\end{equation}
We denote the corresponding Hamiltonians (up to a constant) with 
\begin{equation}
H=\sum_{k} \omega_k a_k^{\dagger} a_k
\end{equation}
where $k$ carries $d$ integer valued components running in $-N/2<k_{\mu} \leq N/2$ and the creation and annihilation operators satisfy $[a_k,a^\dagger_{k'}]=\delta_{k,k'}$. The filed and the momentum operators are expanded in terms of these operators as 
\begin{equation}\label{eq:expansion}
\begin{split}
&\phi_{r} = \frac{1}{N^{d/2}} \sum_{k} \frac{1}{\sqrt{2\omega_k}} 
\left[a_k e^{\frac{2\pi i kr}{N}} + a_k^{\dagger} e^{-\frac{2\pi i kr}{N}}  \right], \\
&\pi_{r} = -\frac{i}{N^{d/2}} \sum_{k} \sqrt{ \frac{\omega_k}{2} }
\left[a_k e^{\frac{2\pi i kr}{N}}  - a_k^{\dagger} e^{-\frac{2\pi i kr}{N}}  \right].
\end{split}  
\end{equation}

\subsubsection{Correlator Method}
We consider the post-selected generalization of the expectation value of an operator, called weak value \cite{Aharonov:1988xu, Dressel2013}, defined as 
\be\label{eq:trace}
_2\langle \mathcal{O} \rangle_1 \equiv \frac{\langle \psi_2 | \mathcal{O} | \psi_1 \rangle}{\langle \psi_2 |\psi_1 \rangle}=\mathrm{Tr}
\left[\mathcal{O} \,\tau^{1|2}\right].
\ee
The goal is to work out the reduced transition matrix in its decoupled basis for Gaussian states $|\psi_{1} \rangle$ and $|\psi_{2} \rangle$. To this end we will follow the steps similar to the correlator method which leads to the decoupled modes of the reduced density matrix \cite{Mollabashi:2020yie}. 
The methodology is following these steps
\begin{enumerate}
\item
Work out the left-hand side of \eqref{eq:trace} for all quadratic operators, namely $\mathcal{O}=\{\phi\phi,\pi\pi,\phi\pi\}$. 
\item
Choose a Gaussian ansatz for the reduced transition matrix and work out the left-hand side.
\item
Construct a composite operator out of the aforementioned quadratic operators from whose spectrum we can work out the normal modes of the transition matrix.     
\end{enumerate}

In order to work out the left-hand side of \eqref{eq:trace}, we need to expand $|\psi_{1,2} \rangle$ in the same basis. More precisely, let us consider the case when $|\psi_{1,2} \rangle$ are vacuum states corresponding to different dispersion relations. In this case we have $a_{1,k}|\psi_1 \rangle=a_{2,k}|\psi_2 \rangle=0$ which are related to each other with the following Bogoliubov transformations
\begin{align}\label{eq:bog1}
\begin{split}
a_{2,k}&=\alpha_k\, a_{1,k} + \beta_k\, a^{\dagger}_{2,-k}
\\
a^{\dagger}_{2,-k}&=\beta_k\, a_{1,k} + \alpha_k\, a^{\dagger}_{1,-k}
\end{split}
\end{align}
where
\be\label{eq:ab}
\alpha_k=\frac{1}{2}\left(\sqrt{\frac{\omega_{2,k}}{\omega_{1,k}}}+\sqrt{\frac{\omega_{1,k}}{\omega_{2,k}}}\right)
\;\;
,
\;\;
\beta_k=\frac{1}{2}\left(\sqrt{\frac{\omega_{2,k}}{\omega_{1,k}}}-\sqrt{\frac{\omega_{1,k}}{\omega_{2,k}}}\right)
\ee
thus we can define $|\psi_2 \rangle$ in terms of the eigenvectors of the number operator $n_{1,k}=a^{\dagger}_{1,k}a_{1,k}$ as $|\psi_2 \rangle=\sum_{n=0}^\infty c_{2n}(k)\,| 2n\rangle_{1}$
where
\be\label{eq:scalarcn}
c_{2n}(k)=\left(-\frac{\beta_k}{\alpha_k}\right)^n\sqrt{\frac{(2n-1)!!}{2n!!}}\;c_0    \,.
\ee
With these in mind, we can calculate the left-hand side of \eqref{eq:trace} which are given by 
\begin{align}\label{eq:cor1}
\begin{split}
&X_{rs}\equiv
{}_2\langle \phi_r\phi_s \rangle_1
=
\frac{1}{N^d}\sum_{k} \frac{1}{\omega_{1,k}+\omega_{2,k}}
e^{\frac{2\pi i k (r-s)}{N}},
\\
&P_{rs}\equiv
{}_2\langle \pi_r\pi_s \rangle_1
=
\frac{1}{N^d}\sum_{k} \frac{\omega_{1,k}\omega_{2,k}}{\omega_{1,k}+\omega_{2,k}}
e^{\frac{2\pi i k (r-s)}{N}},
\\
&R_{rs}\equiv
\frac{{}_2\langle \{\phi_r,\pi_s\} \rangle_1}{2}
=
\frac{i}{2N^d}\sum_{k} \frac{\omega_{2,k}-\omega_{1,k}}{\omega_{1,k}+\omega_{2,k}}
e^{\frac{2\pi i k (r-s)}{N}}.
\end{split}
\end{align}

To calculate the right-hand side of \eqref{eq:trace}, we need a suitable transformation that brings the transition matrix of the subregion of interest with $N_A$ site, to its diagonal form as
\be
\tau^{1|2}=\bigotimes_{k=1}^{N_A}\left(1-e^{-\epsilon_k}\right)e^{-\epsilon_k n_{A,k}},
\ee
where $\{\epsilon_k\}$ are the normal modes of the reduced transition matrix, we need a transformation which preserves the commutation relations (in the following of this section we keep the index $A$ to refer to operators which are localized in this subregion). This is a well-known problem in calculation of entanglement entropy, specially used for non-static cases such as considering quantum quenches, where the $R$ correlators become non-trivial. To this end, we need to consider a generalized vector of canonical variables, the fields and their conjugate momenta, as $r=\left( \phi_1,\cdots,\phi_{N_A},\pi_1,\cdots,\pi_{N_A} \right)^T$. So the canonical commutation relations read
\be   
[r_B,r_C]=i J_{BC}
\;\;\;\;\;\;\;,\;\;\;\;\;\;\;
J=
\begin{pmatrix}
0 & \mathbf{1} \\ -\mathbf{1} & 0
\end{pmatrix},
\ee
where $B,C=1,2,\cdots,2N_A$, and we define a correlator matrix as
\be\label{eq:gammadef}
\gamma_{BC}
\equiv
\frac{1}{2}\langle \{r_B,r_C\} \rangle
=
\begin{pmatrix}
X_{bc} & R_{bc} \\ R_{bc}^T & P_{bc}
\end{pmatrix}.
\ee
We consider the following transformations
\begin{align}\label{eq:bogs1}
\begin{split}
\phi_r &= \alpha^*_{rk} a_{A,k}^\dagger+ \alpha_{rk} a_{A,k}\;, 
\\
\pi_r &= -i\,\beta^*_{rk} a_{A,k}^\dagger + i\,\beta_{rk}a_{A,k}\;,
\end{split}
\end{align}
where from the commutation relations we find
\be
\alpha^*\cdot\beta^T+\alpha\cdot\beta^\dagger=-1
\;,
\ee
as well as $\alpha^*\cdot\alpha^T=\alpha\cdot\alpha^\dagger$ and $\beta^*\cdot\beta^T=\beta\cdot\beta^\dagger$. Using these transformations lead to the following expressions for the correlators
\begin{align}
\begin{split}
X&=\alpha^*\cdot\nu\cdot\alpha^T+\alpha\cdot\nu\cdot\alpha^\dagger\;,
\\
P&=\beta^*\cdot\nu\cdot\beta^T+\beta\cdot\nu\cdot\beta^\dagger\;,
\\
R&=i\left(\alpha^*\cdot\nu\cdot\beta^T-\alpha\cdot\nu\cdot\beta^\dagger\right)\;.
\end{split}
\end{align}
Due to Williamson's theorem \cite{Williamson}, for any symmetric positive definite $\gamma$, there always exists a symplectic transformation, $S\in$ Sp($2N,\mathbb{R}$), with $r'=S\cdot r$  and $J=S\cdot J\cdot S^T$ which acts on the canonical variables such that
\be  
\gamma
=
\begin{pmatrix}
\mathrm{diag}(\nu_k) & \mathbf{0} \\ \mathbf{0} & \mathrm{diag}(\nu_k)
\end{pmatrix}\;.
\ee
In our case, the $R$ correlator is pure imaginary. Therefore, the original form of the Williamson's theorem does not apply. However, we can consider its analytic continuation which is non-singular in our criteria of interest and we will provide several justifications for this analytic continuation in the following. Given that, an easy way to work out $\{\epsilon_k\}$ is to find the spectrum of $(i J\cdot\gamma)$ denoted by $\{\nu_k\}$, which gives a double copy of $\{\epsilon_k\}$ as
\be\label{eq:iJgspec}
\nu_k=\pm \frac{1}{2}\coth\left(\frac{\epsilon_k}{2}\right).
\ee
Plugging this $\{\nu_k\}$ into the following formula one can find the von Neumann and R\'{e}nyi pseudo entropies
\begin{align}\label{eq:RPEcorr}
\begin{split}
S(\tau_A^{1|2})
&=
\sum_{k=1}^{N_A}\Bigg[\left(\nu_k+\frac{1}{2}\right)\log\left(\nu_k+\frac{1}{2}\right)
\\&\hspace{15mm}
-\left(\nu_k-\frac{1}{2}\right)\log\left(\nu_k-\frac{1}{2}\right)\Bigg]\;,
\\
S^{(n)}(\tau_A^{1|2})&=\frac{1}{n-1}\sum_{k=1}^{N_A}\log\left[\left(\nu_k+\frac{1}{2}\right)^n-\left(\nu_k-\frac{1}{2}\right)^n\right]\;.
\end{split}
\end{align}

In the next part of this section we work out another method to calculate the pseudo entropy from the correlators inside the subregion, this time \textit{without} assuming a) diagonalizability of the transition matrix and b) the analytic continuation for the Williamson's theorem. We show that using this operator method one can arrive at the formula \eqref{eq:iJgspec} and \eqref{eq:RPEcorr} for pseudo entropy.

\subsubsection{Operator Method}
The basic idea of expressing $\mathrm{Tr}[\rho_A^n]$ to calculate R\'{e}nyi entropies has been considered by \cite{Cardy:2013nua, Calabrese:2010he, Headrick:2010zt}. Based on this idea the operator method has been developed to calculate the von Neumann and R\'{e}nyi entropies in \cite{Shiba2014,Shiba2020} where a systematic method has been introduced to find such a local operator which we will denote it by the \textit{gluing operator}.

Here we show that this method can be generalized to calculate von Neumann and R\'{e}nyi pseudo entropies. This might be in principle extended to non-Gaussian states in interacting theories, although we develop the generalization for Gaussian states in quadratic theories. As mentioned previously we finally recover the \eqref{eq:RPEcorr} with this method, so one can consider the method as an indirect proof for the existence of the aforementioned analytic continuation in the correlator method.

Let us first quickly review the operator method for calculation of R\'{e}nyi entropies introduced in \cite{Shiba2014}.
We consider $(d+1)$ dimensional spacetime and $n$ copies of the scalar fields. We denote the $i$-th copy by $\phi^{(i)}$. 
The Hilbert space of each copy is denoted by $H$, though the total Hilbert space, denoted by $H^{(n)}$, is the tensor product of the $n$ copies 
$H^{(n)}= H \otimes H \dots \otimes H$.
We define the density matrix $\rho^{(n)}$ in  $H^{(n)}$ as 
\begin{equation}
\rho^{(n)} \equiv \rho \otimes \rho \otimes \dots \otimes \rho
\end{equation}
where $\rho$ is an arbitrary density matrix in $H$. 

The key point is that we can express $\mathrm{Tr}\left[ \rho_{A}^n\right]$ as
\begin{equation}\label{eq:expintro}
\mathrm{Tr} \rho_{A}^n=\mathrm{Tr} \left[\rho^{(n)} E_{A}\right],  
\end{equation}
where we refer the interested reader for the details of the gluing operator, $E_{A}$, to consult with \cite{Shiba2014, supplimentary}. 
For pure states where $\rho=|\Psi\rangle\langle\Psi|$, \eqref{eq:expintro} becomes 
\begin{equation}\label{eq:formIntro}
\mathrm{Tr} \left[\rho_{A}^n\right] = \langle\Psi^{(n)}| E_{A} | \Psi^{(n)}\rangle  \end{equation}
where 
\begin{equation}
| \Psi^{(n)}\rangle = | \Psi\rangle \otimes | \Psi\rangle \otimes\cdots \otimes| \Psi\rangle .
\end{equation}
Let us now consider the case of pseudo entropy. The gluing operator has an important property that for $n$ arbitrary operators $F_{j} (j=1,2,\cdots n)$ on $H$, one can verify that 
\begin{equation}\label{eq:property}
\mathrm{Tr}\left[F_1 \otimes F_2 \otimes \cdots \otimes F_n \cdot E_{A}\right] 
= \mathrm{Tr}\left[F_{1 A}  F_{2 A}  \cdots  F_{n A} \right] 
\end{equation}
where $F_{j A}\equiv \mathrm{Tr}_{A^{c}} \left[F_{j}\right]$. 
Plugging this property into the transition matrix
\begin{equation}
{\tau^{2|1} \equiv \frac{|\psi_2\rangle\langle\psi_1|}{\langle\psi_1|\psi_2\rangle},}
\end{equation}
where $\ket{\psi}_1$ and $\ket{\psi}_2$ are two different ground states associated to two different Hamiltonian,
we find 
\begin{equation}\label{eq:trace1}
\mathrm{Tr} \left[\left(\tau_{A}^{2|1}\right)^n\right]
= \frac{\langle \psi^{(n)}_1| E_{A}|\psi^{(n)}_2\rangle}{\langle \psi_1|\psi_2\rangle^n}
\end{equation}
where $|\psi^{(n)}_{i}\rangle = |\psi_i\rangle \otimes |\psi_i\rangle \cdots |\psi_i\rangle$ and $i=1,2$.

After a quite long algebra with regard to the details of the gluing operator, which have been discussed in \cite{supplimentary}, we can show that $\mathrm{Tr} \left[\left(\tau_{A}^{2|1}\right)^n\right]=\left(\det\, 2 \bar{S}_{n}\right)^{-\frac{1}{2}}$ can be expressed as an analytic function of $n$. We find 
\begin{equation}\label{eq:det S_n}
\begin{split}
\det 2\bar{S}_{n} &= \left(\det iJ \right)^{n-1} \det \left[\left(iJ\cdot \gamma+\frac{1}{2}\right)^n -\left(iJ\cdot \gamma-\frac{1}{2}\right)^n \right] \\
&= \prod_{i=1}^{N_A} \left[\left(\nu_i+\frac{1}{2}\right)^n -\left(\nu_i-\frac{1}{2}\right)^n \right] 
\times \\ &
\;\;\;\;\;\;
\prod_{i=N_A+1}^{2N_A} \left[\left(-\nu_i+\frac{1}{2}\right)^n -\left(-\nu_i-\frac{1}{2}\right)^n \right]  ,
\end{split}
\end{equation}
where $\gamma$ is defined in \eqref{eq:gammadef}, $J$ is the symplectic matrix, $N_A$ is the number of sites in the subsystem $A$ and $\nu_i$'s are the eigenvalue of $iJ\cdot\gamma$.
Using the characteristic equation $\mathrm{det} (\nu-iJ\gamma) =0$, it is not hard to see that when some $\nu$ is in the spectrum of $iJ\cdot\gamma$, then $-\nu$ is also in the spectrum. 
We sort $\nu_i$ as $\nu_{N_A+i}=-\nu_{i}$ which leads to
\begin{equation}\label{eq:det S_n 2}
\mathrm{det}\, 2\bar{S}_{n} = \prod_{i=1}^{N_A} \left[\left(\nu_i+\frac{1}{2}\right)^n -\left(\nu_i-\frac{1}{2}\right)^n \right]^{2}. 
\end{equation}
Plugging this into the R\'{e}nyi entropies formula leads to \eqref{eq:RPEcorr},
which we had already found by analytic continuation of the correlator method in the previous section, but this time without assuming any property for the transition matrix. 

\subsection{Fermionic Theories}\label{sec:fercorr}
In this part we develop a correlator method to calculate pseudo entropy in free fermionic theories. We will later on use this method to study the XY spin chain by mapping it to fermionic models through the Jordan-Wigner transformation. In the following we essentially generalize the methods introduced in \cite{Latorre:2003kg, Vidal:2002rm} for pseudo entropy.

\subsubsection{Correlator Method}
Here we consider generic quadratic Hamiltonians constructed out of fermionic operators $c$ and $c^\dagger$ with $\{c^\dagger_r,c_s\}=\delta_{rs}$, including Hamiltonians with pair creation and annihilation terms. We do not get into the details of any model through this part. Instead, we will focus on developing a correlator method for pseudo entropy. This method is also valid in any spacetime dimensions. 
Similar to the scalar case, we assume the following diagonalized form for the transition matrix
\be
\tau^{1|2}_A=\bigotimes_{k=1}^{N_A}\frac{e^{-\epsilon_k n_{A,k}}}{\left(1+e^{-\epsilon_k}\right)},
\ee
where $n_A$ is the number operator constructed from the decoupled basis in the subregion $A$. The method resembles its scalar counterpart when we use the Majorana representation for the fermionic operators defined as
\be
\tilde{c}_{2m}=(c^\dagger_m+c_m)
\;\;\;\;\;,\;\;\;\;\;
\tilde{c}_{2m-1}=i(c^\dagger_m-c_m)
\ee
where $\{\tilde{c}_{m},\tilde{c}_{n}\}=2\delta_{mn}$. The mode expansion of these operators are given by
\begin{align}
\tilde{c}_{2n}&=\frac{1}{N^{d/2}}
\sum_k
\left(d^\dagger_k\,e^{\frac{2\pi i n k}{N}}+d_k\,e^{-\frac{2\pi i n k}{N}}\right)
\\
\tilde{c}_{2n-1}&=\frac{i}{N^{d/2}}
\sum_k
\left(d^\dagger_k\,e^{\frac{2\pi i n k}{N}}-d_k\,e^{-\frac{2\pi i n k}{N}}\right),
\end{align}
which is similar to the field and conjugate momenta in the scalar case. Using this representation, we should find the spectrum of $\Gamma$ which is a $2N_A\times 2N_A$ matrix defined as
\be\label{eq:cc}
{}_1\langle\mathbf{\tilde{c}}\mathbf{\tilde{c}}\rangle_2=\mathbf{1}+i\,\Gamma
\ee
where the explicit form of the left-hand side of \eqref{eq:cc} is given by 
\begin{align}
\begin{split}
&{}_1\langle\mathbf{\tilde{c}}\mathbf{\tilde{c}}\rangle_2
=
{}_1\langle\begin{pmatrix}\tilde{c}_{2m}\\\tilde{c}_{2m-1}\end{pmatrix} \begin{pmatrix}\tilde{c}_{2n}&\tilde{c}_{2n-1}\end{pmatrix}\rangle_2
\\&=
\begin{pmatrix}\mathbf{C}_1+\mathbf{C}_2+\mathbf{C}_3+\mathbf{C}_4
&
i(-\mathbf{C}_1+\mathbf{C}_2-\mathbf{C}_3+\mathbf{C}_4)
\\
i(-\mathbf{C}_1-\mathbf{C}_2+\mathbf{C}_3+\mathbf{C}_4)
&
-\mathbf{C}_1+\mathbf{C}_2+\mathbf{C}_3-\mathbf{C}_4
\end{pmatrix}
\end{split}
\end{align}
where
\begin{align}\label{eq:fercorr}
\begin{split}
\mathbf{C}_1\equiv{}_1\langle c_mc_n \rangle_2
\;\;\;\;,\;\;\;\;
\mathbf{C}_2\equiv{}_1\langle c_mc^\dagger_n \rangle_2\;,
\\
\mathbf{C}_3\equiv{}_1\langle c^\dagger_mc_n \rangle_2
\;\;\;\;,\;\;\;\;
\mathbf{C}_4\equiv{}_1\langle c^\dagger_mc^\dagger_n \rangle_2\;.
\end{split}
\end{align}
Very similar to the case of the reduced density matrix in calculation of the von Neumann entropy, introduced in \cite{Latorre:2003kg}, the spectrum of this $i\Gamma$ matrix is related to a double copy of the spectrum of the transition matrix as
\be
\nu_i=\pm  \tanh\frac{\epsilon_i}{2}
\ee
and pseudo entropy is given by
\begin{align}
\begin{split}
S(\tau^{1|2}_A)&=-\sum_{i=1}^{N_A}\left(\frac{1+\nu_i}{2}\log\frac{1+\nu_i}{2}+\frac{1-\nu_i}{2}\log\frac{1-\nu_i}{2}\right)
\\
S^{(n)}(\tau^{1|2}_A)&=\frac{1}{n-1}\sum_{i=1}^{N_A}\log\left[\left(\frac{1+\nu_i}{2}\right)^n-\left(\frac{1-\nu_i}{2}\right)^n\right].
\end{split}
\end{align}

\section{Pseudo Entropy in Free Scalar Field Theory}\label{sec:freescalar}
In this section, we study pseudo entropy in Lifshitz scalar field theories. We basically study pseudo entropy between vacuum states corresponding to a pair of parameters denoted by $(m,z)$, which are the mass and the dynamical exponent respectively. The theory is defined such that it is invariant under Lifshitz scaling,
\bea\label{lifscaling}
t\to\lambda^z t\;\;\;\;\;,\;\;\;\;\;\vec{x}\to\lambda \vec{x},
\eea
in the $m\to0$ limit. Such free scalar theories are given by
\bea\label{action}
S=\frac{1}{2}\int dt d\vec{x} \left[\dot{\phi}^2-\sum_{i=1}^{d}(\partial_i^z \phi)^2-m^{2z} \phi^2\right].
\eea
In our analysis, we will consider both $m=0$ and $m>0$ cases. We will always consider integer values for the dynamical exponent $z\geq 1$. 

\subsection*{Lifshitz Harmonic Lattice Model}
Although the method defined in the previous section applies to any spacetime dimension, in this paper we will consider $d=1$. 
To perform concrete calculations, we consider the aforementioned theories regularized on a lattice known as Lifshitz harmonic lattice models whose Hamiltonians are given by \cite{MohammadiMozaffar:2017nri, He:2017wla, MohammadiMozaffar:2017chk, MohammadiMozaffar:2018vmk}

\begin{align}\label{eq:LifH}
\begin{split}
H=\sum_{n=1}^{N}\bigg[&\frac{p_n^2}{2M}+\frac{M m^{2z}}{2}q_n^2\\
&+\frac{K}{2}\left(\sum_{k=0}^z(-1)^{z+k}{{z}\choose{k}} q_{n-1+k}\right)^2 \bigg].
\end{split}
\end{align}
where we set $M=K=1$ without loss of generality. The $z=1$ case is the standard harmonic lattice model, namely the discrete version of Klein-Gordon theory.

The diagonalized Hamiltonian in $d$-dimensions on a square lattice is given by
\bea
H=\sum_{k}\omega_{k}\left(a_{k}a^\dagger_{k}+\frac{1}{2}\right)
\eea
where $k$ is a $d$-dimensional vector and
\bea\label{dispersion}
\omega^2_{k}=m^{2z}+\sum_{i=1}^d\left(2\sin\frac{\pi k_{i}}{N_{x_i}}\right)^{2z}.
\eea

\subsection{Single Intervals}\label{sec:scalarsingle}
In this part, we show several basic properties of pseudo entropy. 
We will discuss (1) some analytic expressions of pseudo entropy which also guarantee the area law behavior, (2) nonpositivity of $\Delta S_{12}\equiv S(\tau^{1|2}_A) - \frac{1}{2}\left(S(\rho_A^1)+S(\rho_A^2)\right)$, (3) saturation behavior for Lifshitz parameters and (4) perturbation of the difference between pseudo entropy and averaged value of entanglement entropy. 

\subsubsection*{Almost Massless Regimes}
First, we consider a periodic system with length $L$ and ``almost massless'' scalar fields with mass $m_i L\ll1$. In the following sections, we take the lattice spacing unit so that we can simply identify the number of sites and length of a given subsystem. Let $\rho^{(i)}_{A}$ be a reduced density matrix for an almost massless scalar field in a single interval $A=[0,\ell]$. It is known that the R\'{e}nyi entropy for $\rho^{(i)}_{A}$ is schematically given by
\begin{align}
S^{(n)}(\rho^{(i)}_{A})=\frac{1}{6}\left(1+\frac{1}{n}\right)&\log\left[\frac{L}{\pi \epsilon}\sin\left(\frac{\pi\ell}{L}\right)\right]\nn\\
&-\frac{1}{2}\log (m_iL)+c^{(n)}_i, \label{eq:nthRenyi}
\end{align}
where $c^{(n)}_i$ is a non-trivial function which includes small position dependence due to the small mass-term. 

\begin{figure}[t]
\includegraphics[scale=.4]{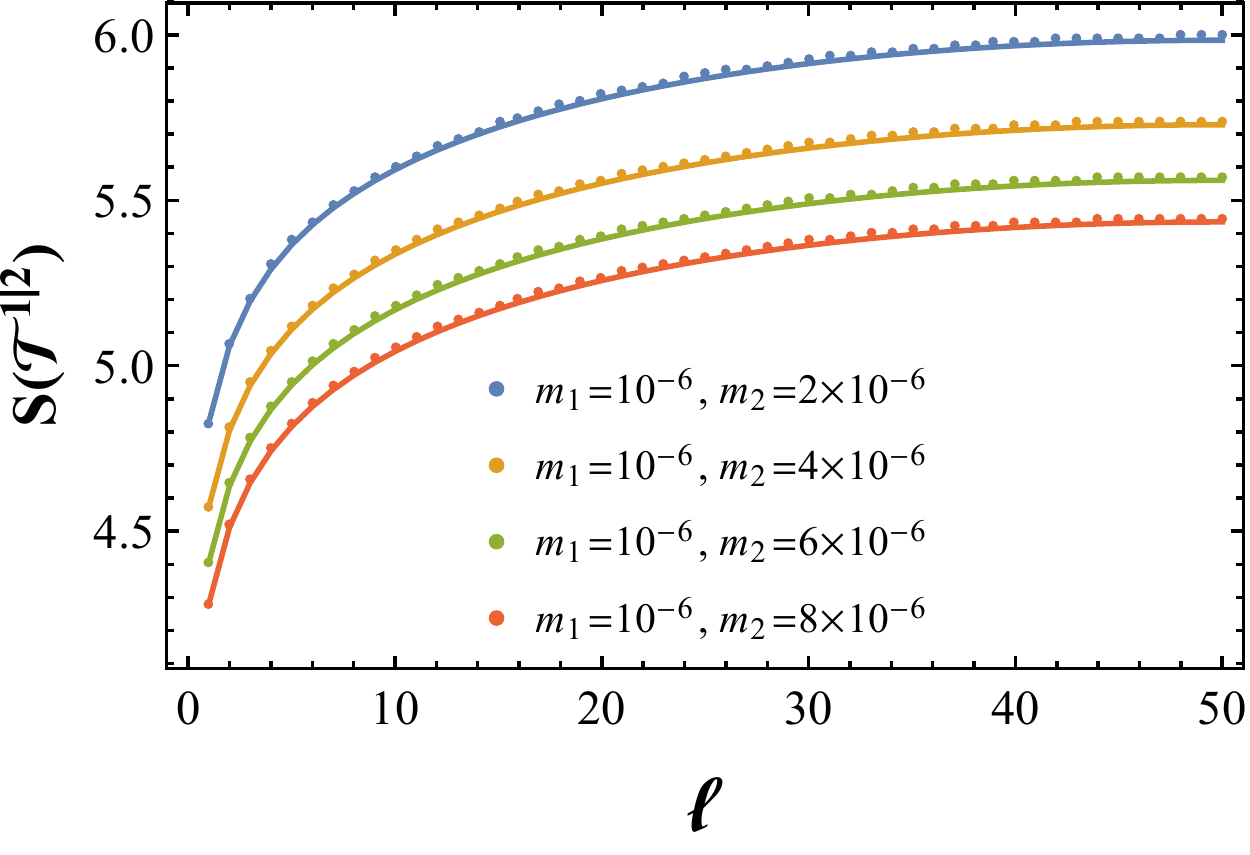}
\caption{$S(\tau^{1|2}_A)$ as a function of the size of the subsystem $\ell$. We set $L=100$ and $z_1=z_2=1$. The solid curves are $(1/3) \log[(L/\pi)\sin(\pi \ell/L)]+\textrm{const.}$, where the constant term is defined in \eqref{eq:PEscalar}.
}
\label{GraphPEvsSizeN200Finite01}
\end{figure}

On the other hand, as for the pseudo entropy for two almost massless scalar fields with mass $m_1$ and $m_2$, we numerically confirmed
\be
S(\tau^{1|2}_A)=\frac{1}{3}\log\left[\frac{L}{\pi \epsilon}\sin\left(\frac{\pi\ell}{L}\right)\right]-\frac{1}{2}\log \left[\frac{m_1+m_2}{2} L\right]+c^{(1)}_{12}, \label{eq:PEscalar}
\ee
where $c^{(n)}_{12}$ is again some mass-dependent function which is less important. See Figure \ref{GraphPEvsSizeN200Finite01}. 

\begin{figure}[t]
 \begin{center}
\resizebox{90mm}{!}{
\includegraphics[scale=.55]{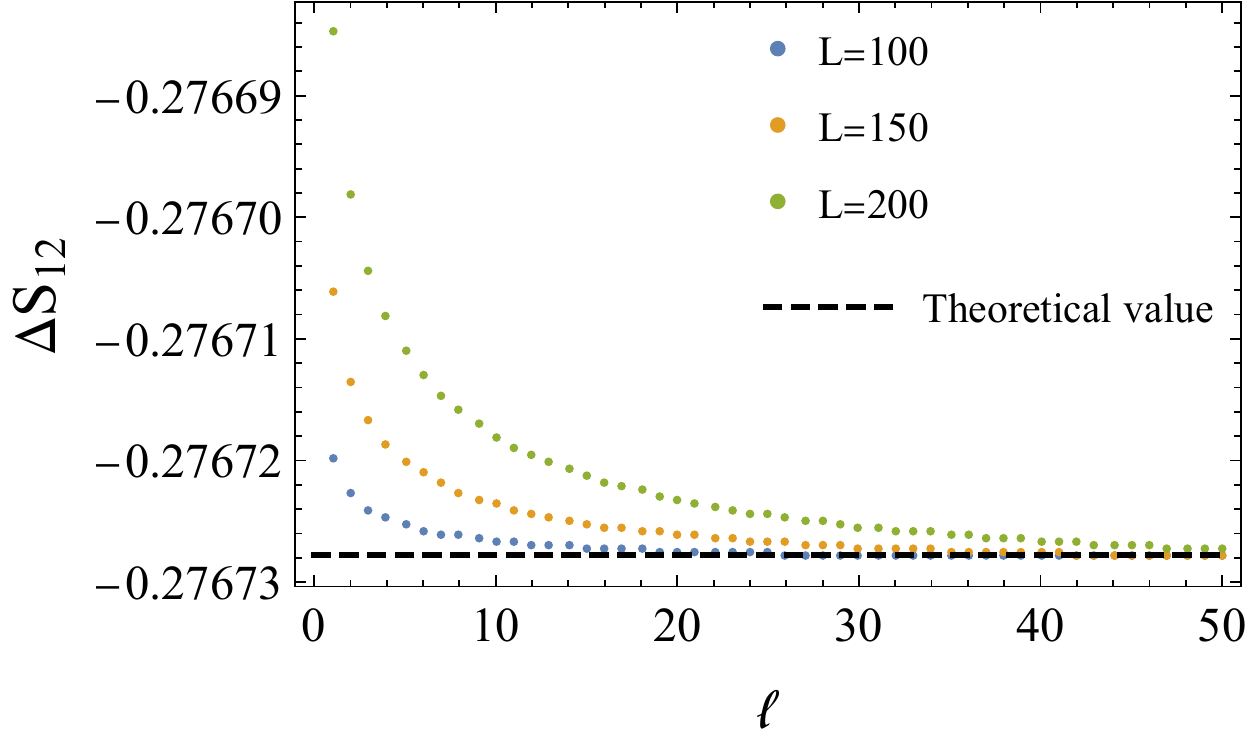}
\includegraphics[scale=.5]{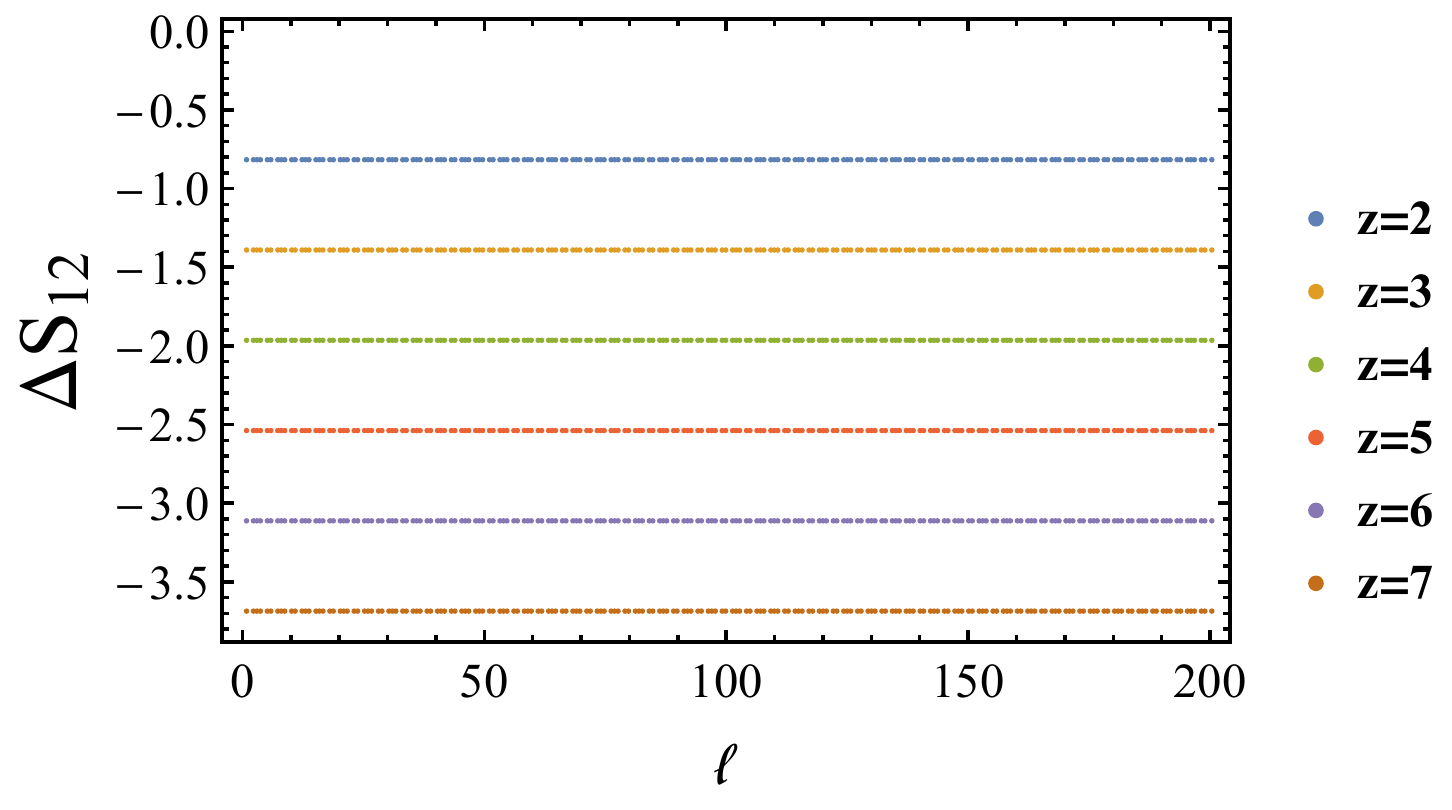}}
 \end{center}
 \caption{Left: $\Delta S_{12}$ for free scalar with $z_1=z_2=1$. Here we set $m_1=10^{-5}$ and $m_2=10^{-4}$. We have small $\ell$ and $L$-dependence but it is negligible up to $3$ or $4$ digit. Therefore, the second term of \eqref{eq:PEscalar} essentially explains this negative value. Right: The $z$-dependence of $\Delta S_{12}$ for $z>1$. Here we set $L=2000, m_1= 10^{-7}, m_2=10^{-8}$ and $z_1=z_2\equiv z$. It can be perfectly explained by the equation \eqref{eq:liflifzz}. Note that we can also produce the same results from smaller total systems and see the perfect agreement in the whole subregions.}\label{fig:barediff}
\end{figure}

In particular, we numerically study the difference between the pseudo entropy and the averaged entanglement entropy,
\be
\Delta S_{12}\equiv S(\tau^{1|2}_A)-\frac{S(\rho^{(1)}_{A})+S(\rho^{(2)}_{A})}{2}.
\ee
Interestingly, it can be well-approximated using the mass terms introduced above as
\begin{align}
\Delta S_{12}
&\simeq-\dfrac{1}{4}\log\left[\dfrac{(m_1+m_2)^2}{4m_1m_2}\right]\leq 0,
\end{align}
which does not depend on the system size. See the left panel of Figure \ref{fig:barediff}. Notice that it is always negative in our almost massless regimes. It means that these mass terms (the second term of \eqref{eq:nthRenyi} and \eqref{eq:PEscalar}) essentially explain the nonpositivity of $\Delta S_{12}$. 

\begin{figure}[ht]
\begin{center}
\includegraphics[scale=0.4]{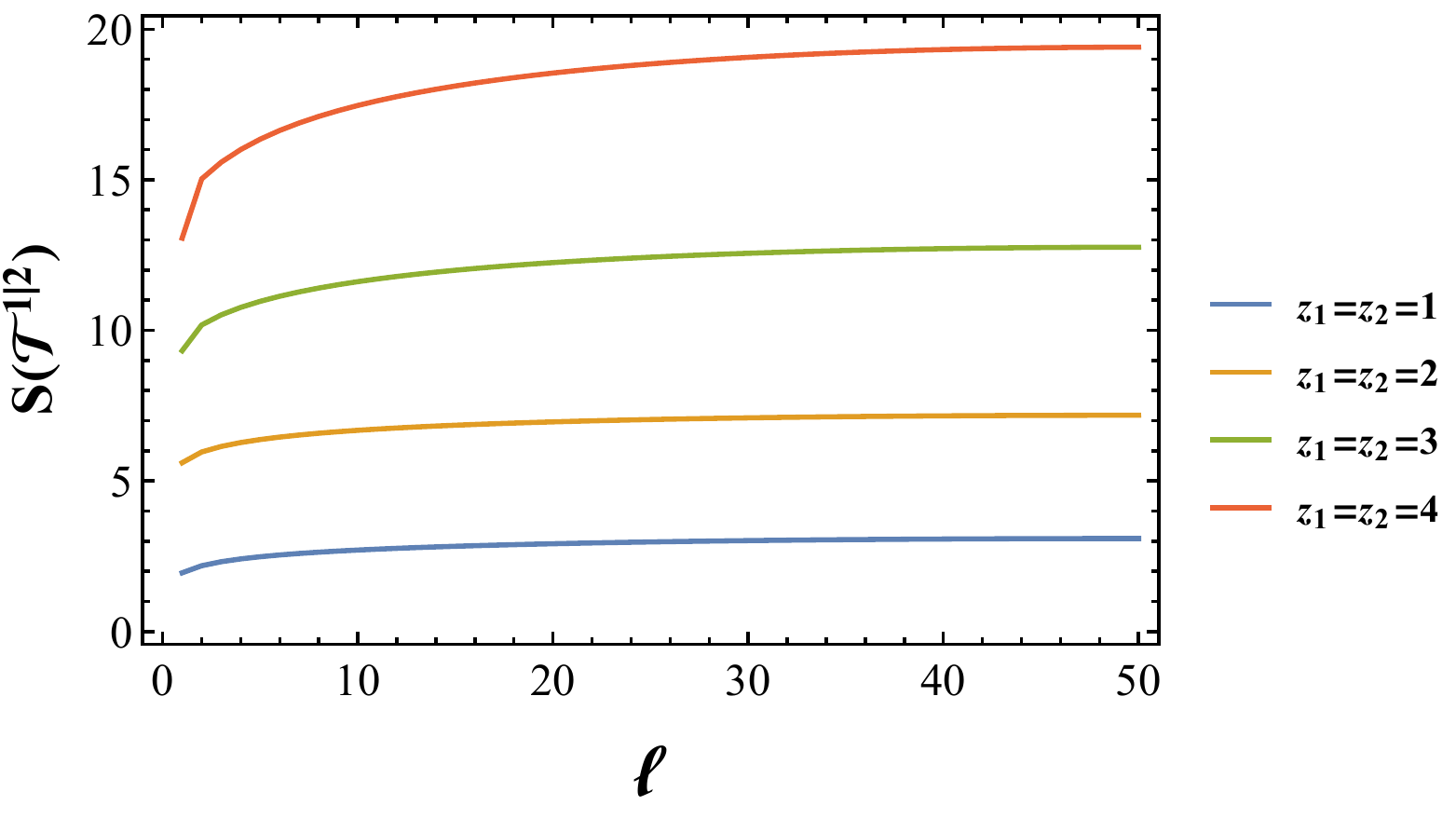}
\includegraphics[scale=0.4]{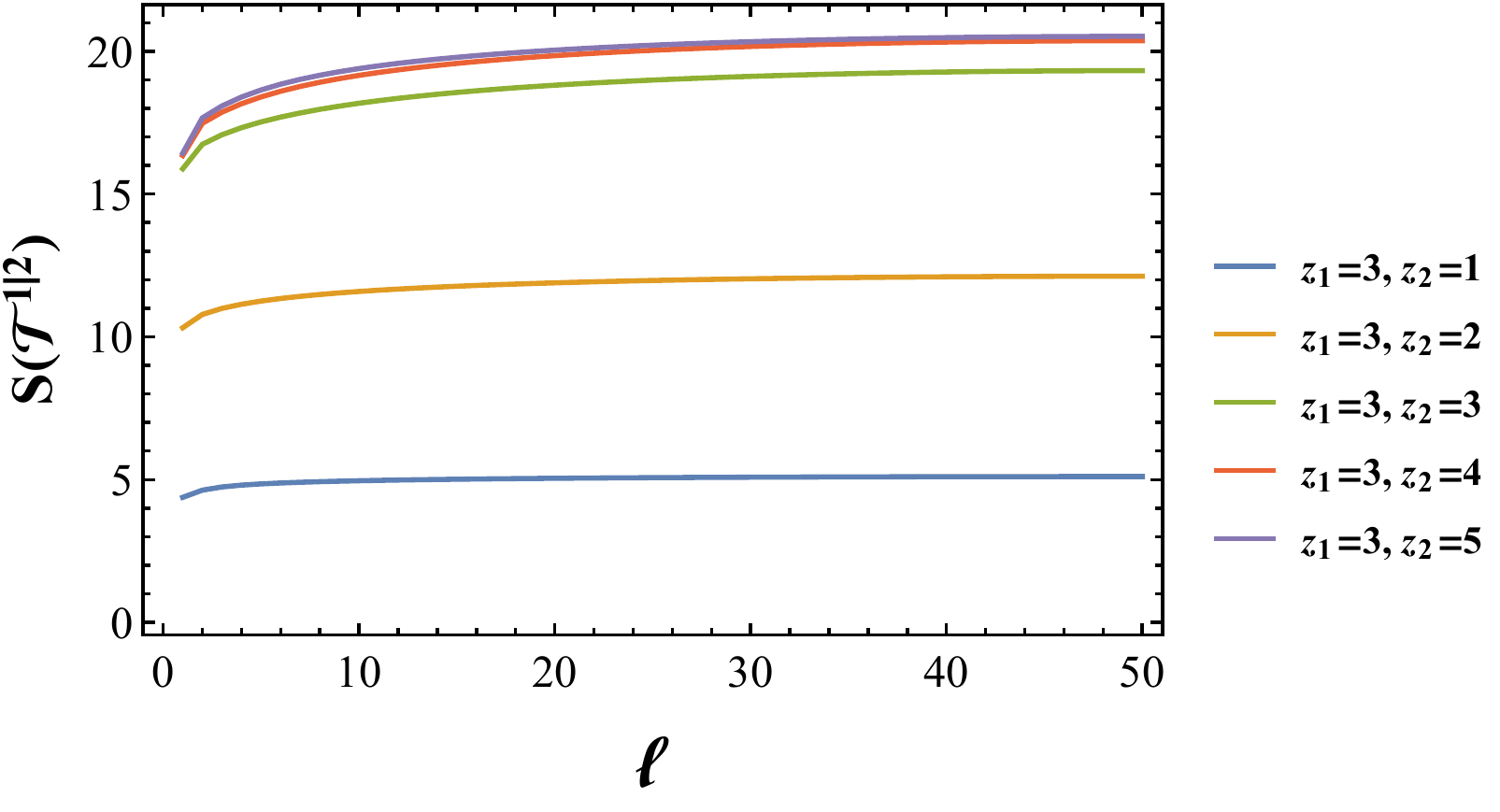}
\end{center}
\caption{
The upper plot shows the pseudo entropy as a function of the subsystem size $\ell$ when we chose  $m_1=10^{-3}$ and $m_2=10^{-5}$
for various values of $z_1=z_2$. The lower plot shows the pseudo entropy when we set $z_1=3$ and $m_1=m_2=10^{-5}$. We chose the total system $L=100$.
}
\label{fig:ccresultas}
\end{figure}
We repeat the same analysis for $z_1=z_2\equiv z>1$ cases. See Figure \ref{fig:ccresultas}. We can also find how $\Delta S_{12}$ depends on $m_1, m_2$ and $z$. 
We have numerically confirmed
\be
\Delta S_{12}\simeq -\dfrac{1}{4}\log\left[\dfrac{(m_1^z+m_2^z)^2}{4(m_1m_2)^z}\right]\leq 0. \label{eq:liflifzz}
\ee
In other words, the correct expression is given by replacing $m_iL$ to $(m_iL)^z$. 
See the right panel of Figure \ref{fig:barediff}. We stress that the $z$-dependence does not show up as an overall factor. A remarkable thing is that the agreement is now exact up to the current numerical precision. This would suggest that the mass-corrections for $z>1$ are quite different from the $z=1$ one. 
\subsubsection*{Dynamical exponent dependence: Saturation Behavior}
\begin{figure}[t]
\begin{center}
\resizebox{60mm}{!}{\includegraphics[scale=0.4]{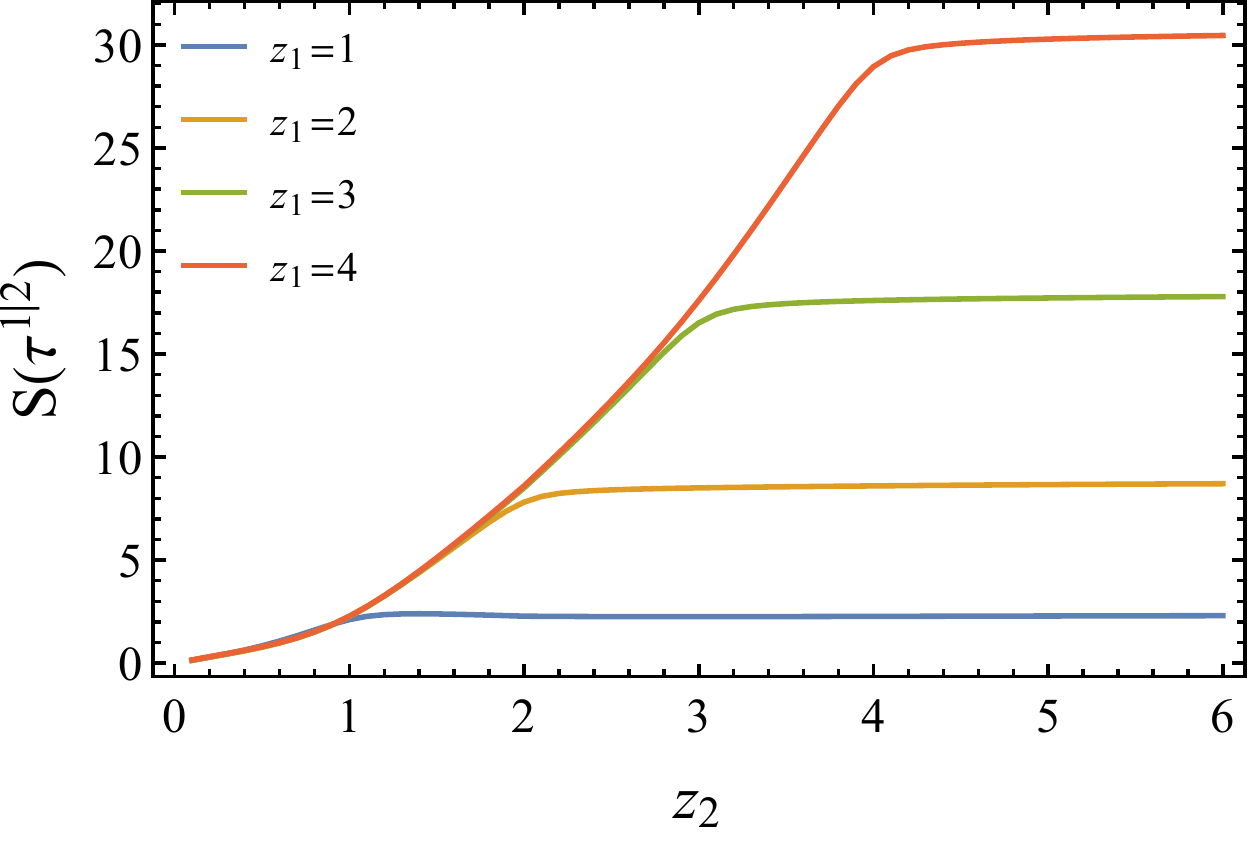}
}
\end{center}
\caption{Dynamical exponent dependence of the pseudo entropy. We set $m_1=m_2=10^{-5}$ and $\ell=30$ on an infinite lattice. The pseudo entropy saturates at the value of the larger $z$.}
\label{fig:ccresultss}
\end{figure}

So far, we have discussed the case with $z_1=z_2$. Next, we study the dynamical exponent dependence of pseudo entropy for more general $z_1$ and $z_2$. See Figure \ref{fig:ccresultss}. A remarkable feature of pseudo entropy in this case is that as we increase one of the two dynamical exponents (and fix all other parameters), the pseudo entropy quickly converges to a fixed value. We call this as saturation behavior of pseudo entropy. We can also understand this behavior analytically by using the periodic subsystems in section  \ref{subsec:periodic_sub}.

\subsubsection*{Massive Regimes}
In this section, we study the pseudo entropy for massive scalar fields. In contrast to the almost massless regimes, our result here is based on an analytical approach. For the detail of the derivation, please refer to the supplemental material in \cite{supplimentary}.
We obtain a mass-correction formula of the pseudo entropy for scalar fields as
\begin{align}
S(\tau^{m_1,m_2}_{A_\ell})-S(\tau^{m_1,m_2}_{A_{\ell_0}})&=f^{m_1,m_2}(\ell)-f^{m_1,m_2}(\ell_0),\label{eq:regPE}
\end{align}
where
\begin{align}
f^{m_1,m_2}(\ell)&=\frac{1}{3}\log\left[\frac{L}{\epsilon \pi}\sin\left(\frac{\pi \ell}{L}\right)\right]\nn\\
&+\frac{1}{2}\log\left[-\frac{m_1^2\log[m_1\ell]-m_2^2\log[m_2\ell]}{m_1^2-m_2^2}\right]. 
\end{align}
Here $S(\tau^{m_1,m_2}_{A_\ell})$ gives the pseudo entropy for a single interval $A_\ell=[0,\ell]$ between two vacuum states with different mass $m_1$ and $m_2$. We also subtract the value at $\ell_0$ as a reference point in order to get rid of irrelevant contributions. Note that this formula is a leading order approximation. Namely, it is valid only for small interval size such that $m_1\ell, m_2\ell\ll1$. Under $ m_1\rightarrow m_2$ limit, it reduces to the famous result for the entanglement entropy for a massive scalar field \cite{Casini:2005zv}. It is also worth noting that the $f^{m_1,m_2}(\ell)$ is symmetric, {\it i.e.} $f^{m_1,m_2}(\ell)=f^{m_2,m_1}(\ell)$ which is also guaranteed by our numerical results. 

\begin{figure}[t]
 \begin{center}
  \resizebox{80mm}{!}{
 \includegraphics{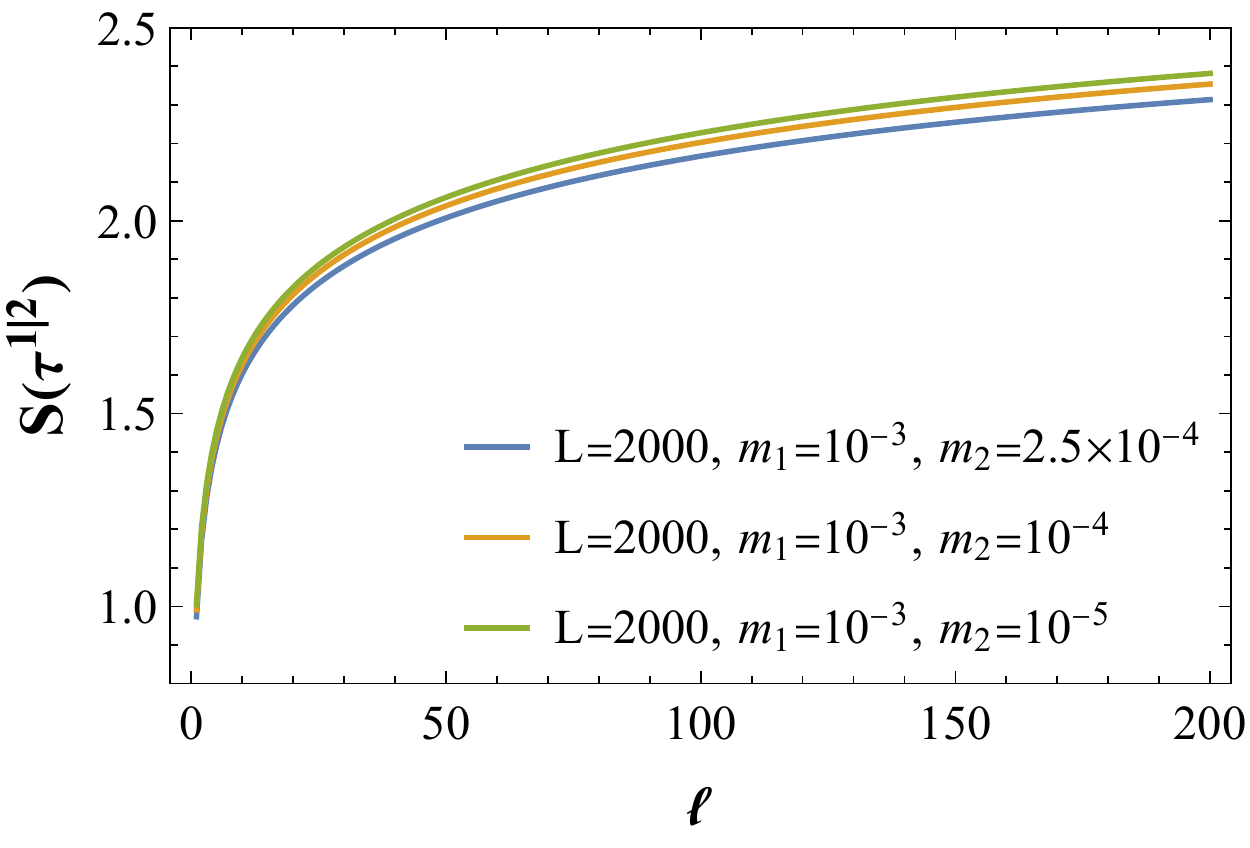}
 \includegraphics{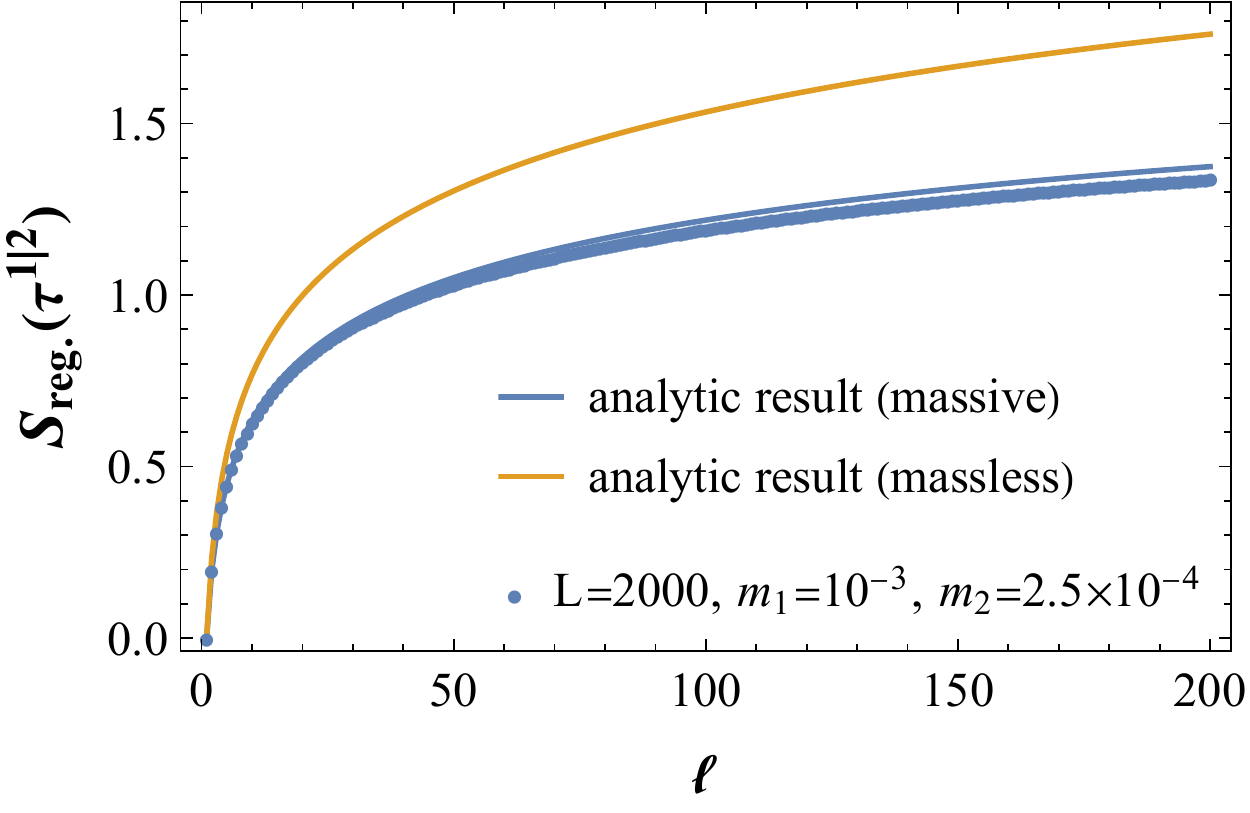}}
 \resizebox{80mm}{!}{
 \includegraphics{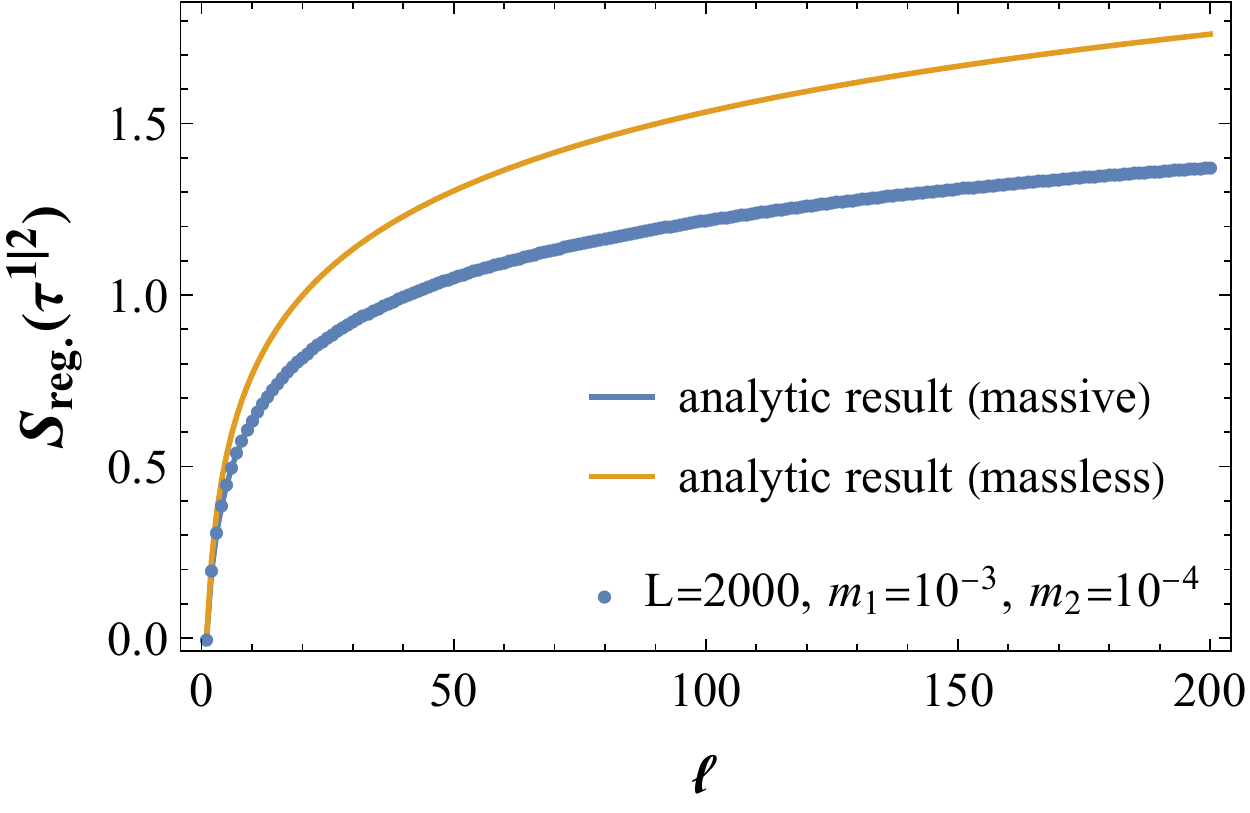}
 \includegraphics{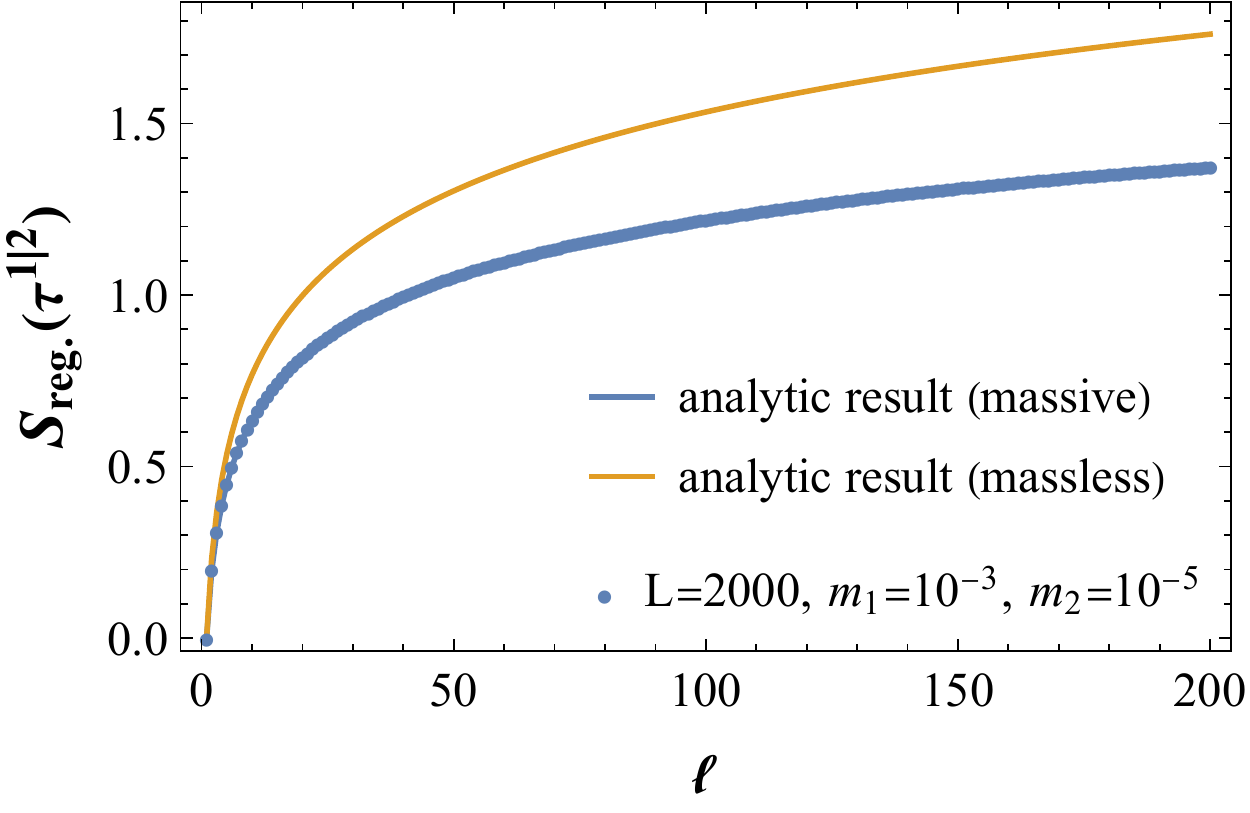}}
 \end{center}
 \caption{The pseudo entropy and regularized pseudo entropy for fixed $m_1=10^{-3}$ with various mass $m_2$. As a reference, we also plot the entanglement entropy for vacuum with $c=1$ (orange curve). Note that this formula is valid only in the small subsystem such that $m_i\ell\ll1$. Out of this regime, as we can see from the right-top figure, there is a small deviation. }
 \label{fig:dep_mass}
\end{figure}

For convenience, we define a  regularized pseudo entropy as
\begin{equation}
    S_{\textrm{reg.}}(\tau^{m_1,m_2}_{A_\ell})=S(\tau^{m_1,m_2}_{A_\ell})-S(\tau^{m_1,m_2}_{A_{\epsilon}}),
\end{equation}
which corresponds to the left-hand side of \eqref{eq:regPE} with $\ell_0=\epsilon$. In Figure \ref{fig:dep_mass}, we plotted pseudo entropy and the regularized pseudo entropy for fixed $m_1$ with various $m_2$. These figures are numerically consistent with the above mass-corrected formula. 

\subsubsection*{Perturbation of \texorpdfstring{$\Delta S_{12}$}{the difference}}
\begin{figure}[h!]
\begin{center}
\resizebox{80mm}{!}{
\includegraphics[scale=0.5]{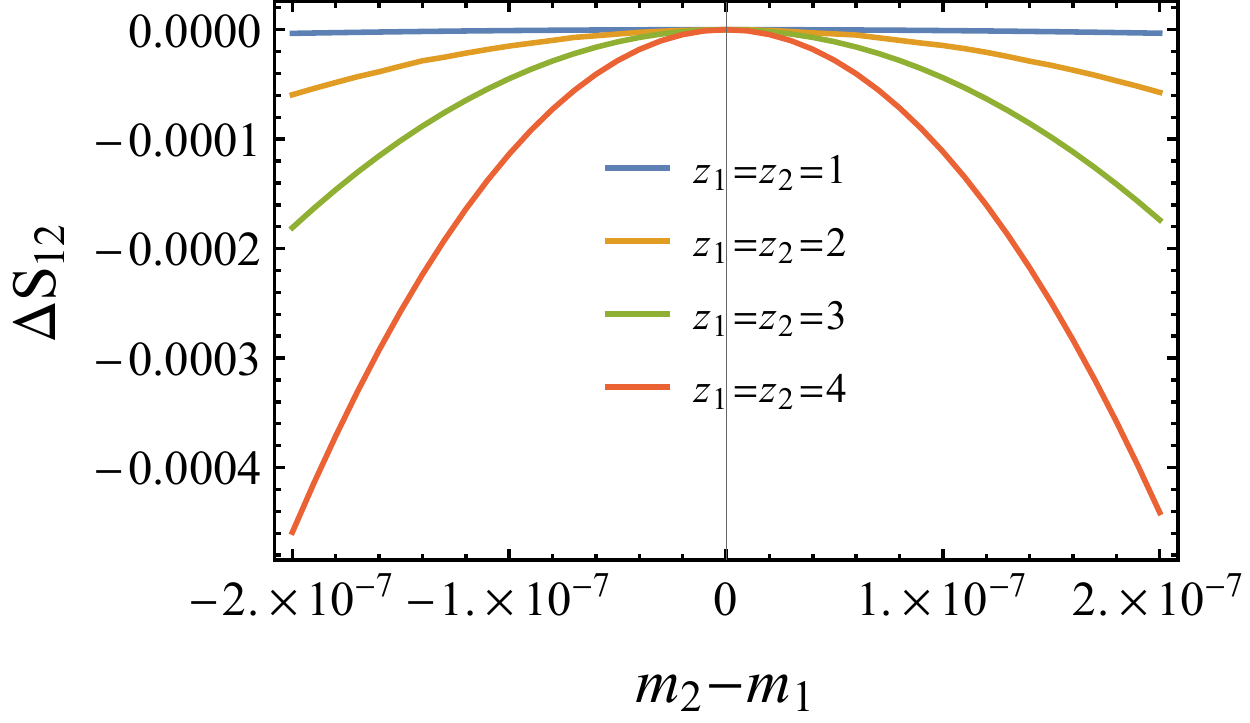}
\includegraphics[scale=0.45]{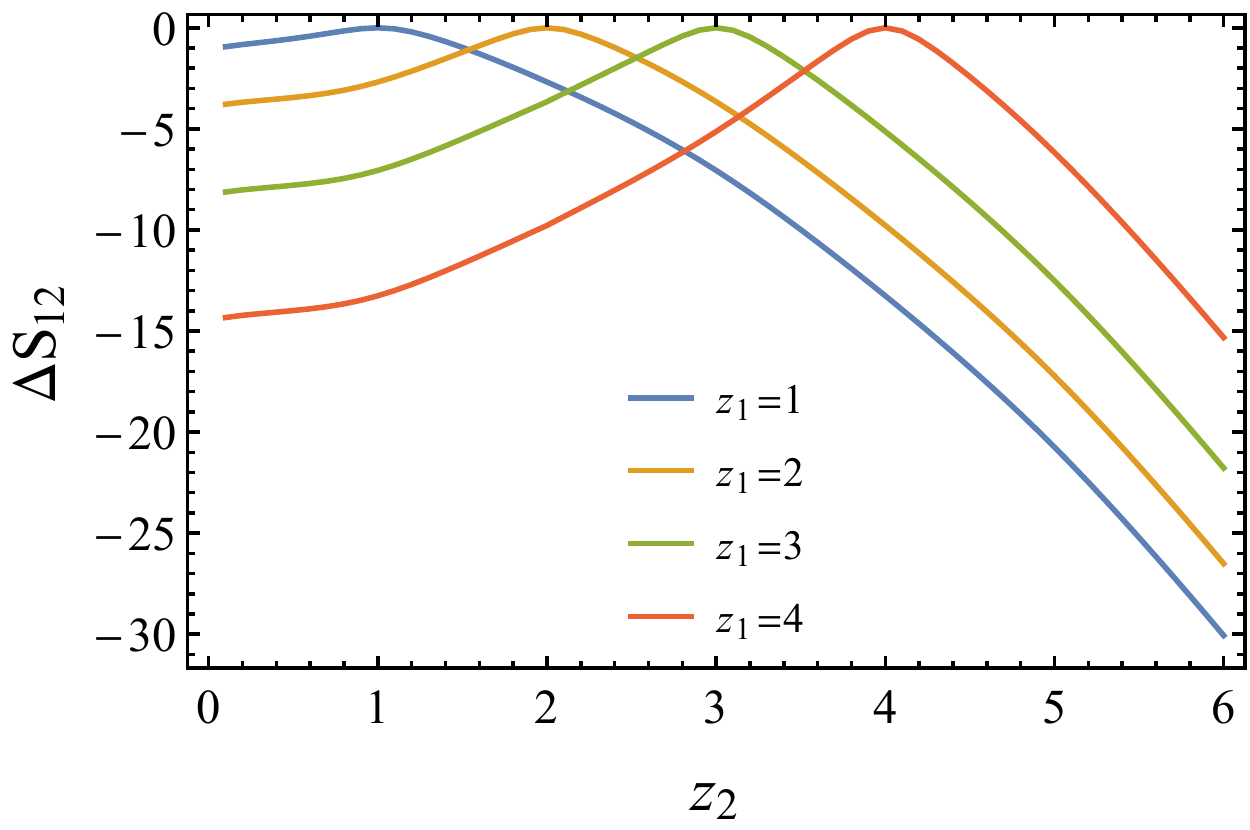}}
\end{center}
\caption{Mass and dynamical exponent dependence of $\Delta S_{12}$. Here we set $\ell=30$ on an infinite lattice. Left: we also set $m_1=10^{-5}$. Right: $m_1=m_2=10^{-5}$. These figures ensure the quadratic behavior of $\Delta S_{12}$ when the parameter is small. }
\label{fig:ccresults}
\end{figure}

It is well-known that the entanglement entropy satisfies the first-law like relation under small-perturbation\cite{Blanco:2013joa,Bhattacharya:2012mi}. Here we treat mass and dynamical exponents as perturbation parameters and numerically study a difference $\Delta S_{12}$. 
The Figure \ref{fig:ccresults} shows the universal quadratic behavior with respect to the small parameters $\delta=m_2-m_1$ or $z_2-z_1$, 
\be
\Delta S_{12}\simeq -\gamma \delta^2<0,
\ee
where $\gamma$ is a positive constant which depends on the other fixed parameters. This behavior is true only when $\delta$ is small enough. These results would suggest that $\Delta S_{12}$ is always negative under the small perturbation.

\subsection{Disconnected Subsystems}\label{subsec:ssa}
It is interesting to ask if the pseudo entropy satisfies the subadditivity,
\begin{align}
S(\tau^{1|2}_{A})+S(\tau^{1|2}_{B})&\overset{?}{\geq} S(\tau^{1|2}_{AB}),
\end{align}
and strong subadditivity,
\begin{align}
S(\tau^{1|2}_{AB})+S(\tau^{1|2}_{BC})&\overset{?}{\geq} S(\tau^{1|2}_{ABC})+S(\tau^{1|2}_{B}),
\end{align}
in the present setup. To this end, we continue our study of pseudo entropy to two-interval cases as this is the simplest setup to check these inequalities. 

For the later convenience, we define {\it pseudo mutual information} (PMI),
\be
I^{1|2}(A:B)\equiv S(\tau^{1|2}_{A})+S(\tau^{1|2}_{B})-S(\tau^{1|2}_{AB}). \label{eq:pmi_amassless}
\ee
In the language of PMI, the subadditivity of pseudo entropy is equivalent to the positivity of PMI,
\begin{align}
I^{1|2}(A:B)&\overset{?}{\geq} 0,
\end{align}
and the strong subadditivity of pseudo entropy is to the monotonicity of PMI,
\begin{align}
I^{1|2}(A:BC)&\overset{?}{\geq} I^{1|2}(A:B).
\end{align}
Besides its convenience, it is intriguing to study the PMI itself. 

In summary, we have numerically observed that the subadditivity of pseudo entropy (the positivity of PMI) is always satisfied, whereas the strong subadditivity of pseudo entropy (the monotonicity of PMI) can be violated in general. 
\subsubsection*{Subadditivity}
\begin{figure}[ht]
 \begin{center}
 \resizebox{35mm}{!}{
 \includegraphics{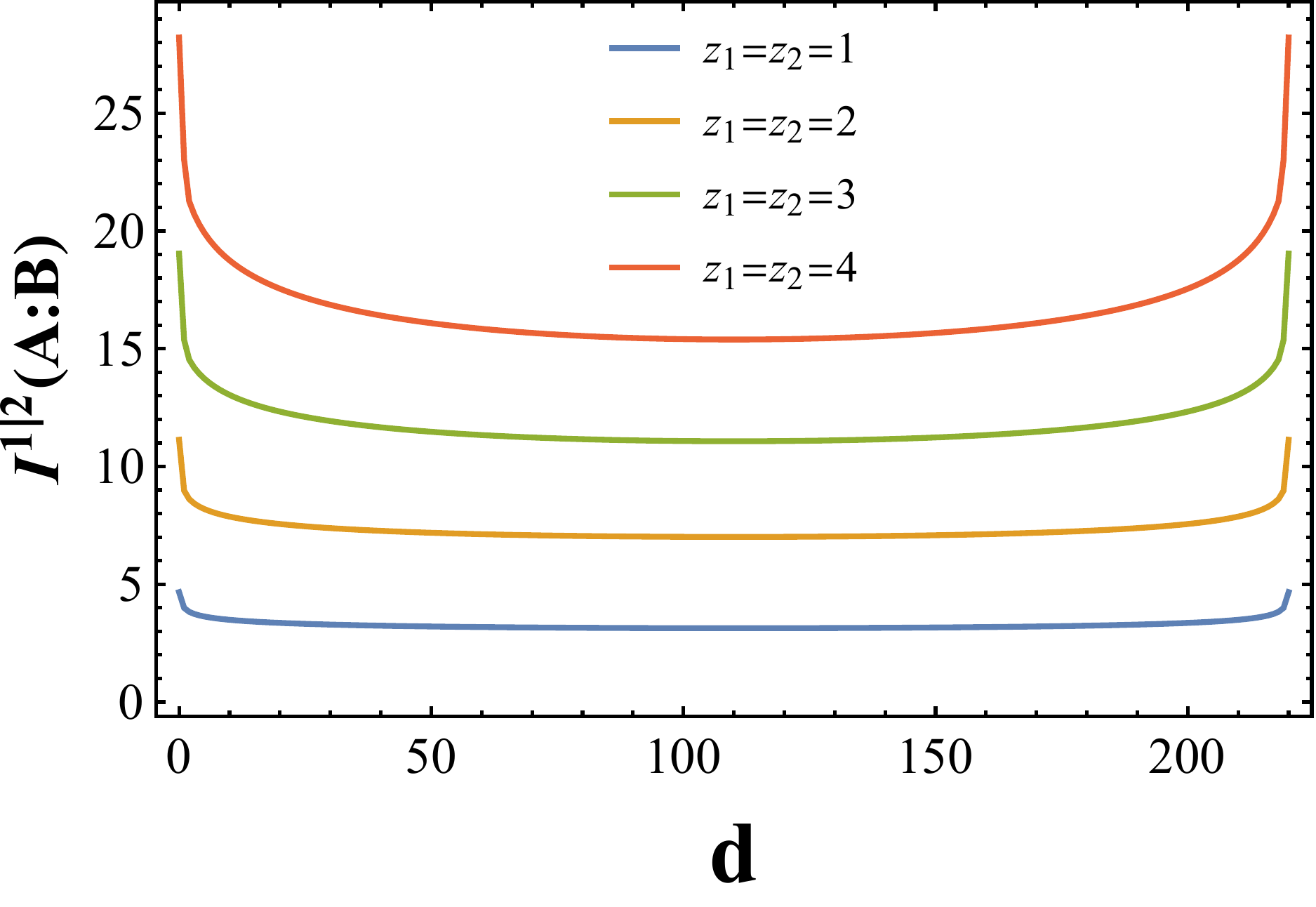}}\hspace{5mm}\resizebox{35mm}{!}{
 \includegraphics{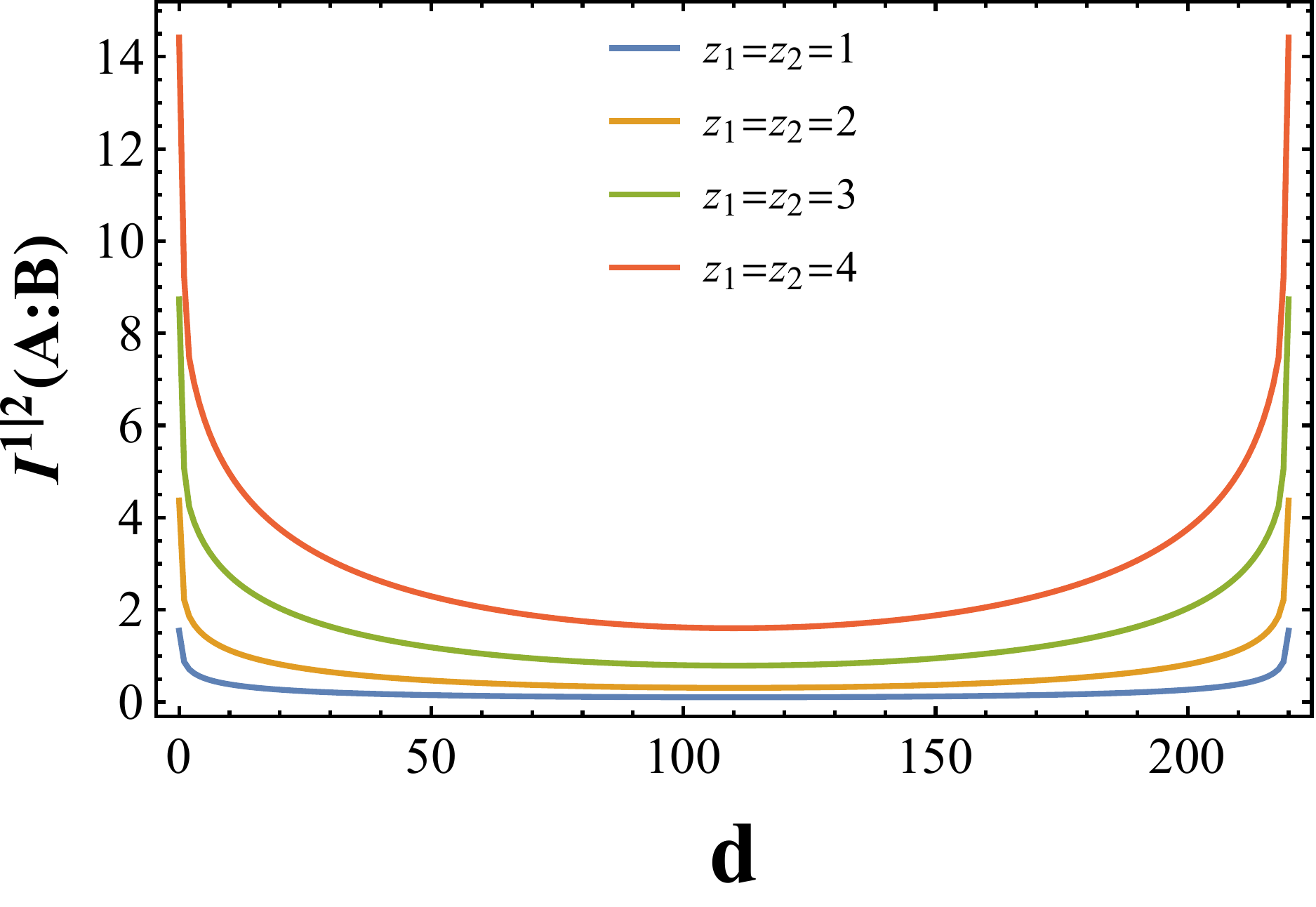}}
 \end{center}
 \caption{Plot for PMI as a function of distance $d$ between two subsystems $A$ and $B$. Left: Almost massless regime. Here we set $L=300, \ell_A=\ell_B=40, m_1=10^{-6}, m_2=10^{-5}$ and $z_1=z_2\equiv z$. Right: Massive regime. Here we set $m_1=10^{-3}, m_2=10^{-2}$ and other parameters are the same as the almost massless ones.}\label{fig:PMImassless}
\end{figure}

One can numerically confirm the positivity of PMI,
\be
I^{1|2}(A:B)\geq 0.
\ee
See Figure \ref{fig:PMImassless}. Although these figures only show limited examples for $z_1=z_2$ but $m_1\neq m_2$ cases, we can see the agreement for more general setup. It suggests that the subadditivity of pseudo entropy hold in the free scalar theories. 

\subsubsection*{\texorpdfstring{$\Delta S_{12}$}{The difference} for two intervals}
\begin{figure}[t]
 \begin{center}
 \resizebox{60mm}{!}{
 \includegraphics{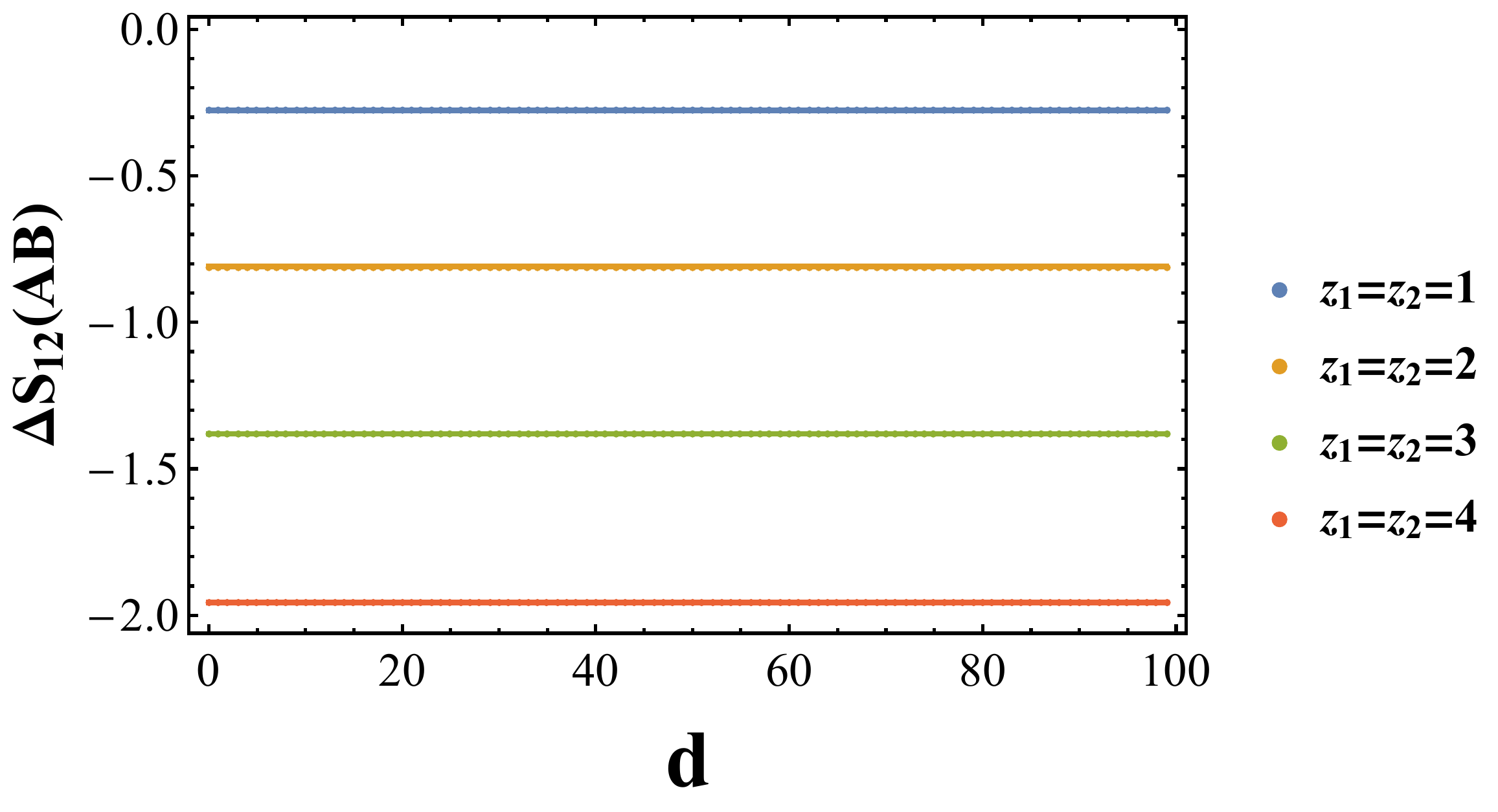}}
 \end{center}
 \caption{The $z$-dependence of $\Delta S_{12}$ for two-interval cases. Here we set $L=300, \ell_A=\ell_B=40, m_1=10^{-6}, m_2=10^{-5}$ and $z_1=z_2\equiv z$. We can see perfect agreement with analytic results. }\label{fig:diff2}
\end{figure}

\begin{figure}[ht]
 \begin{center}
 \resizebox{60mm}{!}{
 \includegraphics{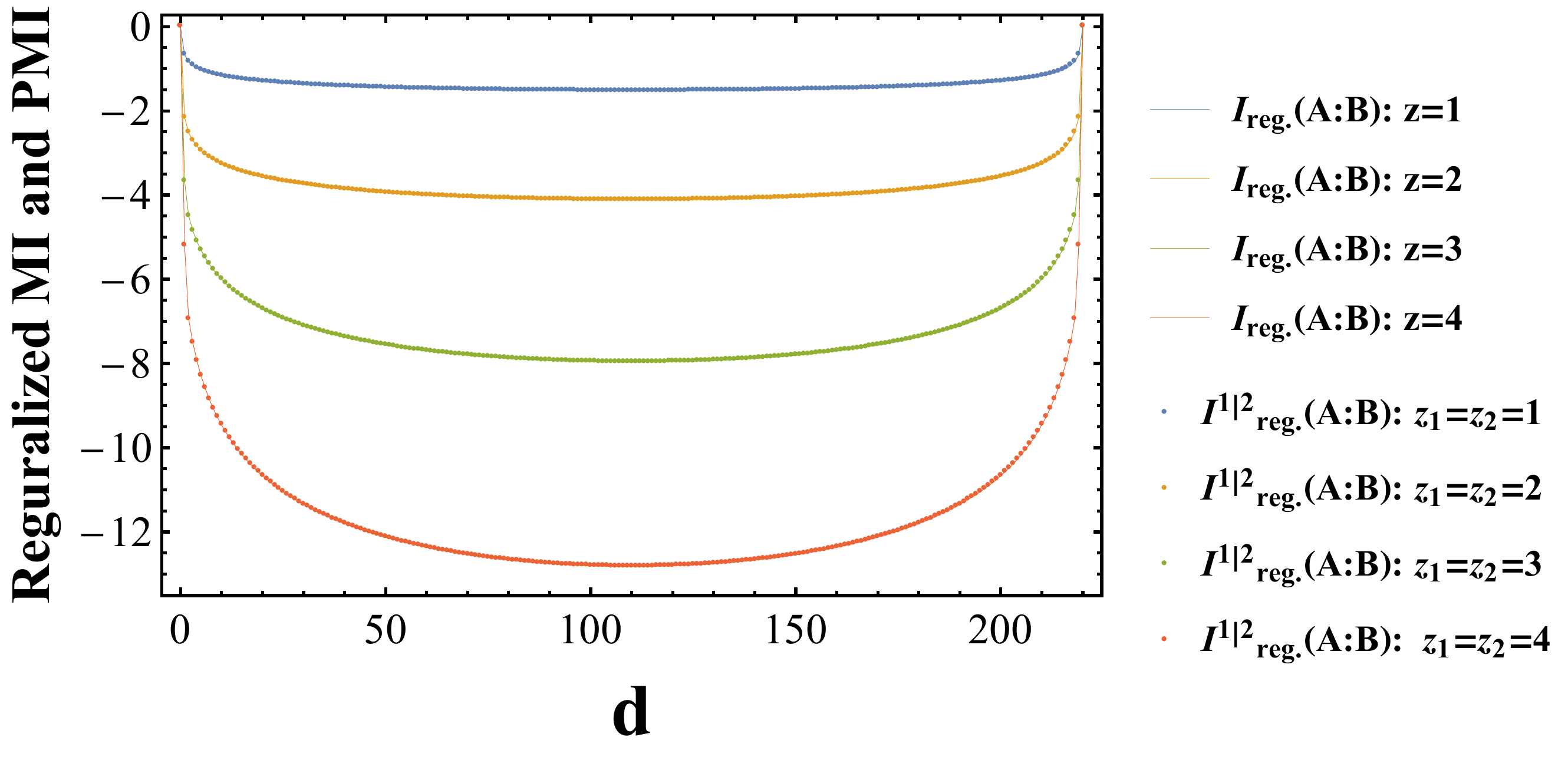}}
 \end{center}
 \caption{Plot for MI and PMI as a function of distance $d$ between two subsystems $A$ and $B$. For clarity, we plot reguralized version, $I_{\textrm{reg.}}=I(d)-I(0)$ and $I^{1|2}_{\textrm{reg.}}=I^{1|2}(d)-I^{1|2}(0)$. Here we set $L=300, \ell_A=\ell_B=40, m_1=10^{-6}, m_2=10^{-5}$ and $z_1=z_2\equiv z$. This plot suggests that $I^{1|2}$ and $I$ are the same up to the constant term as described in \eqref{eq:pmi_amassless}.}\label{fig:PMImassless2}
\end{figure}

Similar to the single interval case, we can introduce the difference between the pseudo entropy and averaged value of the entanglement entropy,
\be
\Delta S_{12}=S(\tau^{1|2}_{AB})-\dfrac{S(\rho^{1}_{AB})+S(\rho^{2}_{AB})}{2}.
\ee
Let us focus on $z_1=z_2$ but $m_1\neq m_2$ cases as we have observed the simple analytical expression in the single interval cases. 
If one focuses on the almost massless regime, we can indeed observe the same relation,
\be
\Delta S_{12}=-\dfrac{1}{4}\log\left[\dfrac{(m_1^z+m_2^z)^2}{4(m_1m_2)^z}\right]. \label{eq:delta_massless}
\ee
See Figure \ref{fig:diff2}. It means that the PMI also follows a similar simple formula,
\begin{align}
I^{1|2}(A:B)&=I(A:B)-\dfrac{1}{2}\log\left[\dfrac{(m_1^z+m_2^z)}{2}\right],\label{eq:pmi_amassless2}
\end{align}
where $I(A:B)$ represents the ordinary mutual information for massless scalar fields with dynamical exponents $z_1=z_2=z$. See Figure \ref{fig:PMImassless2}.

\begin{figure}[ht]
 \begin{center}
 \resizebox{40mm}{!}{
 \includegraphics{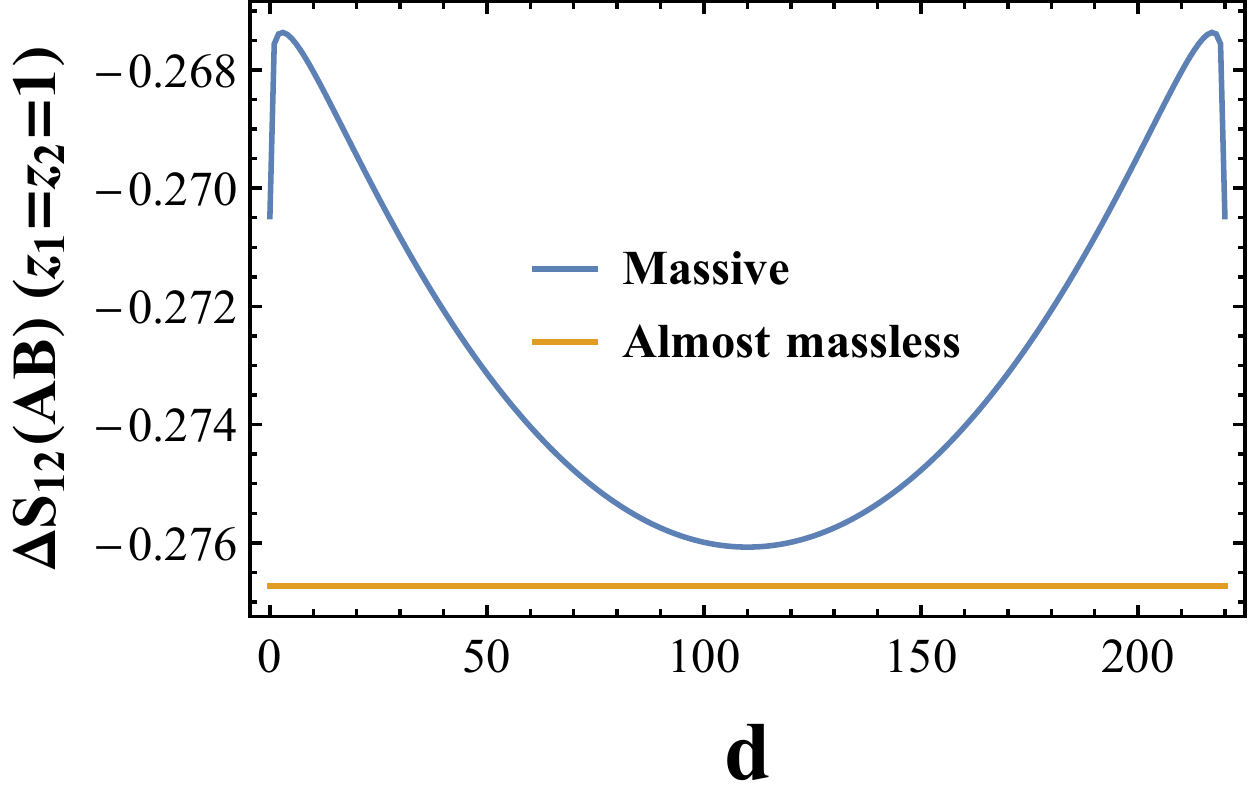}}\resizebox{40mm}{!}{
 \includegraphics{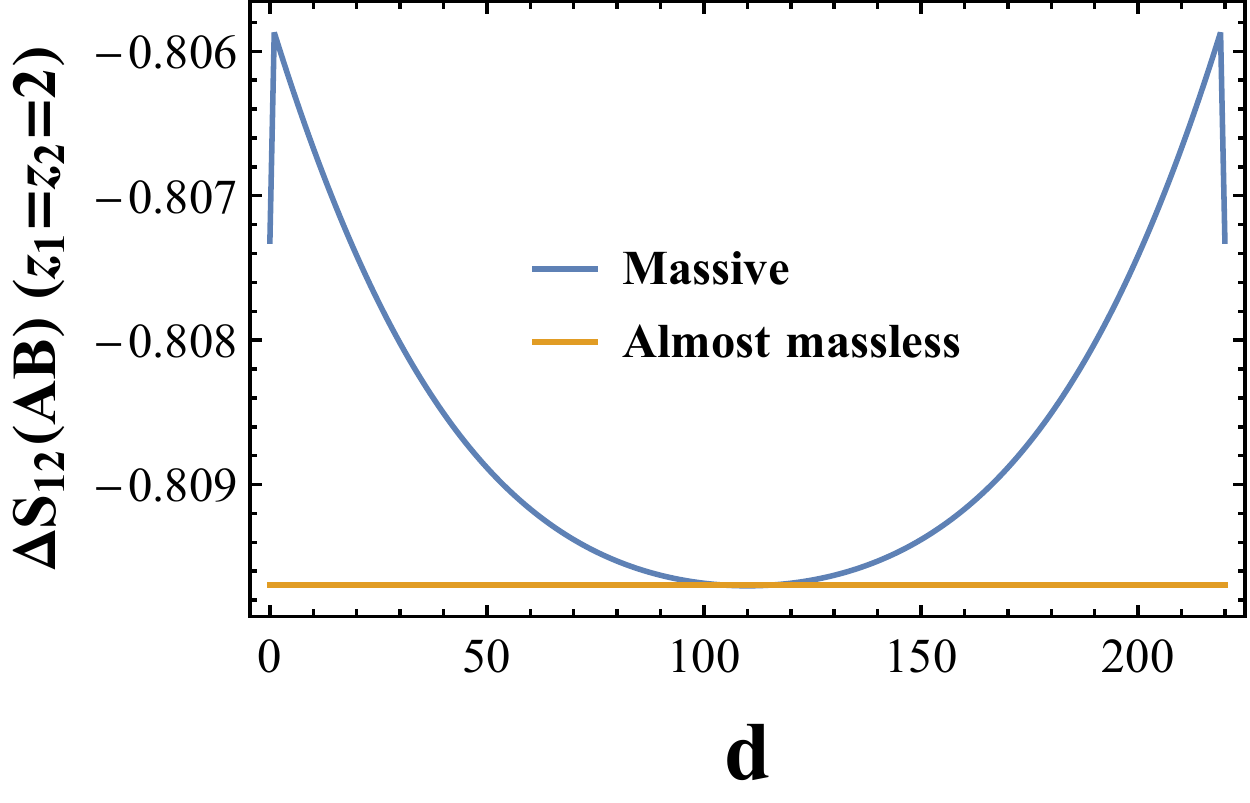}}\\
 \resizebox{40mm}{!}{
 \includegraphics{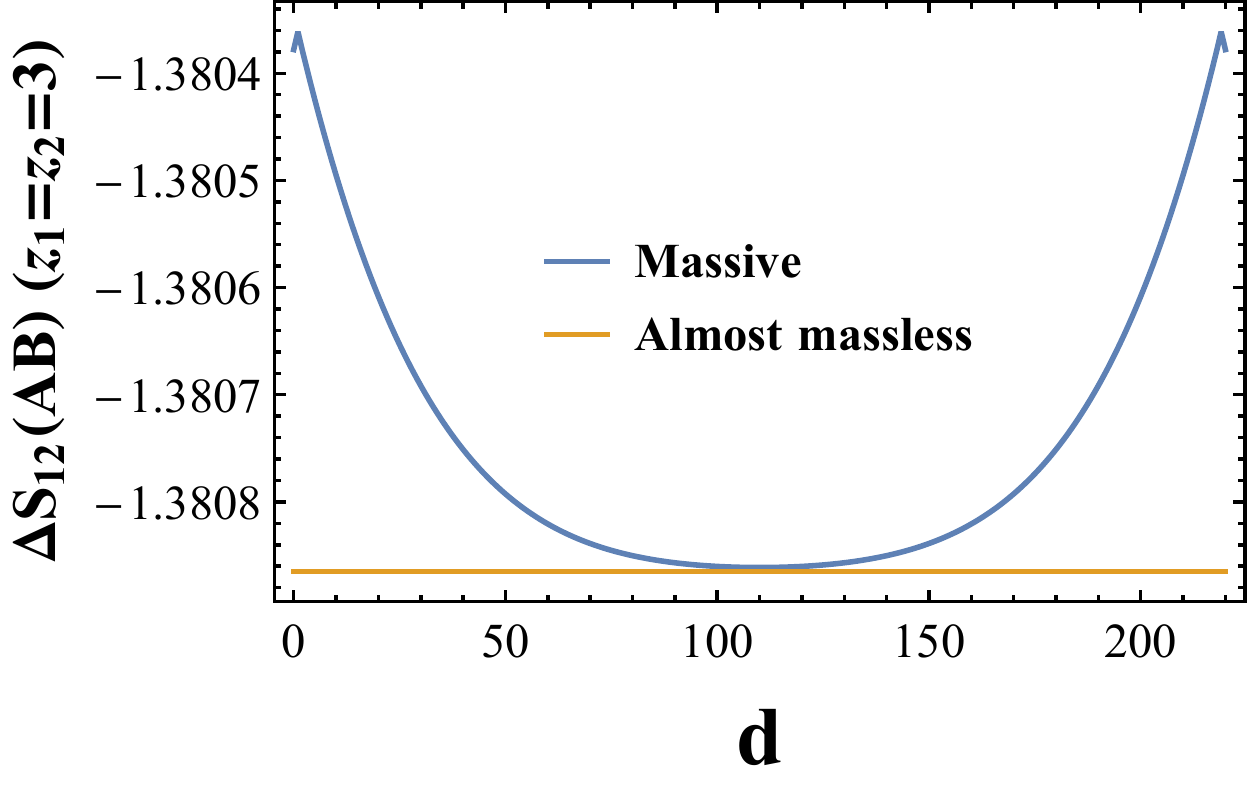}}\resizebox{40mm}{!}{
 \includegraphics{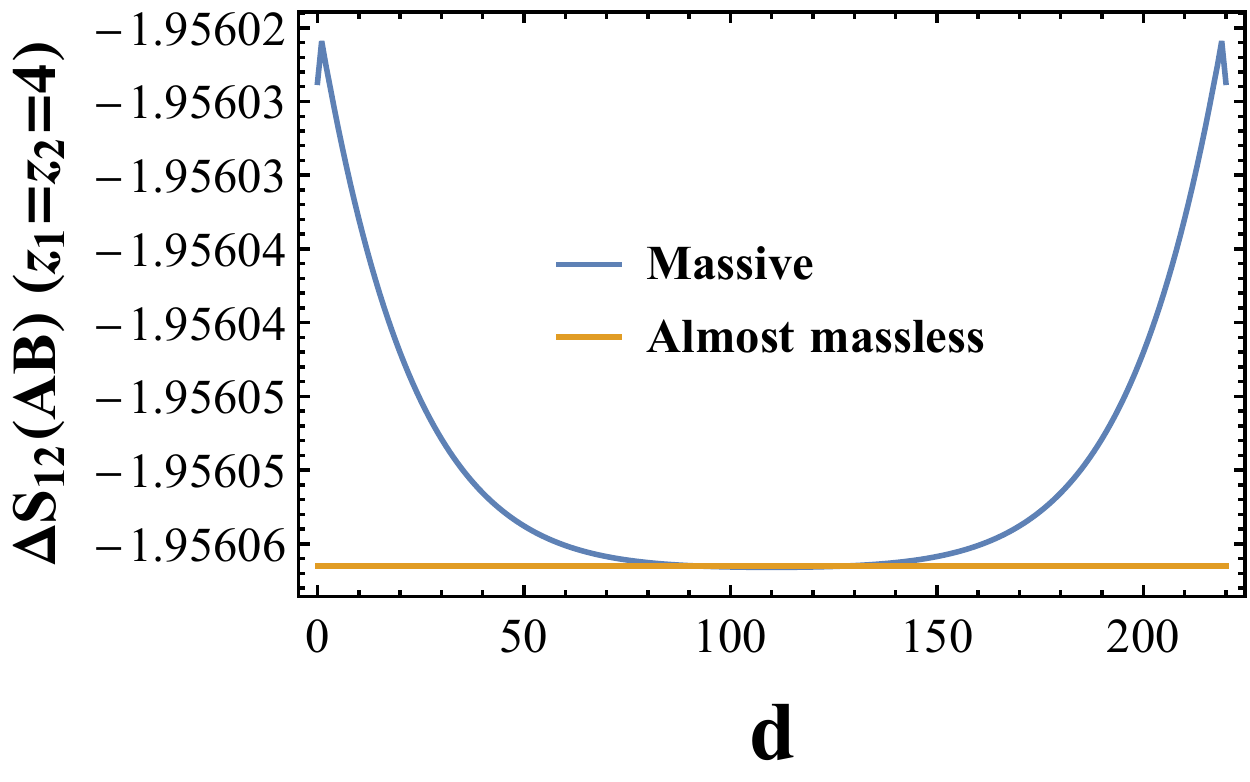}}
 \end{center}
 \caption{The $z$-dependence of the difference between pseudo entropy and averaged entanglement entropies, $\Delta S_{12}$ for two-interval cases. Here we set $L=300, \ell_A=\ell_B=40, m_1=10^{-3}, m_2=10^{-2}$ and $z_1=z_2\equiv z$. For $z>1$ case, the minimum value attaches the one for almost massless regime \eqref{eq:delta_massless}. }\label{fig:zmassive}
\end{figure}

For the massive regime, see Figure \ref{fig:zmassive}. In this regime, we observed
\be
 -\dfrac{1}{4}\log\left[\dfrac{(m_i^z+m_j^z)^2}{4(m_im_j)^z}\right] 
 \lesssim
\Delta S_{12} <0.
\ee

\subsubsection*{Strong Subadditivity}

\begin{figure}[ht]
 \begin{center}
 \resizebox{70mm}{!}{
 \includegraphics{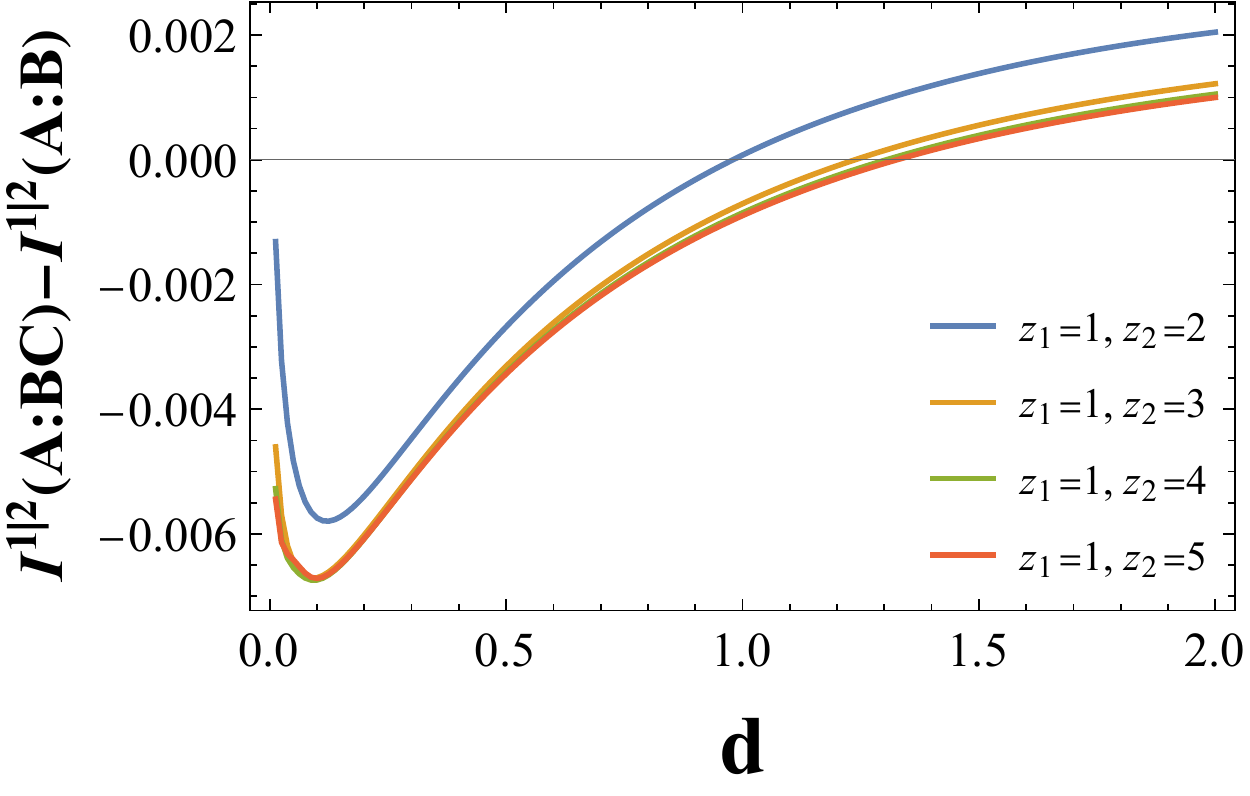}}
 \end{center}
 \caption{The breaking of the monotonicity of PMI or strong subadditivity of PE. Here we set $\ell_A=80, \ell_B=\ell_C=40$ and $m_1=m_2=10^{-4}$ and take infinite size limit. We can see the similar saturation behavior as the pseudo entropy. For the sake of resolution, we did not plot the values at $d=0$ which take positive values. }\label{fig:nmonoPMI_continuum}
\end{figure}

In contrast with the subadditivity of pseudo entropy (or positivity of PMI), we observed a breakdown of strong subadditivity (or monotonicity of PMI). As an example, see Figure \ref{fig:nmonoPMI_continuum}. 
It means that the strong subadditivity of pseudo entropy can be violated in general. We also observed this in various combinations of $(z_1,z_2)$ with $z_1\neq z_2$, whereas some of them do not persist in the continuum and/or infinite system size limit. If $z_1=z_2$ but $m_1\neq m_2$ and if both $m_1$ and $m_2$ belong to the almost massless regime, then the monotonicity of PMI can persist because the equation  \eqref{eq:pmi_amassless2} readily guarantees the monotonicity of PMI (thanks to the monotonicity of the ordinary MI). 

\begin{figure}[ht]
 \begin{center}
 \resizebox{70mm}{!}{
 \includegraphics{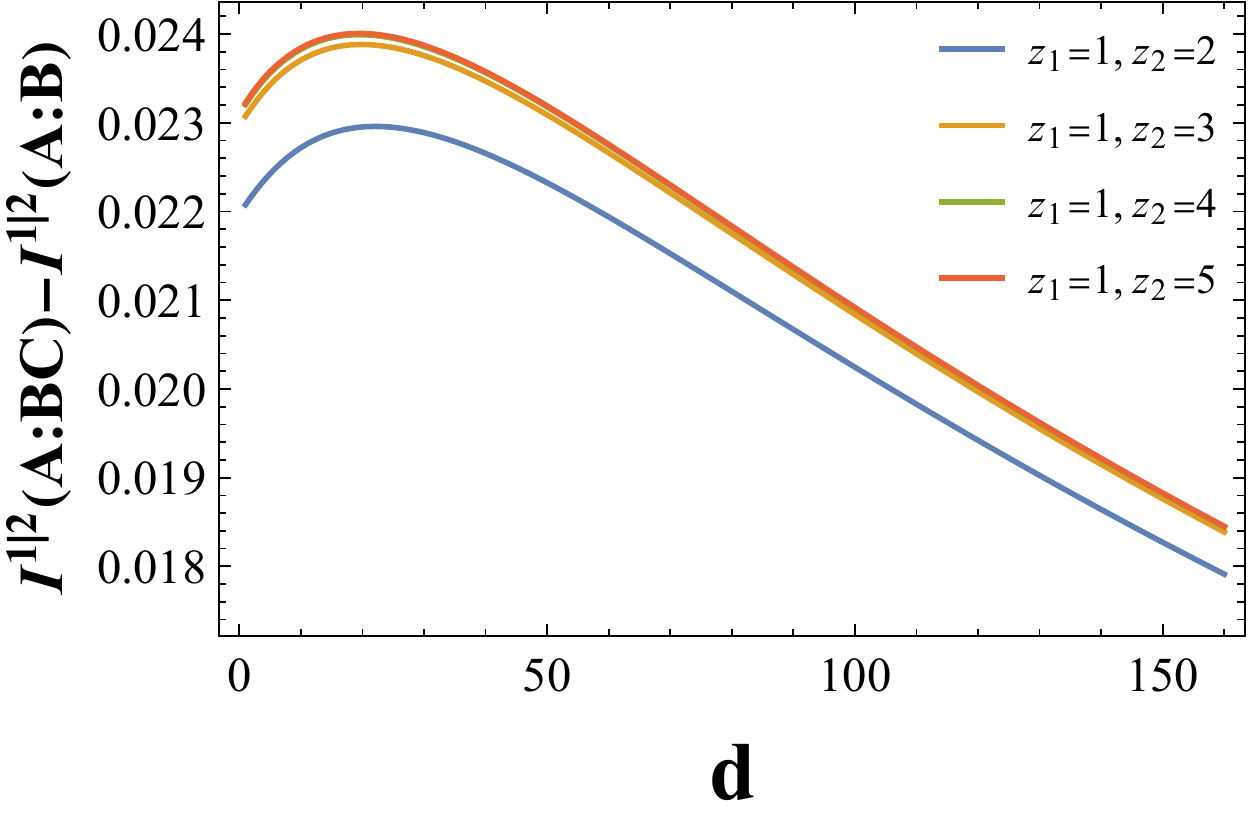}}
 \end{center}
 \caption{The recovery of breaking from the low-energy dispersion relation \eqref{eq:IRdis}. Here we again set $\ell_A=80, \ell_B=\ell_C=40$ and $m_1=m_2=10^{-4}$ and take infinite size limit. }\label{fig:PMI_IR}
\end{figure}

\begin{figure}[ht]
 \begin{center}
 \resizebox{70mm}{!}{
 \includegraphics{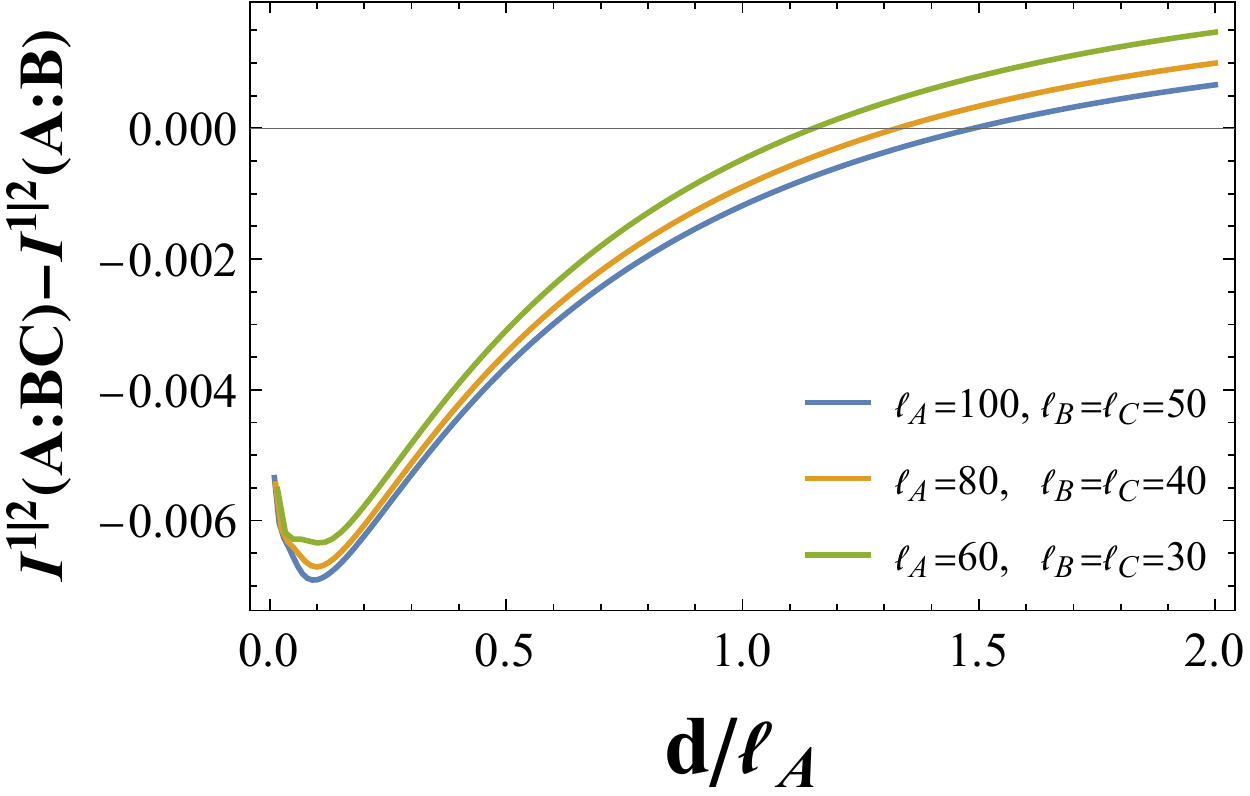}}
 \end{center}
 \caption{This figure suggests the breaking of monotonicity indeed persists even in the continuum limit. Here we set $m_1=m_2=10^{-4}$ and $z_1=1, z_2=5$. The minimum value becomes smaller as we enlarge the number of lattice in the subsystem. }\label{fig:nmonoPMI_continuum2}
\end{figure}

It is intriguing to understand better the origin of this breaking. If we modify the dispersion relation \eqref{dispersion} to
\be
\omega_k^2=m^{2z}+\sum_{i=1}^d \left(\frac{2\pi k_i}{N}\right)^{2z},\label{eq:IRdis}
\ee
so that the high-energy dispersion relation coincides with the field theory limit, we observed such breaking disappears. See Figure \ref{fig:PMI_IR}. In this regard, we can understand the origin of violation as high-energy modes, which may be regarded as lattice artifacts. On the other hand, we also observe that the breaking can persist in the continuum interpolation as Figure \ref{fig:nmonoPMI_continuum2}. It is interesting to understand any sharp criteria when the breaking happens. We leave further investigations as future work. 


\subsection{Periodic Subregions}\label{subsec:periodic_sub}
Following \cite{He:2016ohr} we consider a periodic lattice with $N$ sites and a periodic sub-lattice with $N_A$ sites which
$$N_A=\frac{N}{p}\;\;\;\;\;\;\;,\;\;\;\;\;\;p\in\mathbb{Z}^+\;\;.$$ 
The advantage of this set-up is that the periodicity of the subregion results in a \textit{circulant} structure for the relevant correlators and makes it possible to work-out the entanglement spectrum analytically. In general for a circulant matrix defined with
$$\{c_0,c_1,\cdots,c_{N_A-1}\},$$
the eigenvalues are given by
\be
\lambda_{j}=\sum_{k=0}^{N_A-1}c_k\,e^{-\frac{2\pi i j k}{N_A}}
\ee
and the corresponding eigenvectors are given by
\be
v^{(j)}_k=\frac{1}{\sqrt{N_A}}e^{-\frac{2\pi i j (k-1)}{N_A}}
\ee

For arbitrary periodic lattices and $p$-alternating sub-lattices, one can utilize the above structure for the case of pseudo entropy. As we have discussed in the previous section, the structure for all $X$, $P$, and $R$ correlators is given by
\begin{align}
\begin{split}
\mathcal{F}_{rs}&=
\frac{1}{N}\sum_{k=0}^{N-1} f^{(x/p/r)}(k)\cos\left(\frac{2\pi k (r-s)}{N}\right)
\end{split}
\end{align}
where
\begin{align}\label{eq:fs}
\begin{split}
f^{(x)}(k)&=\frac{1}{\omega_{1,k}+\omega_{2,k}},
\\
f^{(p)}(k)&=\frac{\omega_{1,k}\omega_{2,k}}{\omega_{1,k}+\omega_{2,k}},
\\
f^{(r)}(k)&=\frac{i}{2}\,\frac{\omega_{2,k}-\omega_{1,k}}{\omega_{1,k}+\omega_{2,k}},
\end{split}
\end{align}
which their eigenvalues are given by
\begin{align}
\begin{split}
&\lambda^{(x/p/r)}_{j}=
\frac{1}{2p}\sum_{k=0}^{p-1}f^{(x/p/r)}(j+k\,N_A).
\end{split}
\end{align}
Since we are interested in the eigenvalues of $i\,J\cdot\gamma$, one can check that the eigenvalues are given in terms of the eigenvalues of the $X$, $P$ and $R$ matrices by
\be\label{eq:lambda}
\lambda_j=\pm\sqrt{\lambda^{(x)}_{j} \lambda^{(p)}_{j}-{\lambda^{(r)}_{j}}^2}.
\ee
The expression in the square root is always positive. Now in order to simplify the results in our region of interest, we take $N_A,N\gg1$ while $N/N_A=p$ is held fixed. In this case we find 
\be
\lambda(x)=\pm\frac{1}{2p}\left[\sum_{r,s=0}^{p-1}\frac{4\omega_{1,x+\frac{r}{p}}\omega_{2,x+\frac{r}{p}}+\omega_-(r)\omega_-(s)}{\omega_+(r)\omega_+(s)} \right]^\frac{1}{2}
\ee
where $x=j/N$ and
$$\omega_{\pm}(i)\equiv\left(\omega_{1,x+\frac{i}{p}}\pm\omega_{2,x+\frac{i}{p}}\right).$$
The pseudo entropy is given by
\begin{align}
\begin{split}
&\frac{S_A(\tau)}{N_A}=p\int_0^{\frac{1}{p}}dx\bigg[\left(\lambda(x)+\frac{1}{2}\right)\log\left(\lambda(x)+\frac{1}{2}\right)
\\&\hspace{20mm}
-\left(\lambda(x)-\frac{1}{2}\right)\log\left(\lambda(x)-\frac{1}{2}\right)\bigg].	\end{split}    
\end{align}
Now let us focus on the case that we can compute the $z$ dependence analytically. Let us for simplicity first focus on $p=2$ case. For simplicity we take the limit $m_1,m_2\to 0$. In this limit it is not hard to find that
\begin{align}
\begin{split}
\lambda(x)&=
\frac{1}{2}
\left[\frac{\left(2^{z_2-z_1}+\cos^{z_1}(\pi x)\csc^{z_2}(\pi x)\right)}{\left(2^{z_2-z_1}+\csc^{z_2-z_1}(\pi x)\right)}
\right]^\frac{1}{2}\times
\\&\hspace{12mm}
\left[\frac{\left(2^{z_2-z_1}+\sin^{z_1}(\pi x)\sec^{z_2}(\pi x)\right)}{\left(2^{z_2-z_1}+\sec^{z_2-z_1}(\pi x)\right)}
\right]^\frac{1}{2}
\end{split}
\end{align}
which is symmetric on $z_1$ and $z_2$. One can see that when $z_2\gg z_1$, the eigenvalues approach a constant value. This behavior is similar to the plateau we have found numerically for single interval pseudo entropy.  

Using the analytic expression of the eigenvalues, we can analytically find how pseudo entropy depends on the value of $z$ as well as finding the $z$-dependence of the plateau. For similar analysis in case of von-Neumann entanglement entropy, see \cite{He:2017wla}. To do so we consider the following approximation for the eigenvalues
\be
\lambda(x)\approx
\begin{cases}
\frac{1}{2}
\cot(\pi x)^\frac{z_1}{2}
  &  ~~~~~~ 0<x<1/4 \\
\frac{1}{2}\tan(\pi x)^\frac{z_1}{2}
  & ~~~~~~ 1/4<x<1/2 
\end{cases}
\ee
where the neglected terms are $\mathcal{O}\left((z_2-z_1)^0\right)$. In the $\lambda(x)\gg1$ limit, one can find that using the above approximations leads to
\begin{align}
\begin{split}
\frac{S_A(\tau)}{N_A}
&\approx
2\int_0^{\frac{1}{2}}dx\,\log\left(\lambda(x)\right)
\\
&\approx
\frac{z_1}{2}\int_0^{\frac{1}{4}}dx\,\log\cot(\pi x)
+
\frac{z_1}{2}\int_{\frac{1}{4}}^{\frac{1}{2}}dx\,\log\tan(\pi x)
\\
&=
\frac{G_c}{\pi}z_1
\end{split}
\end{align}
where $G_c$ is the Catalan's constant ($\approx 0.916$). This linear dependence is in agreement with numerical study of these periodic subregions but is different from the small $z$ cases that we have studied with a single connected subregion where the $z$ dependence is quadratic. Other features of these periodic subregions are similar to the single interval case we studied earlier in this section.   

\section{Pseudo Entropy as a Probe for Quantum Phase Transitions}\label{sec:spinex}

In this section, we study the pseudo entropy of the quantum XY model which can be mapped to quadradic fermion via Jordan-Wigner transformation and shows colorful phase transitions. We will focus on the difference $\Delta S_{12} = S(\tau^{1|2}_A) - \frac{1}{2}\left(S(\rho_A^1)+S(\rho_A^2)\right)$. We will see that is $\Delta S_{12}$ is always nonpositive when $\ket{\psi_1}$ and $\ket{\psi_2}$ are in the same quantum phase, while it tends to be positive when $\ket{\psi_1}$ and $\ket{\psi_2}$ are in different quantum phases. An interpretation of this kind of behavior will be discussed in holographic setups in section \ref{sec:holPT}.

\subsection{XY Spin Model}
The XY spin chain or quantum XY model is a nearest neighbour interacting spin chain defined with the following Hamiltonian
\be
H_{\mathrm{XY}}=-\frac{1}{2}\sum\left(\frac{1+\gamma}{2}\sigma^x_n\sigma^x_{n+1}+\frac{1-\gamma}{2}\sigma^y_n\sigma^y_{n+1}+\lambda\sigma^z_n\right),
\ee
where $\{\sigma^x,\sigma^y,\sigma^z\}$ refer to Pauli matrices and $\lambda$ denotes the external magnetic field. The other parameter $\gamma$ controls the interaction in the $xy$-plane. The quantum XY model reduces to the famous transverse Ising model at $\gamma=1$. 

\begin{figure}[h!]
\begin{center}
\includegraphics[width=9.5cm]{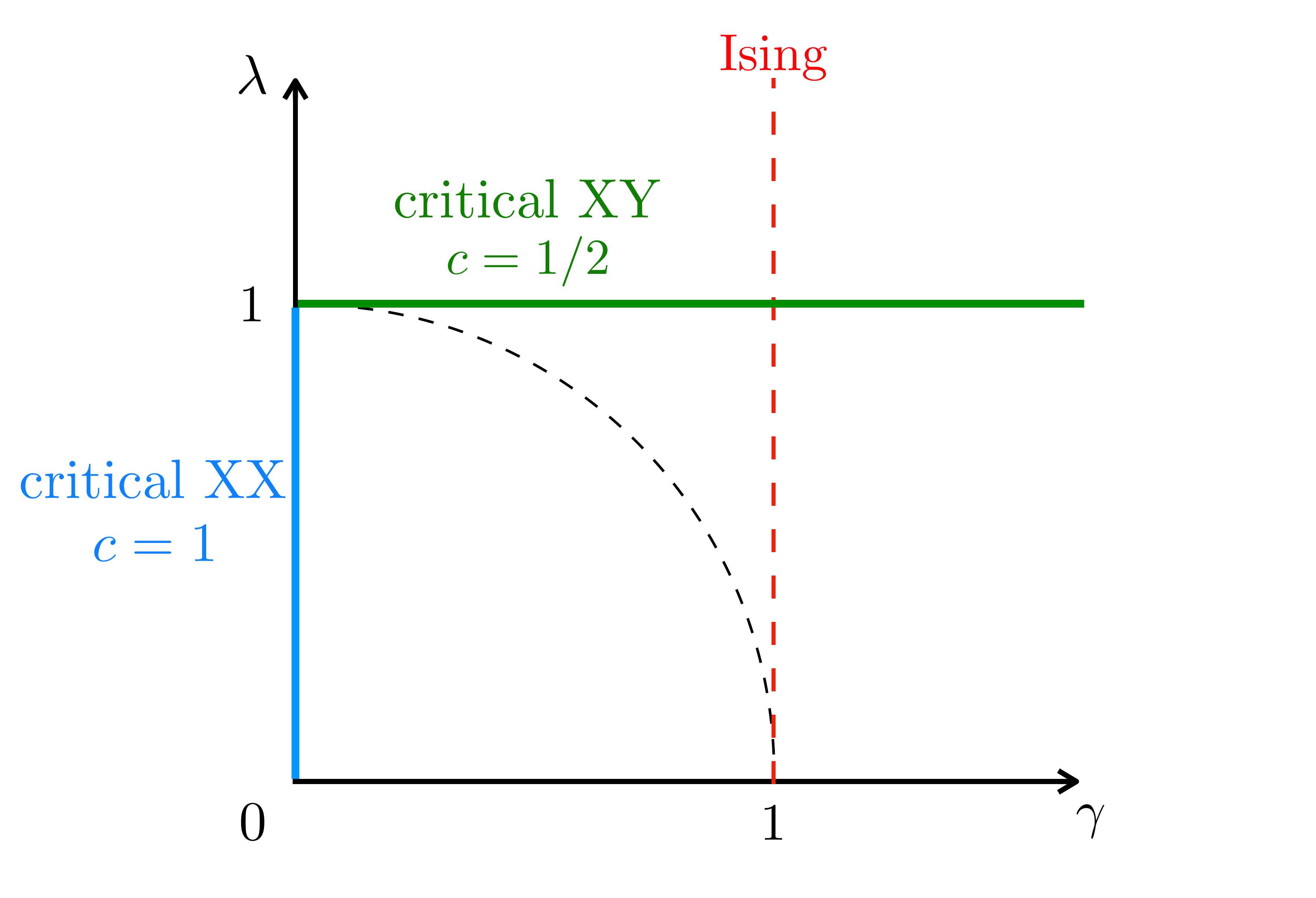}
\caption{The phase structure of the quantum XY model. The bold blue line ($\gamma=0$ and $0\leq\lambda<1$) and the bold green line ($\lambda=1$ and $\gamma>0$) are the two critical lines. The red dashed line ($\gamma=1$) corresponds to the transverse Ising model. The ground state becomes degenerate at the black dashed line ($\lambda^2+\gamma^2=1$).}
\label{fig:XYphase}
\end{center}
\end{figure}

The ground state of this model shows a rich phase structure \cite{Vidal:2002rm, XY1, XY2}. See Fig. \ref{fig:XYphase}. Thanks to the symmetries, it is sufficient to look at $\gamma\geq0$ and $\lambda\geq0$. There are two critical lines. One is $\gamma=0$ and $0\leq\lambda<1$ shown in blue. This critical line is often called the critical XX model and belongs to the universality class with central charge $c=1$. The other critical line is $\lambda=1$ and $\gamma>0$ shown in green. This is called the critical XY model and belongs to the universality class with central charge $c=1/2$. Except these, there are two dashed lines which are also important even though they are not critical. One is the $\gamma=1$ line shown shown in red. The famous transverse Ising model corresponds to this line. The other is the $\gamma^2+\lambda^2=1$ line shown in black. The ground state acquires a 2-degeneracy along this line. 

This rich phase structure makes it possible to testify our conjecture introduced in \cite{Mollabashi:2020yie} on the ability of pseudo entropy to distinguish between states belonging to different/same quantum phases. This conjecture is as follows. Consider the difference $\Delta S_{12} = S(\tau^{1|2}_A) - \frac{1}{2}\left(S(\rho_A^1)+S(\rho_A^2)\right)$. We conjecture that, if $\Delta S_{12}$ is positive, then $\ket{\psi_1}$ and $\ket{\psi_2}$ must lie in different phases. In the following, we will see that numerical results in the quantum XY model support this conjecture.   

To adapt the correlator method for calculation of pseudo entropy corresponding to a single block of spins in these models, we consider the fermionic representation (see for instance \cite{Latorre:2003kg} for the details of the Jordan-Wigner transformation) of the XY chain as
\begin{align}\label{eq:HXY}
\begin{split}
H_{\mathrm{XY}}&=\frac{1}{2}\sum_{n=-\frac{N-1}{2}}^{\frac{N-1}{2}}\Bigg[\left(c^\dagger_{n+1}c_{n}+c^\dagger_{n}c_{n+1}\right)
\\&\;\;\;\;\;\;\;\;\;\;\;\;
+ \gamma\left(c^\dagger_{n}c^\dagger_{n+1}+c_{n+1}c_{n}\right)-\lambda c^\dagger_{n}c_{n}\Bigg]\;.
\end{split}
\end{align}
Considering the Fourier transformation
\be
c_n=\frac{1}{\sqrt{N}}\sum_{k=-\frac{N-1}{2}}^{\frac{N-1}{2}}d_k\,e^{-\frac{2\pi i n k}{N}}\;,
\ee
together with the following Bogoluibov transformations,
\be
d^\dagger_{k}=u_k b^\dagger_{k}-iv_k b_{-k}\;,
\ee
where $u_k=\cos\theta_k$, $v_k=\sin\theta_k$ and
\begin{align}
\begin{split}
\tan(2\theta_k)=\frac{\gamma\sin\frac{2\pi k}{N}}{\cos\frac{2\pi k}{N}-\lambda}\;,
\end{split}
\end{align}
the Hamiltonian can be diagonalized as
\be\label{eq:HXYdiag}
H_{\mathrm{XY}}=\sum_{k=-\frac{N-1}{2}}^{\frac{N-1}{2}}
\Lambda_k b^\dagger_{k}b_{k}\;,
\ee
where
$$\Lambda_k=\sqrt{\left(\cos\frac{2\pi k}{N}-\lambda\right)^2+\gamma^2\sin^2\frac{2\pi k}{N}}\;.$$
Since we are interested in computing the pseudo entropy, here we consider an extra Bogoliubov transformation between vacuum states corresponding to $(\lambda_i,\gamma_i)$ parameters in the Hamiltonian, which are the input states of the pseudo entropy as
\be
b_{2,k}=\alpha_k b_{1,k}+i\beta_k b_{1,-k}^{\dagger},
\ee
where
\be
\alpha_k=\cos(\theta_1-\theta_2)
\;\;\;\;\;,\;\;\;\;\;
\beta_k=\sin(\theta_1-\theta_2).
\ee
The vacuum states $|\psi_i\rangle$ associated to the Hamiltonian with $(\lambda_i,\gamma_i)$ are expressed in terms of each other as
\be
|\psi_2\rangle=\bigotimes_{k>0}\left(\alpha_k-i\beta_k{b_{k}^{(1)}}^{\dagger}{b_{-k}^{(1)}}^{\dagger}\right)|\psi_1\rangle.
\ee
With these in hand, we can find that the explicit expressions for the correlators, defined previously in \eqref{eq:fercorr}, are given by 
\begin{align}
\mathbf{C}_1&=\frac{-2}{N}\sum_{k>0}\left(u_kv_k-\frac{\beta_{k}}{\alpha_{k}}u_k^2\right)\sin\left(\frac{2\pi k(m-n)}{N}\right),
\\
\mathbf{C}_2&=\frac{2}{N}\sum_{k>0}\left(u_k^2+\frac{\beta_{k}}{\alpha_{k}}u_kv_k\right)
\cos\left(\frac{2\pi k(m-n)}{N}\right),
\\
\mathbf{C}_3&=\frac{2}{N}\sum_{k>0}\left(v_k^2-\frac{\beta_{k}}{\alpha_{k}}u_kv_k\right)
\cos\left(\frac{2\pi k(m-n)}{N}\right),
\\
\mathbf{C}_4&=\frac{2}{N}\sum_{k>0}\left(u_kv_k+	\frac{\beta_{k}}{\alpha_{k}}v_k^2\right)\sin\left(\frac{2\pi k(m-n)}{N}\right).
\end{align}
We can plug these expressions in the method introduced in Section \ref{sec:fercorr} to produce the numerical results.

\subsection{Numerical results}

In this section we present two sets of numerical results for the XY model. The first set is produced by the continuum limit of the correlator method, where we report results for large subsystems on an infinite lattice. We also present a second set of results produced by direct diagonalization (with the help of \cite{WB17}) of the spin models on quite smaller systems where the size of the total system is 14 and the subregion is half of the chain. We use periodic boundary conditions.  

We study pseudo entropy by fixing one state and changing the other. More precisely, we consider $\ket{\psi_1}$ to be the ground state at a given fixed point in the phase space and follow a path in the phase space in which we consider $\ket{\psi_2}$ to be the ground state of the model through this path. We show the paths with orange zigzag line on the phase diagram. The paths which we are interested in are those which the path crosses the special lines illustrated in the phase space in Figure \ref{fig:XYphase}. These lines are sometimes borders between two different phases so in part of the path $\ket{\psi_1}$ and $\ket{\psi_2}$ belong to the same phase and in the other part they belong to different phases. 

We will see in the following numerical results that $\Delta S_{12} \leq 0$ if $\ket{\psi_1}$ and $\ket{\psi_2}$ are in the same quantum phase. We will also see that $\Delta S_{12}$ tends to be positive (though not always) when $\ket{\psi_1}$ and $\ket{\psi_2}$ are in different quantum phases.

\subsubsection*{Crossing Critical Lines}
Let us fix $\ket{\psi_1}$ and change $\ket{\psi_2}$ to see how pseudo entropy and $\Delta S_{12}$ change. We will firstly focus on the cases where $\ket{\psi_2}$ crosses critical lines. 

\begin{figure*}[t]
\begin{center}
\includegraphics[width=5.5cm]{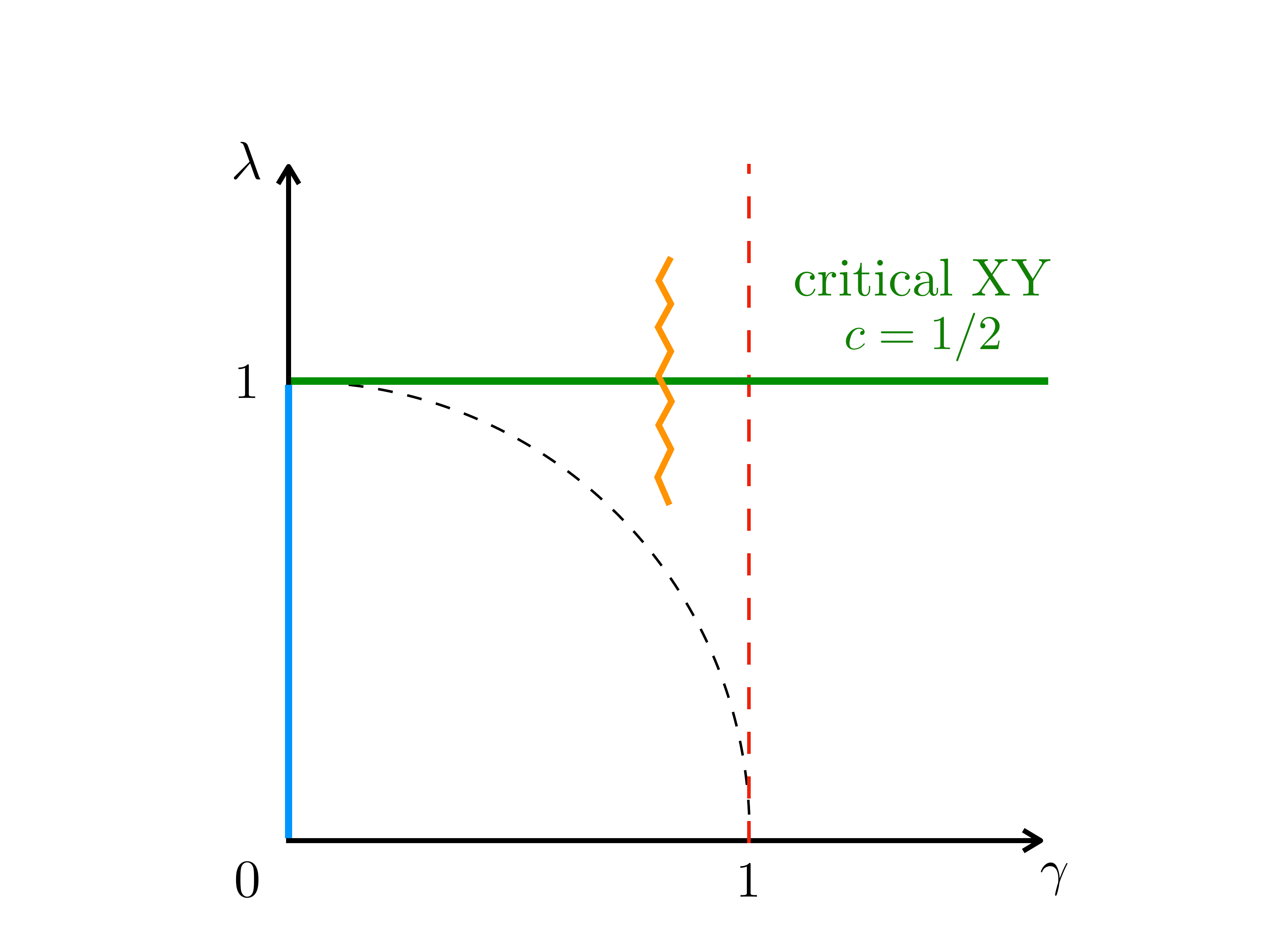}
\includegraphics[scale=0.3]{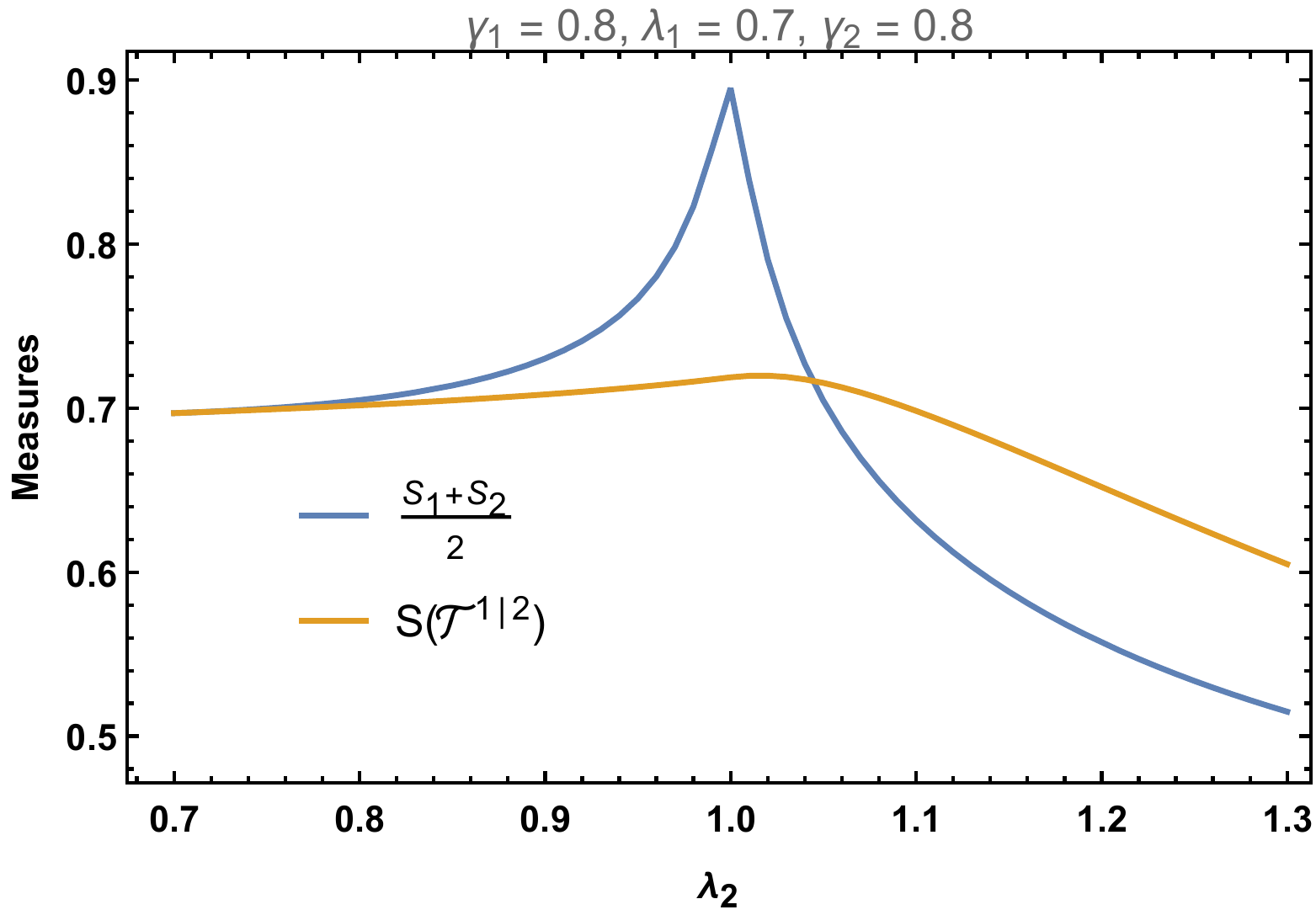}
\includegraphics[scale=0.3]{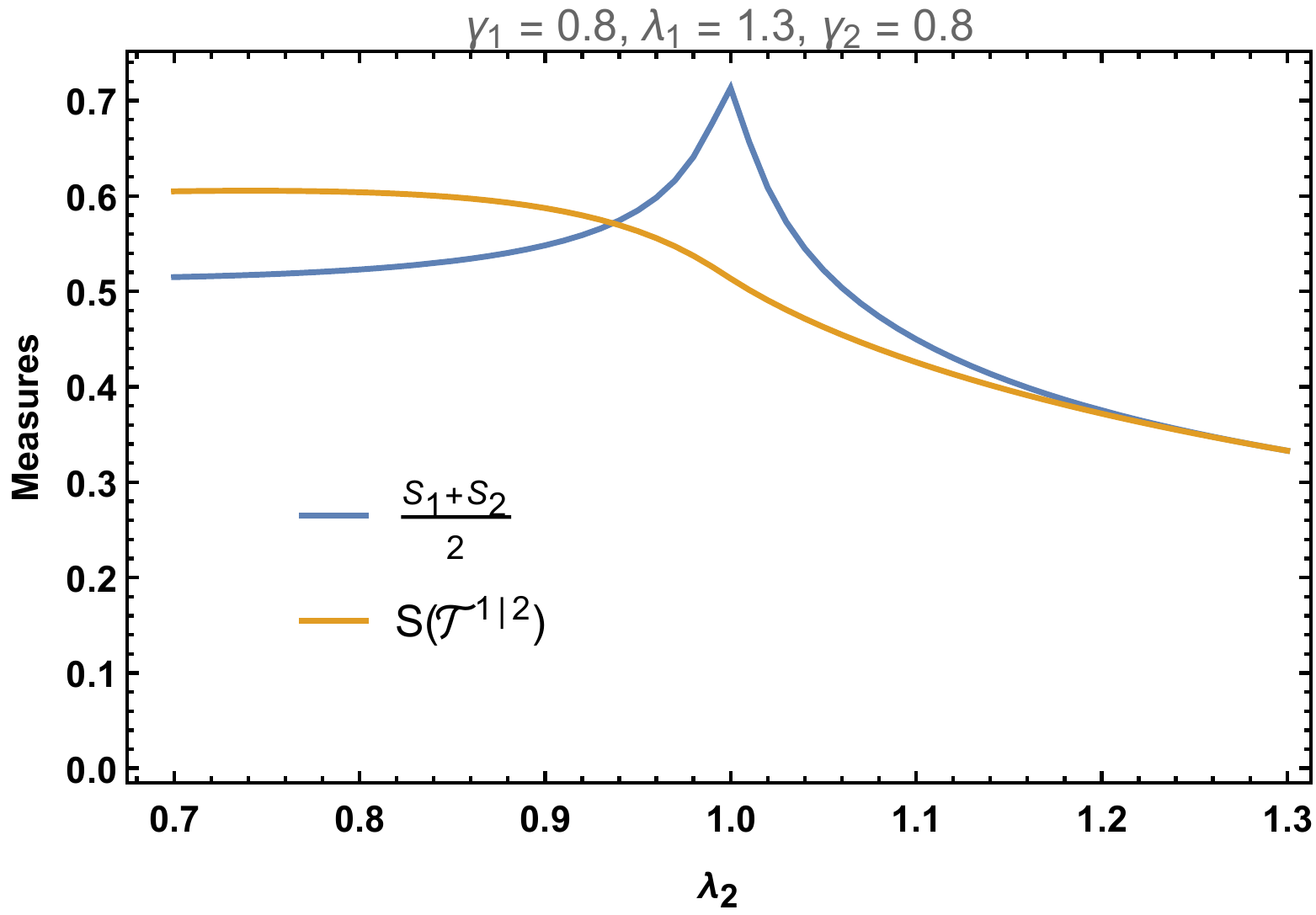}
\\
\hspace{55mm}
\includegraphics[scale=0.35]{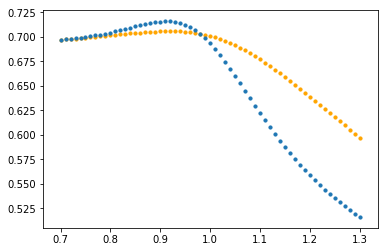}
\includegraphics[scale=0.35]{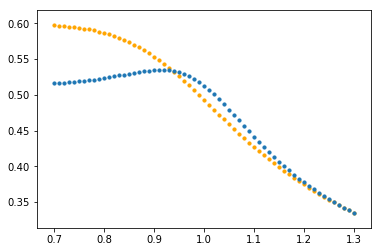}
\caption{Crossing the XY critical line. The upper line curves correspond to correlator method for subregion $\ell_x=50$ on an infinite lattice and the lower line correspond to direct calculation with $\ell_x=7$ on a periodic lattice with 14 sites.}
\label{fig:XYcrossXY}
\end{center}
\end{figure*}

In Figure \ref{fig:XYcrossXY} we show how pseudo entropy behaves while $\ket{\psi_2}$ crosses the XY critical line which belongs to the universality class with central charge $c=1/2$. $\ket{\psi_1}$ is fixed on one of the end points of the path in each panel and we can see how pseudo entropy, entanglement entropy and $\Delta S_{12}$ change across this line. We can see $\Delta S_{12}$ becomes positive only when $\ket{\psi_1}$ and $\ket{\psi_2}$ are in different phases. The spin models calculation is smooth at the critical line due the small size of the system. 

\begin{figure*}[t]
\begin{center}
\includegraphics[width=5.5cm]{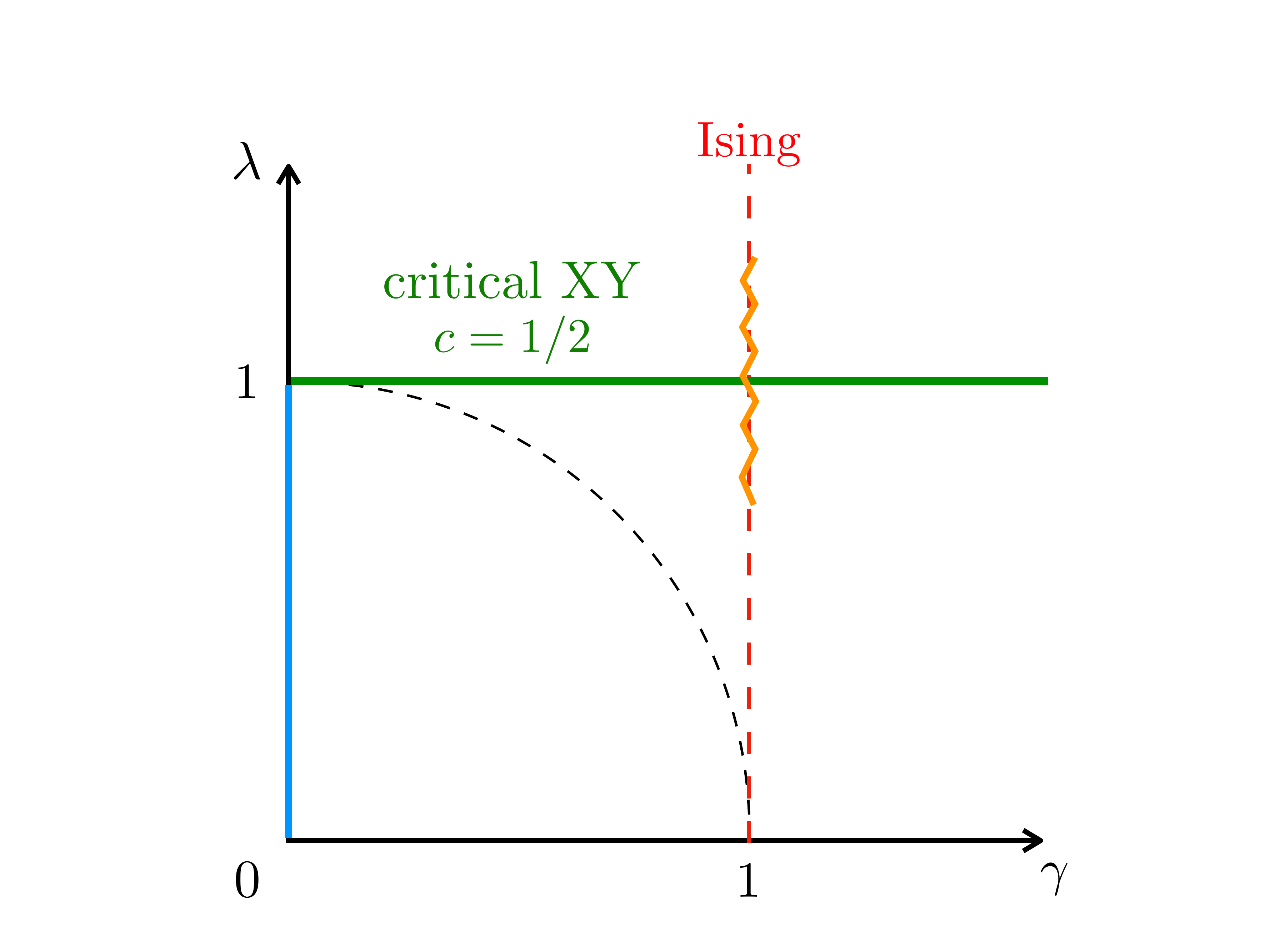}
\includegraphics[scale=0.3]{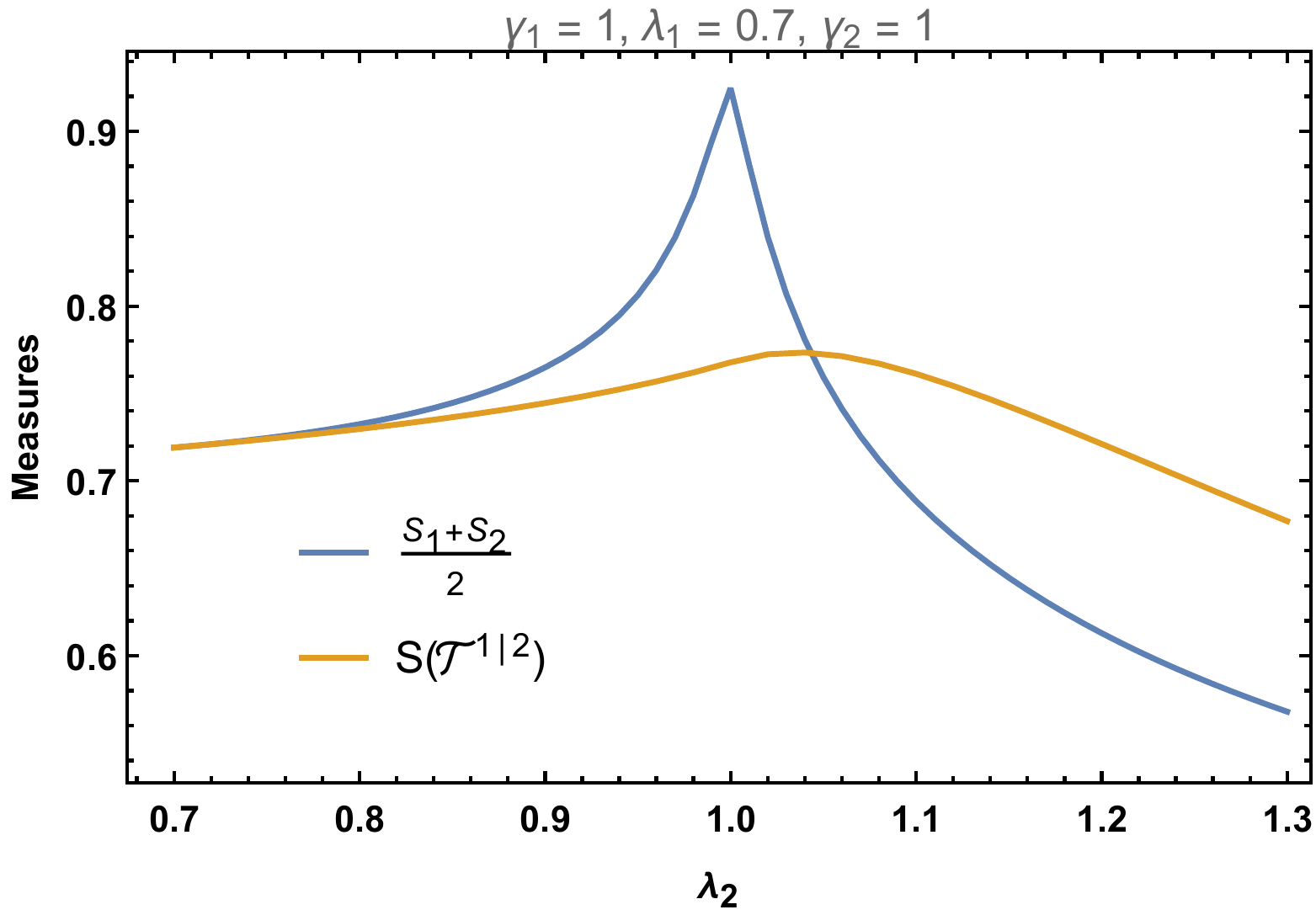}
\includegraphics[scale=0.3]{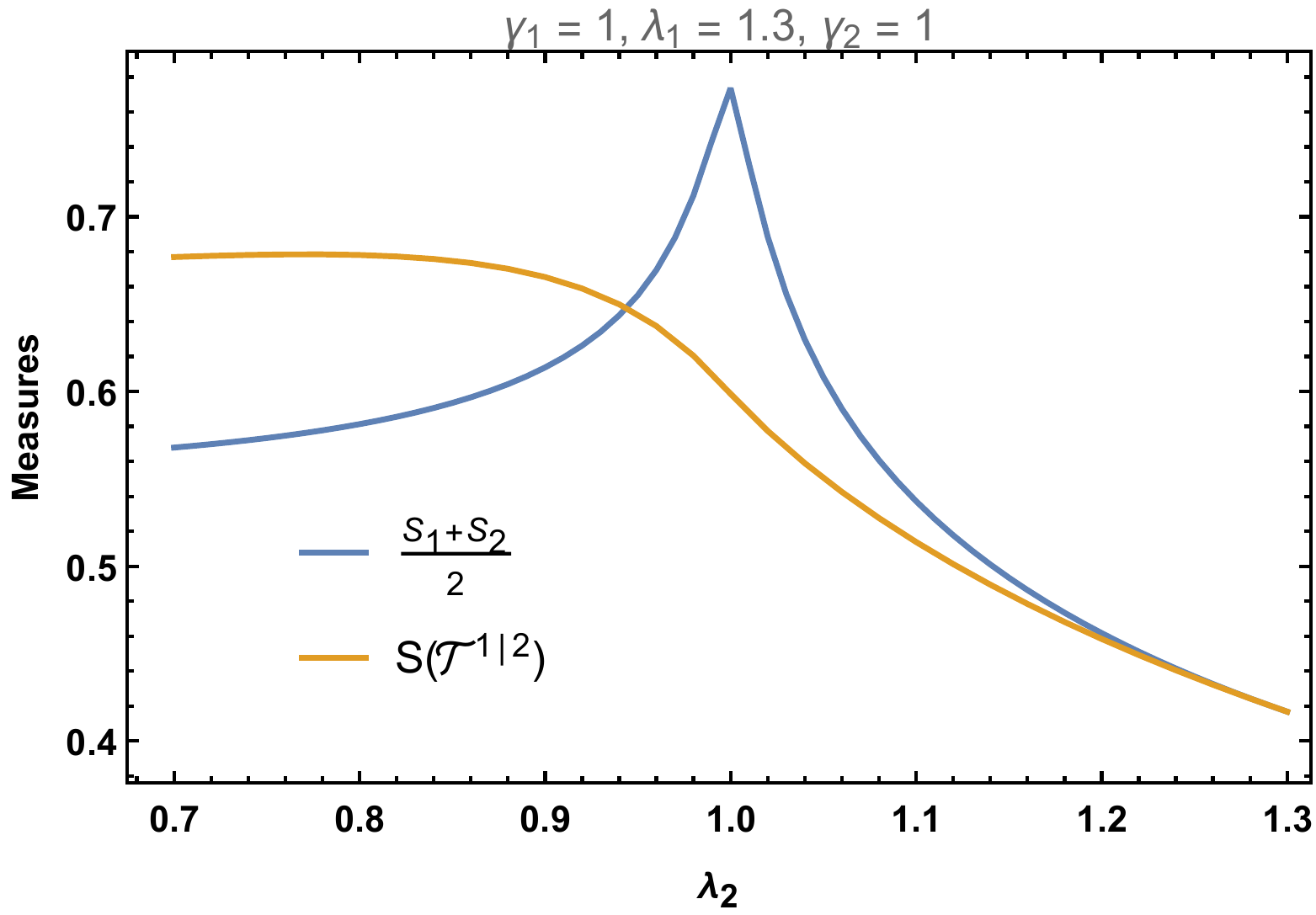}
\\
\hspace{55mm}
\includegraphics[scale=0.35]{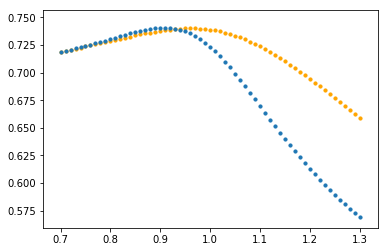}
\includegraphics[scale=0.35]{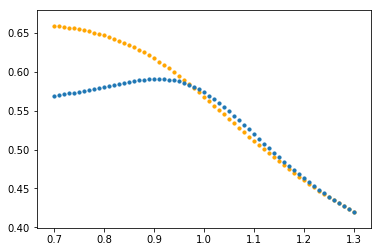}
\caption{Crossing the XY critical line along the Ising line. The upper plots are given by the correlator method for subregion $\ell_x=50$ on an infinite lattice, and the lower plots are given by direct diagonalization with $\ell_x=7$ on a periodic lattice with 14 sites.}
\label{fig:XYcrossXYIsing}
\end{center}
\end{figure*}

In Figure \ref{fig:XYcrossXYIsing} we show how pseudo entropy behaves while it crosses the XY critical line along on the Ising line, though the crossing point is critical Ising with central charge $c=1/2$. This is basically similar to the result presented previously in \cite{Mollabashi:2020yie}. In each panel $\ket{\psi_1}$ is fixed on one of the end points of the path and we again confirm that $\Delta S_{12}$ becomes positive only when $\ket{\psi_1}$ and $\ket{\psi_2}$ are in different phases.

\begin{figure*}[t]
\begin{center}
\includegraphics[width=5.5cm]{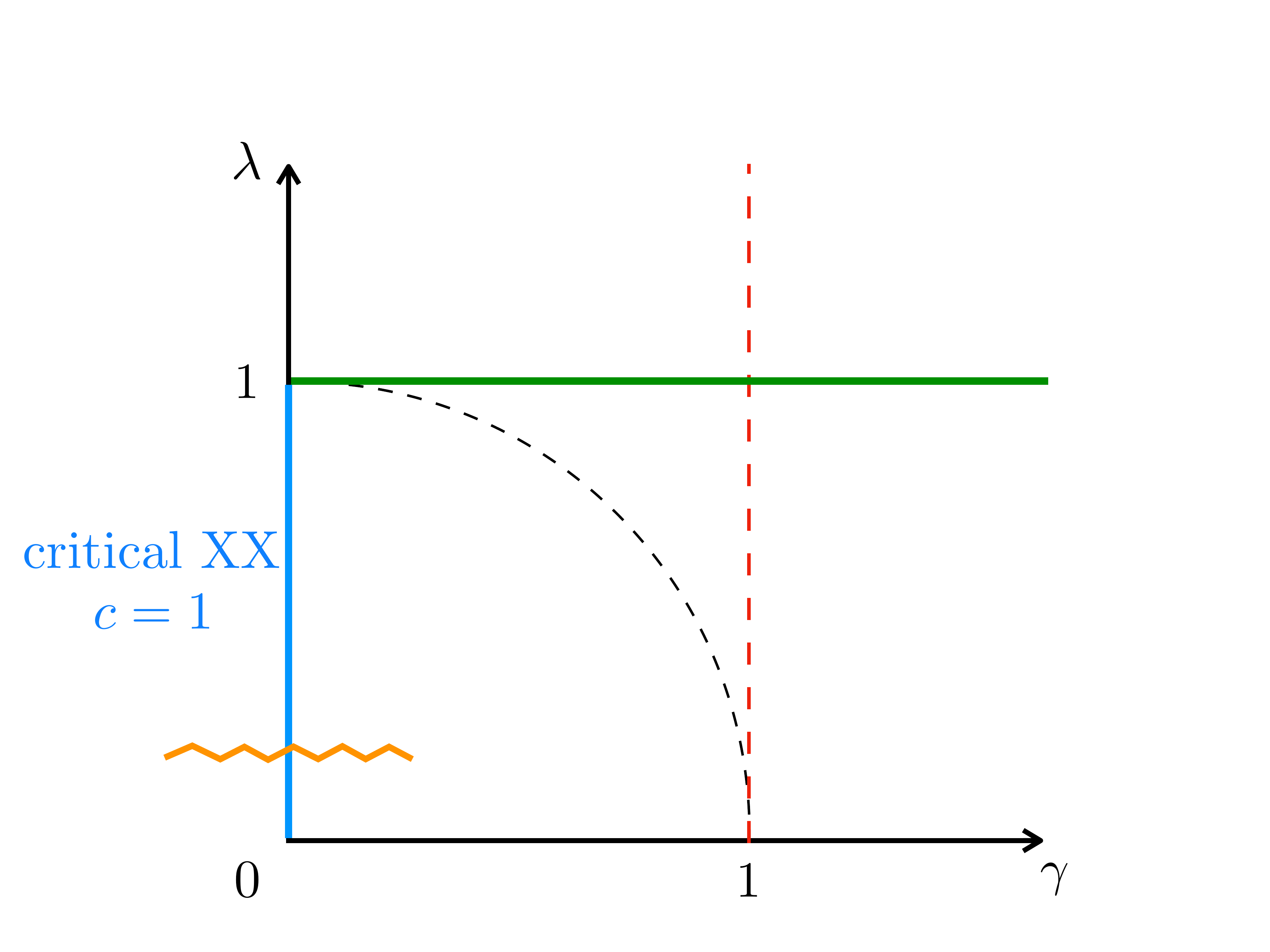}
\includegraphics[scale=0.3]{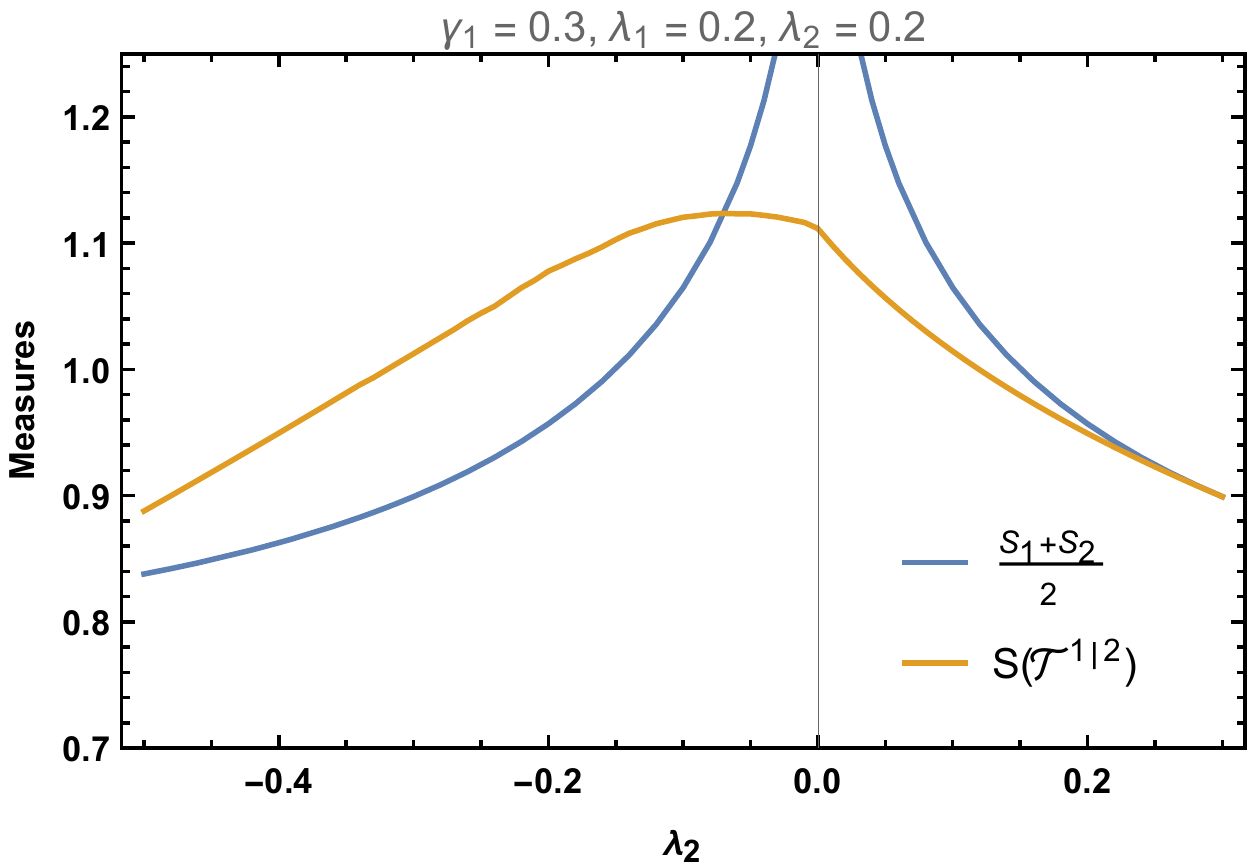}
\includegraphics[scale=0.3]{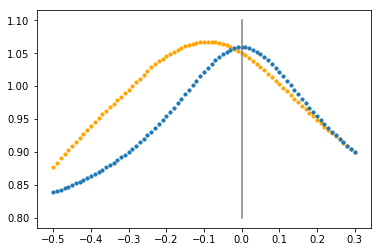}
\caption{Crossing the XX critical line. The middle plots are given by the correlator method for subregion $\ell_x=100$ on an infinite lattice and the right plots are given by direct diagonalization with $\ell_x=7$ on a periodic lattice with 14 sites.}
\label{fig:XYcrossXX}
\end{center}
\end{figure*}

In Figure \ref{fig:XYcrossXX} we show how pseudo entropy behaves while it crosses the XX critical line which belongs to the universality class with central charge $c=1$. Here we have set $\ket{\psi_1}$ to be the right endpoint $\gamma>0$ of the path. We can again confirm that the entanglement entropy shows a sharp behavior at the critical line for large systems and $\Delta S_{12}$ becomes positive only when $\ket{\psi_1}$ and $\ket{\psi_2}$ are in different phases.

\subsubsection*{Crossing Non-Critical Lines}
Although we have already seen how $\Delta S_{12}$ changes when $\ket{\psi_2}$ crosses critical lines, we would like to present more examples confirming $\Delta S_{12}\leq0$ when $\ket{\psi_1}$ and $\ket{\psi_2}$ are in the same phase. 

\begin{figure*}[t!]
\begin{center}
\includegraphics[width=5.5cm]{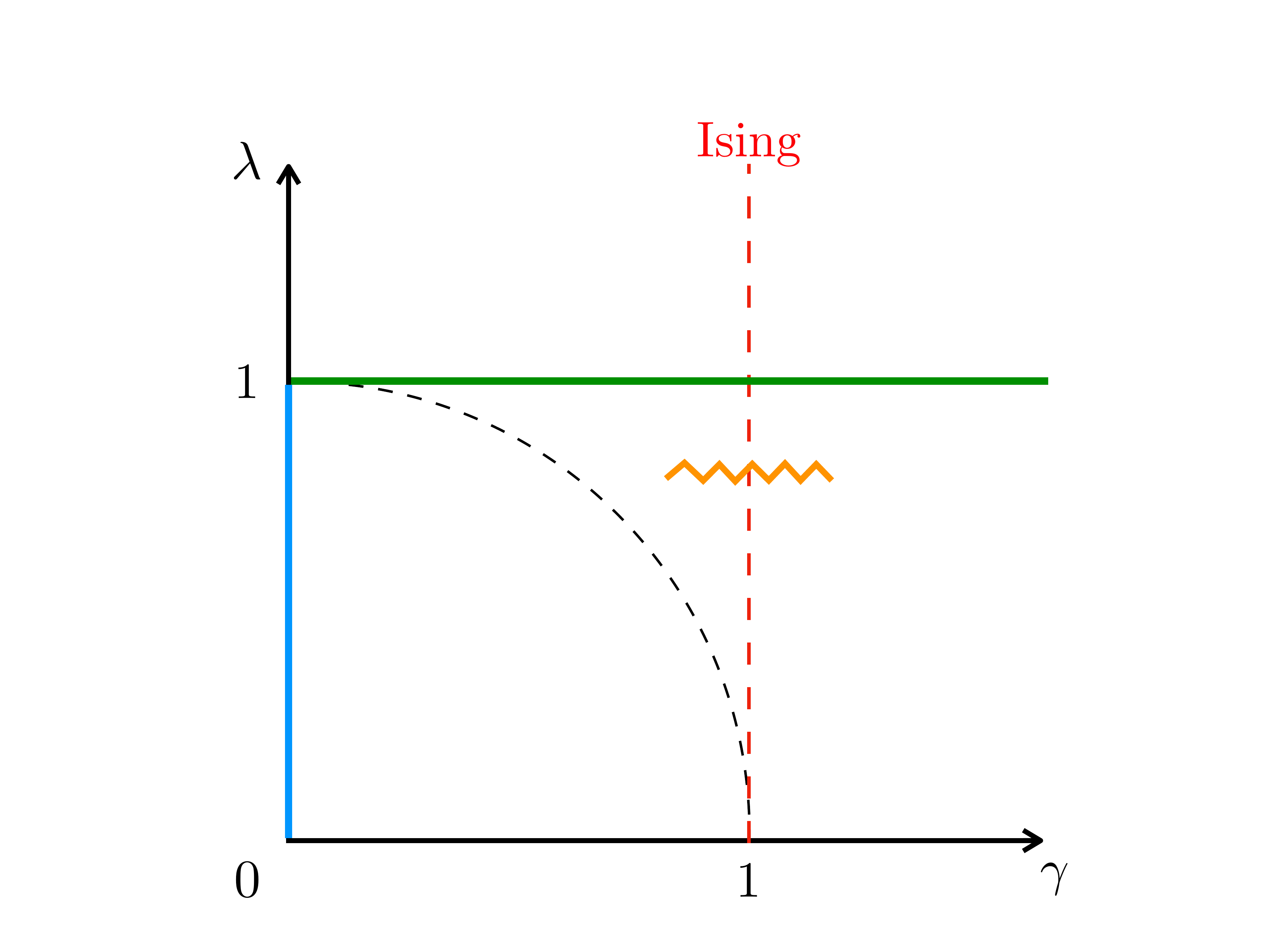}
\includegraphics[scale=0.3]{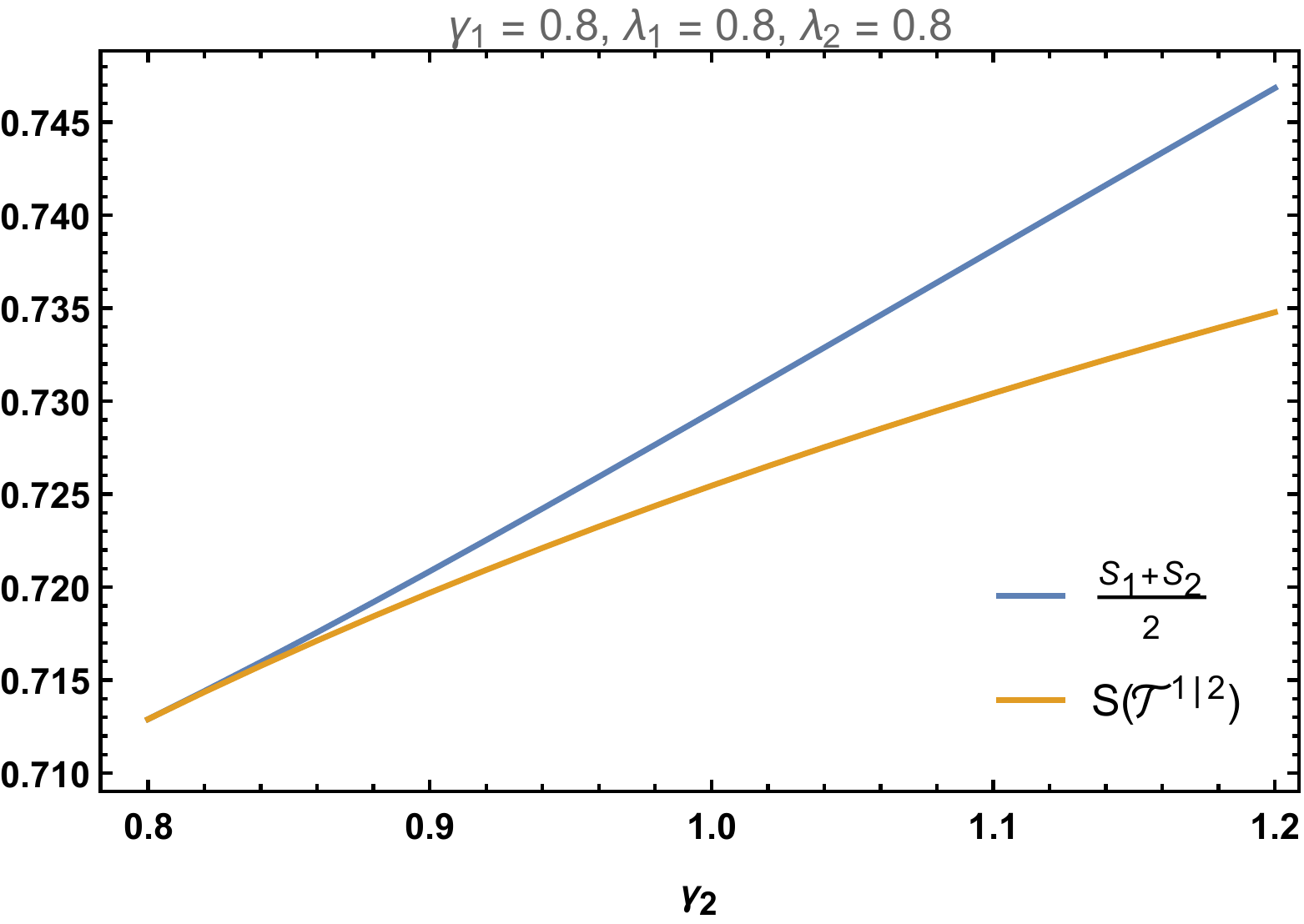}
\includegraphics[scale=0.3]{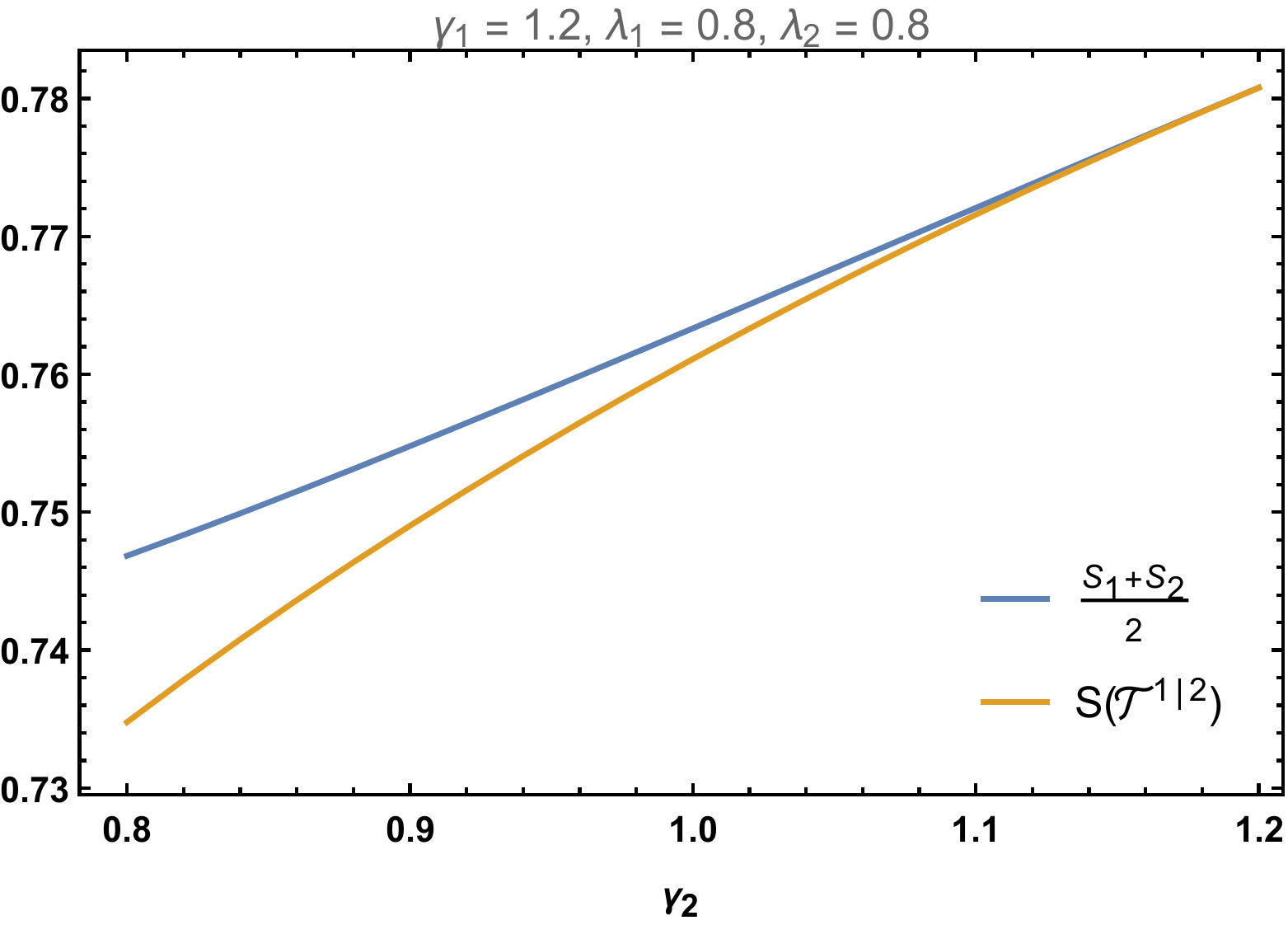}
\\
\hspace{55mm}
\includegraphics[scale=0.35]{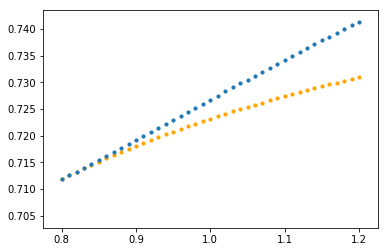}
\includegraphics[scale=0.35]{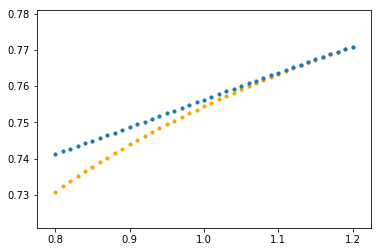}
\caption{Crossing the Ising line. The upper plots are given by the correlator method for subregion $\ell_x=50$ on an infinite lattice, and the lower plots are given by direct diagonalization with $\ell_x=7$ on a periodic lattice with 14 sites.}
\label{fig:XYcrossIsing}
\end{center}
\end{figure*}

\begin{figure*}[t]
\begin{center}
\includegraphics[width=4.0cm]{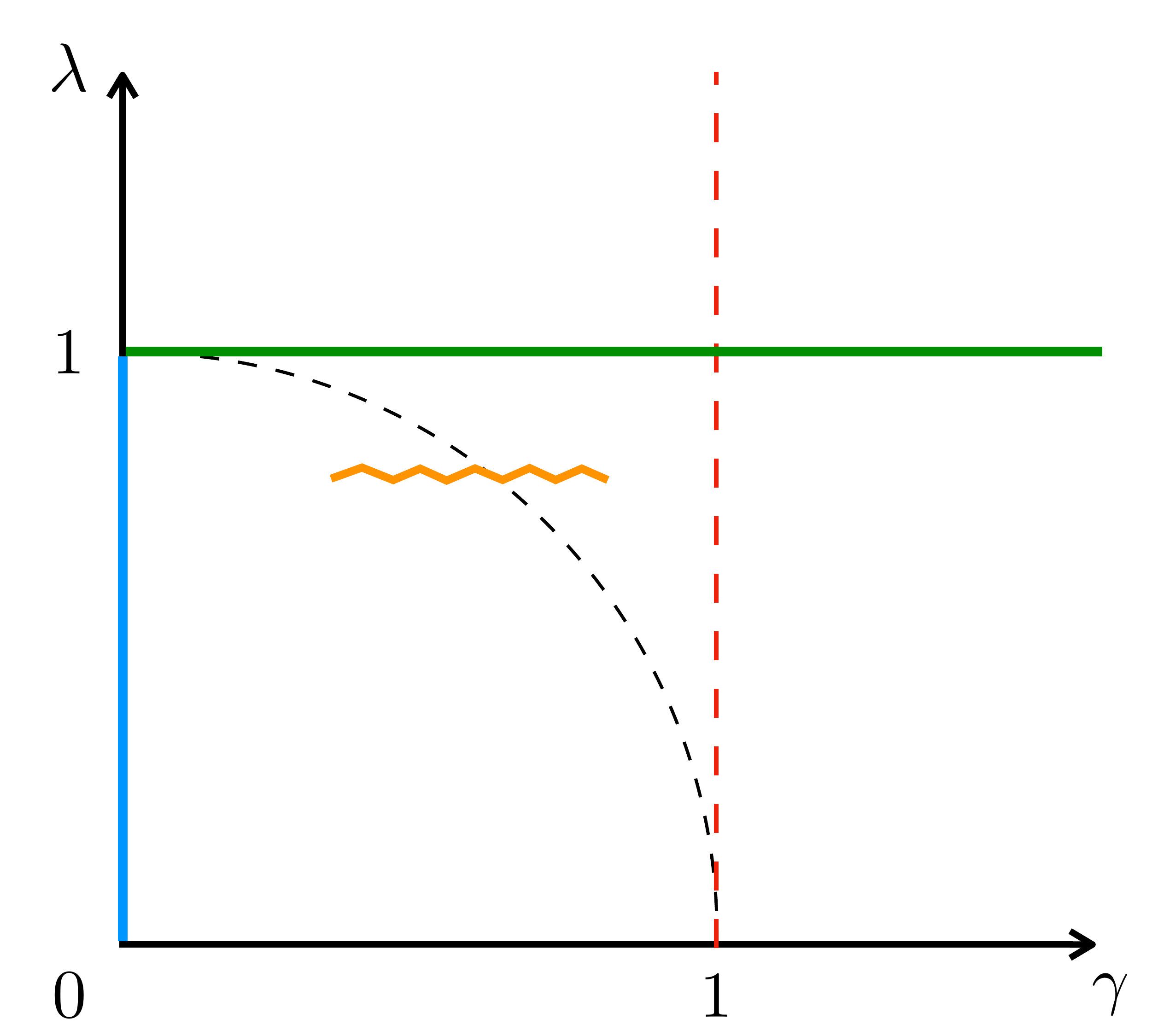}
\includegraphics[scale=0.28]{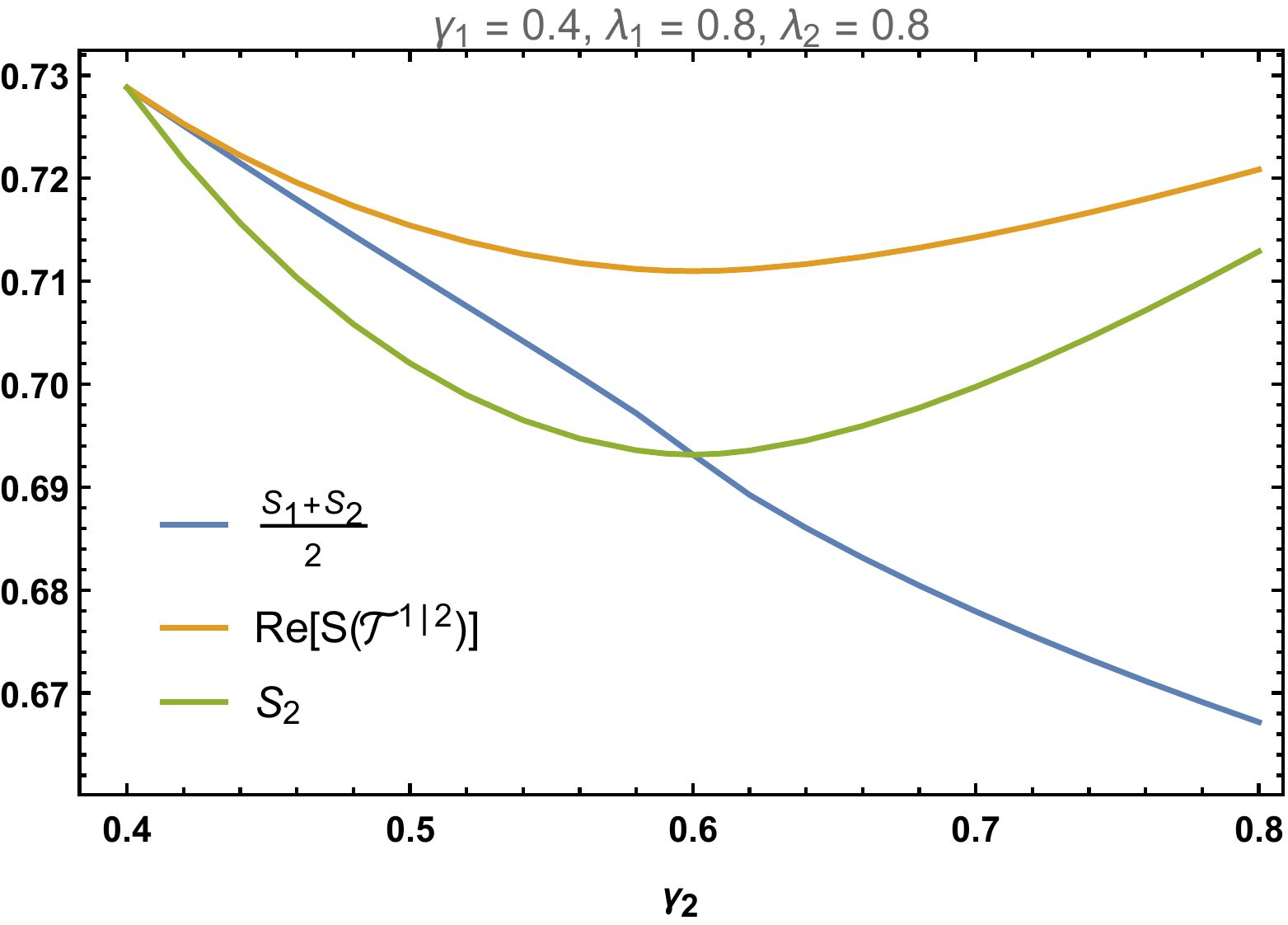}
\includegraphics[scale=0.33]{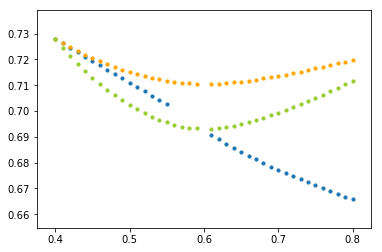}
\includegraphics[scale=0.27]{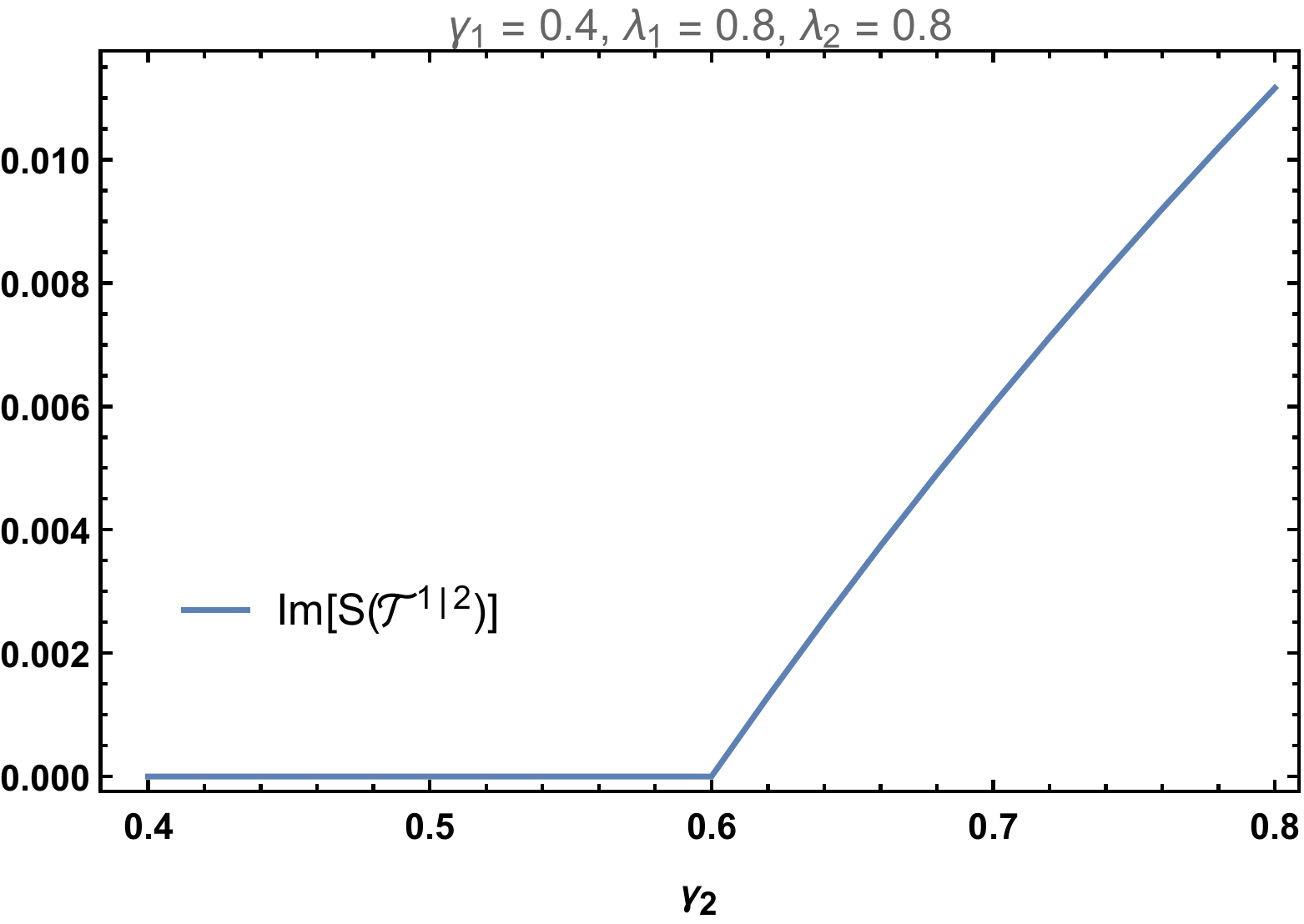}
\caption{Crossing the degenerate semi circle. The second panel is given by the correlator method for subregion $\ell_x=100$ on an infinite lattice and the third panel is given by direct diagonalization with $\ell_x=7$ on a periodic lattice with 14 sites. The right panel shows to the imaginary part of the pseudo entropy.}
\label{fig:XYcrossC}
\end{center}
\end{figure*}

In Figure \ref{fig:XYcrossIsing} we show the case in which $\ket{\psi_2}$ crosses the Ising line. Note that since the Ising line is not a critical line, $\ket{\psi_2}$ lies in the same phase from the start to the end. We can see that $\Delta S_{12}\leq0$ always holds in this case. 

In Figure \ref{fig:XYcrossC} we show how pseudo entropy behaves while $\ket{\psi_2}$ crosses the $\gamma^2+\lambda^2=1$ line where the ground state of the XY model is a product state but doubly degenerate \cite{Mueller}. In this case as we can see the entanglement entropy of $\ket{\psi_2}$ is $\ln 2$ when it is on this line from the green curve. Due to the degeneracy of the states numerical, computation becomes very costy in the vicinity of this line. Again as can been seen from the two middle panels, $\Delta S_{12}\leq0$. This is again consistent with the conjecture, since $\gamma^2+\lambda^2=1$ is not a critical line. The other specific property of this line is that while $\ket{\psi_2}$ passes through the degenerate line (the two states belong to two sides of a degenerate point), the pseudo entropy takes an imaginary value. Similar observation for qubit systems can be found in \cite{Nakata:2021ubr}.


\section{Global Quantum Quenches}\label{sec:globalquench}
In this section, we investigate the time evolution of pseudo entropy after a global quantum quench. 

A global quench is to firstly prepare an initial state (which is usually translational invariant) with only short-range correlation in it, and then let it evolve using a massless Hamiltonian. Tracking the time evolution of this state, one can see how the long-range correlation is created. For example, entanglement entropy is one of the most typical quantities which capture the features. Under the time evolution, the entanglement entropy of a finite interval firstly grows linearly and then saturate to a value which is proportional to its length.

To study the pseudo-entropy associated to global quenches, we let {\it two different initial states} evolve under {\it the same Hamiltonian}. In other worlds, we consider a transition matrix in the following form
\begin{align}\label{eq:TMquench}
    \tau^{1|2} = \frac{|\psi_1(t)\rangle\langle\psi_2(t)|}{\langle\psi_2(t)|\psi_1(t)\rangle},
\end{align}
where
\begin{align}
    |\psi_1(t)\rangle &= e^{-iHt}|\psi_1(0)\rangle, \\
    |\psi_2(t)\rangle &= e^{-iHt}|\psi_2(0)\rangle.
\end{align}
Note that, in the $|\psi_2(0)\rangle\to|\psi_1(0)\rangle$ limit, it reduces to the standard density matrix after a global quench. Of course in the case of pseudo entropy one can consider different types of quenches where in general $|\psi_1(0)\rangle$ and $|\psi_2(0)\rangle$ are evolved with different Hamiltonians. We will not consider this general case here.

In the following, we firstly give a formalism which allows one to compute the pseudo-entropy after global quenches analytically in a CFT and show the results for the free Dirac fermion as a concrete example in Sec. \ref{sec:quenchCFT}. After that, we independently compute the same quantity in the free scalar theory using the correlator method in Sec. \ref{sec:quenchscalar}. We show that the spectrum of the reduced transition matrix becomes imaginary. 

Finally, in Sec. \ref{sec:quenchGraDual}, we give the gravity dual of the transtion matrix (\ref{eq:TMquench}) in a holographic CFT. We show that the gravity dual contains an end-of-the-world brane whose location is given by complexified coordinates. This reflects the imaginary nature of the transition matrix. 

\subsection{CFT}\label{sec:quenchCFT}
A global quench in a CFT can be simulated by evolving a boundary state $\ket{B}$ \cite{Cardy:2004hm} as following \cite{Calabrese:2006rx,Calabrese:2007rg,Calabrese:2016xau} 
\begin{align}
    e^{-itH} e^{-\alpha H} \ket{B}.
\end{align}
Here, $e^{-\alpha H}$ is a smearing factor introduced to avoid divergence. As we will see below, this factor actually gives an effective temperature $\beta=4\alpha$ to the quenched state. 

\begin{figure}[H]
    \centering
    \includegraphics[width=6cm]{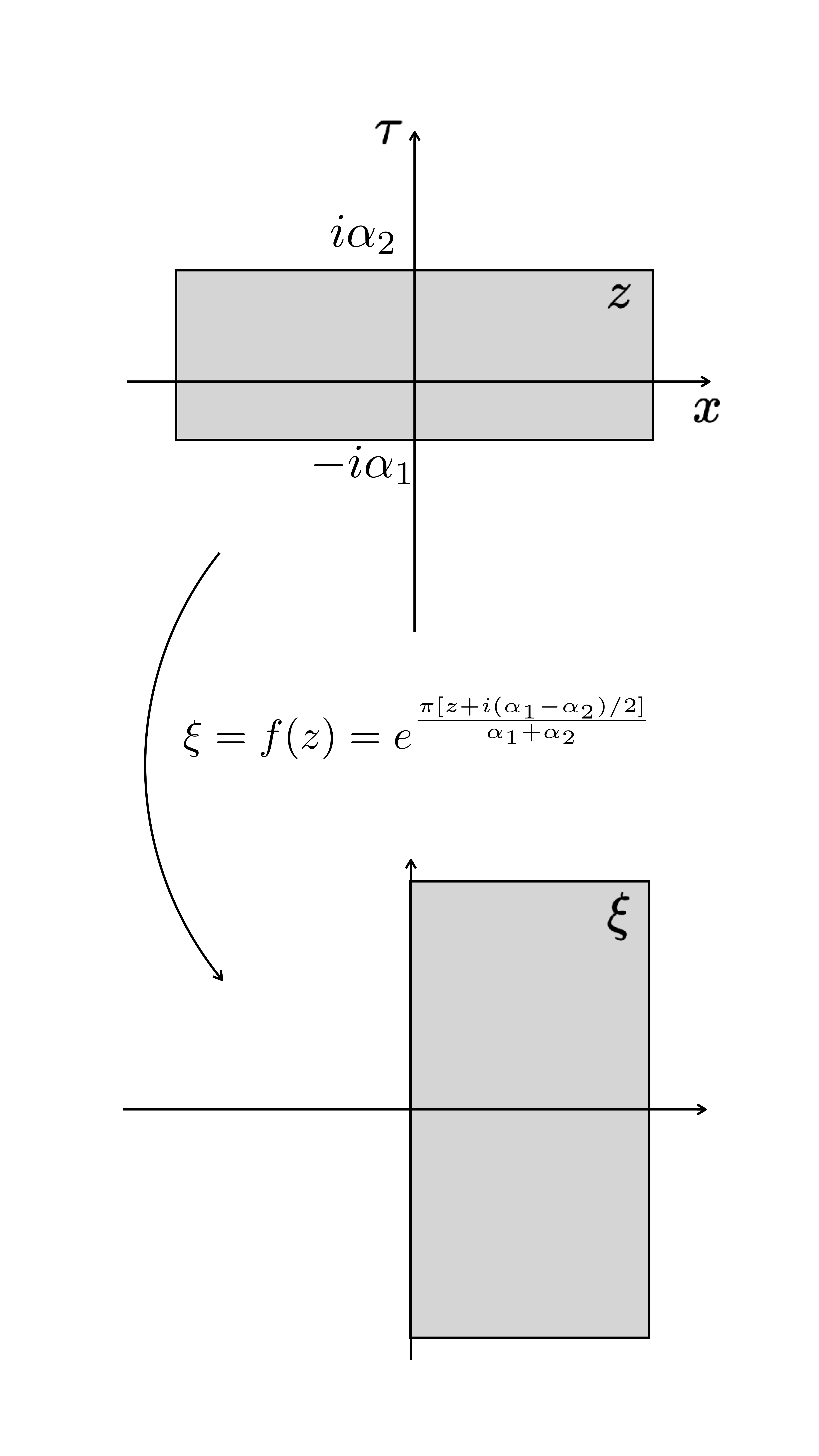}
    \caption{The Euclidean path integral corresponding to $\tau^{1|2}(t)$ (upper panel, parameterized by $z=x+i\tau$) and the conformal transformation mapping it to a right half plane (lower panel, parameterized by $\xi$).}
    \label{fig:quenchPI}
\end{figure}

Let us consider a transition matrix with two different initial states $e^{-\alpha_1 H} \ket{B}$ and $e^{-\alpha_2 H} \ket{B}$ evolving under the same Hamiltonian $H$:
\begin{align}\label{eq:quenchTMCFT}
    \tau^{1|2}(t) = \frac{e^{-itH} e^{-\alpha_1 H} |B\rangle \langle B| e^{-\alpha_2 H} e^{itH}}{\braket{B|e^{-(\alpha_1+\alpha_2) H}|B}}.
\end{align}
The corresponding Euclidean path integral, whose coordinates are parameterized by $z=x+i\tau$, is a strip shown in the upper half of Fig. \ref{fig:quenchPI}. 

As shown in the lower half of Fig. \ref{fig:quenchPI}, it is convenient to map the strip to a right half plane (RHP) parameterized by $\xi$ using the following conformal transformation:
\begin{align}\label{eq:map}
    \xi = f(z) = e^{\frac{\pi [z+ i(\alpha_1 - \alpha_2)/2]}{\alpha_1+\alpha_2}}.
\end{align}
Therefore, a one-point function of a primary operator $\CO(z,\bar{z})$ with conformal weight $(h,\bar{h})$ is evaluated as 
\begin{align}\label{eq:1p-function}
    \braket{\CO(z,\bar{z})}_{\rm strip} &= 
    f'(z)^{h} \bar{f}'(\bar{z})^{\bar{h}} \braket{\CO(\xi,\bar{\xi})}_{\rm RHP} \nonumber \\
    &= C f'(z)^{h} \bar{f}'(\bar{z})^{\bar{h}} \braket{\CO(\xi,\bar{\xi})\CO(-\bar{\xi},-\xi)}_{\mathbb{C}},
\end{align}
where a mirror trick is performed in the second line, and $C$ is a constant which depends on the details of the boundary condition. For simplicity, we will take $C = 1$ in the following. Generic correlation functions can be computed in a similar manner. 

\subsubsection*{Pseudo Entropy for an Infinite Interval}
Let us firstly take the subsystem $A$ to be a half of the whole system (i.e. $A=\{x~|~0<x\}$) and consider its pseudo-entropy. 

Thanks to the conformal symmetry, the computation of $n$-th pseudo-R\'{e}nyi entropy reduces to evaluating correlation functions of twist operators $\sigma_n$ inserted in $\partial A$ \cite{CC04,Nakata:2021ubr}. The conformal weight of $\sigma_n$ is given by 
\begin{align}
    h_n = \bar{h}_n = \frac{c}{24}\left(n-\frac{1}{n}\right)
\end{align}
where $c$ is the central charge of the CFT. 

In this case, the $n$-th pseudo-R\'{e}nyi entropy is 
\begin{align}
    S^{(n)}(\tau^{1|2}_A(t)) &= \frac{1}{1-n} \log \braket{\sigma_n(i\tau,-i\tau)}_{\rm strip} \Big|_{\tau = it}~.
\end{align}
Using Eq. (\ref{eq:map}) and (\ref{eq:1p-function}), it is straightforward to find that 
\begin{align}
    &S(\tau^{1|2}_A(t))  \nonumber\\
    = & \frac{c}{6} \log \left(\frac{2\alpha_+}{\pi\epsilon} \cos\frac{\pi (2it+\alpha_-)}{2\alpha_+} \right) \nonumber \\
    = &\frac{c}{6} \log \Biggl(\frac{2\alpha_+}{\pi\epsilon} 
    \Bigl[\cosh\frac{\pi t}{\alpha_+} \cos\frac{\pi\alpha_-}{2\alpha_+} - i~\sinh\frac{\pi t}{\alpha_+} \sin\frac{\pi\alpha_-}{2\alpha_+}\Bigr]\Biggr)
\end{align}
by taking $n\rightarrow1$. Here, we introduced $\alpha_{\pm} = \alpha_1 \pm \alpha_2$. $\ep$ is a UV cutoff corresponding to the lattice distance.

By replacing $\alpha_+$ with $2\alpha_i~(i=1,2)$ and $\alpha_-$ with $0$, we can recover the standard entanglement entropy for each state: 
\begin{align}
    S(\rho^{i}_A(t)) 
    = \frac{c}{6} \log \left(\frac{4\alpha_i}{\pi\epsilon} \cosh\frac{\pi t}{2\alpha_i} \right).
    \label{eq:EEinf}
\end{align}
The key feature of this entanglement entropy is that it has a linear time evolution at late time:
\begin{align}
    S(\rho^i_A(t)) = \frac{\pi c}{12} \frac{t}{\alpha_i} + \cdots .
\end{align}

Note that the results in this section is universal in any CFT, since we only used two point functions of primary operators. 

\subsubsection*{Finite Single Interval in Dirac Free Fermion}

Let us then consider a finite interval with length $l$ as the subsystem $A$. In this case, we need to compute the two-point function of the twist operator on the strip. This computation can be done by evaluating a four-point function of twist operators on the $\mathbb{C}\simeq\mathbb{R}^2$. Therefore, we need to consider some solvable theories. In this part, we consider Dirac free fermion as a concrete example. Note that $c=1$ in this case.

To begin with, in free Dirac fermion CFT, the 2-point function of twist operators on the right half plane (RHP) is given by \cite{Casini:2009sr}
\begin{align}
    &\left\langle{\sigma}_{n}\left(\xi_1, \bar{\xi}_1\right) \bar{\sigma}_{n}\left(\xi_2, \bar{\xi}_2\right)\right\rangle_{\mathrm{RHP}} \nonumber\\
    =& \sqrt{\left\langle\sigma_n\left(-\bar{\xi}_1, -{\xi}_1\right) \bar{\sigma}_{n}\left(-\bar{\xi}_2, -{\xi}_2\right)\sigma_n\left(\xi_1, \bar{\xi}_1\right) \bar{\sigma}_{n}\left(\xi_2, \bar{\xi}_2\right)\right\rangle_{\mathbb{C}}} \nonumber\\
    \propto &\left(\frac{\left(\xi_1+\bar{\xi}_2\right)\left(\bar{\xi}_1+\xi_{2}\right)}{\left({\xi}_1-\xi_{2}\right)\left(\bar{\xi}_1-\bar{\xi}_{2}\right)\left({\xi}_1+\bar{\xi}_{1}\right)\left({\xi}_2+\bar{\xi}_{2}\right)}\right)^{2 h_{n}}. 
\end{align}
Let us use $z_1=i\tau$ and $z_2=l+i\tau$ to denote the two edges of the subsystem $A$. 

The $n$-th pseudo-R\'{e}nyi entropy is 
\begin{align}\label{eq:freefermionPRE}
    &S(\tau^{1|2}_A(t)) \nonumber\\ 
    = &\lim_{n\rightarrow1}\frac{1}{1-n} \log \braket{\sigma_n(i\tau,-i\tau)\tilde{\sigma}_n(z+i\tau,z-i\tau)}_{\rm strip} \Big|_{\tau = it} \nonumber\\
    = &\frac{1}{6}
\log\Bigg[\left(\frac{2\alpha_+}{\pi\epsilon}\right)^2 \nonumber\\
&\qquad\quad \frac{\cosh^2\left(\frac{\pi}{\alpha_+}(t-\frac{i}{2}\alpha_-)\right)\sinh^2\left(\frac{\pi\ell}{2\alpha_+}\right)}{\cosh\left(\frac{\pi}{\alpha_+}(t-\frac{i}{2}\alpha_--\frac{\ell}{2})\right)\cosh\left(\frac{\pi}{\alpha_+}(t-\frac{i}{2}\alpha_-+\frac{\ell}{2})\right)}\Bigg]~.
\end{align}
The entanglement entropy for each state is  
\begin{align}\label{eq:freefermionEE}
    &S(\rho^{i}_A(t)) \nonumber\\
    = &\frac{1}{6} \log \left(\left(\frac{4\alpha_i}{\pi\epsilon}\right)^2 \frac{\cosh^2\left(\frac{\pi t}{2\alpha_i}\right) \sinh^2\left(\frac{\pi l}{4\alpha_i}\right)}{\cosh\left(\frac{\pi }{2\alpha_i}(t-\frac{l}{2})\right) \cosh\left(\frac{\pi }{2\alpha_i}(t+\frac{l}{2})\right)} \right). 
\end{align}
For $\alpha_i \ll t \ll l$, 
\begin{align}
    S(\rho^i_A(t)) &= \frac{\pi}{6} \frac{t}{\alpha_i} + \cdots .
\end{align}
On the other hand, for $\alpha_i \ll l \ll t$,
\begin{align}
    S(\rho^i_A(t)) &= \frac{\pi}{12} \frac{l}{\alpha_i} + \cdots .
\end{align}
From this behavior, we can see that at late time, the state mimics a thermal state with temperature $\beta = 4\alpha$.

\subsection{Free Scalar Theory}\label{sec:quenchscalar}

\begin{figure*}[t]
\begin{center}
\includegraphics[scale=0.3]{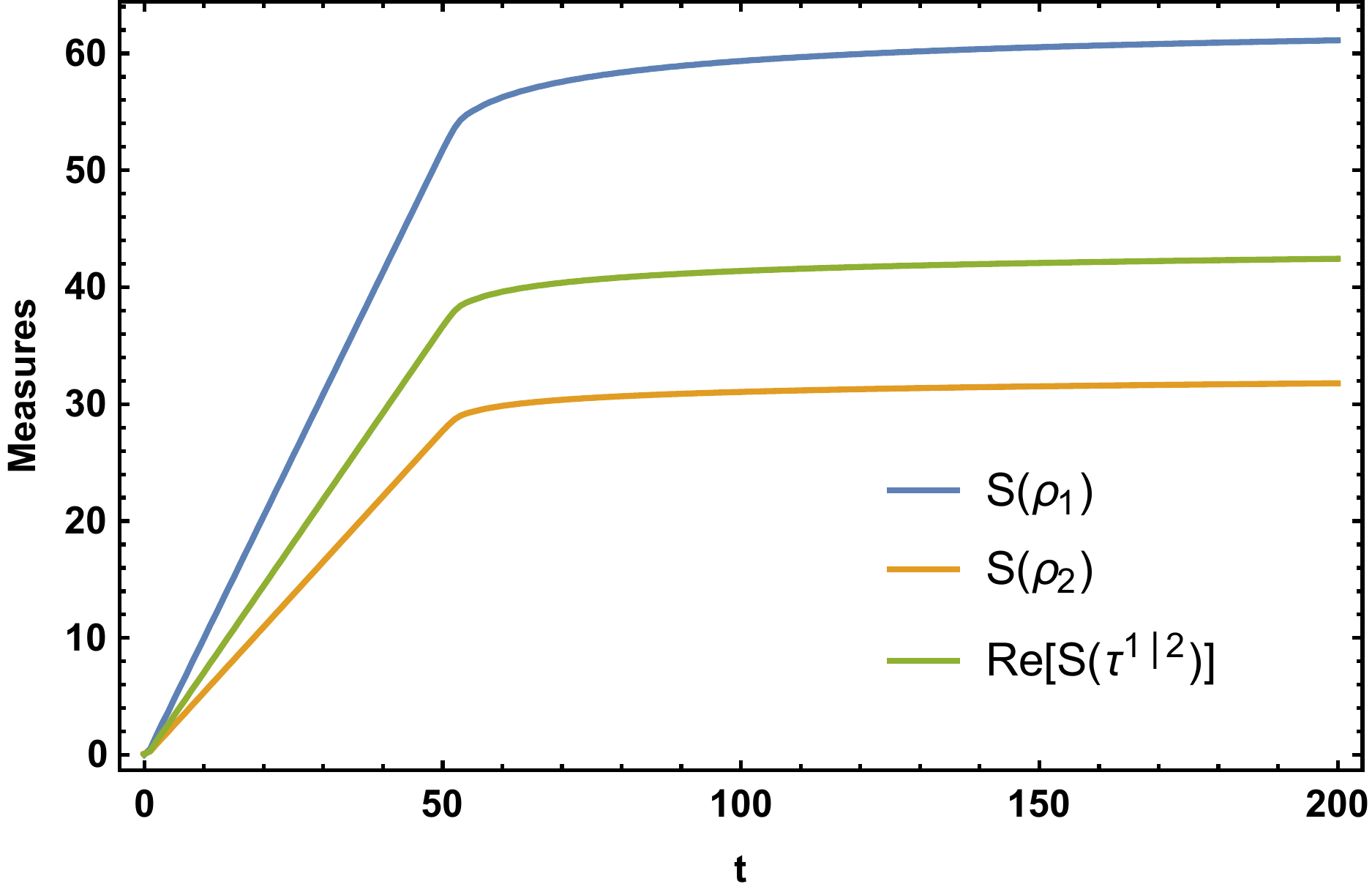}
\includegraphics[scale=0.35]{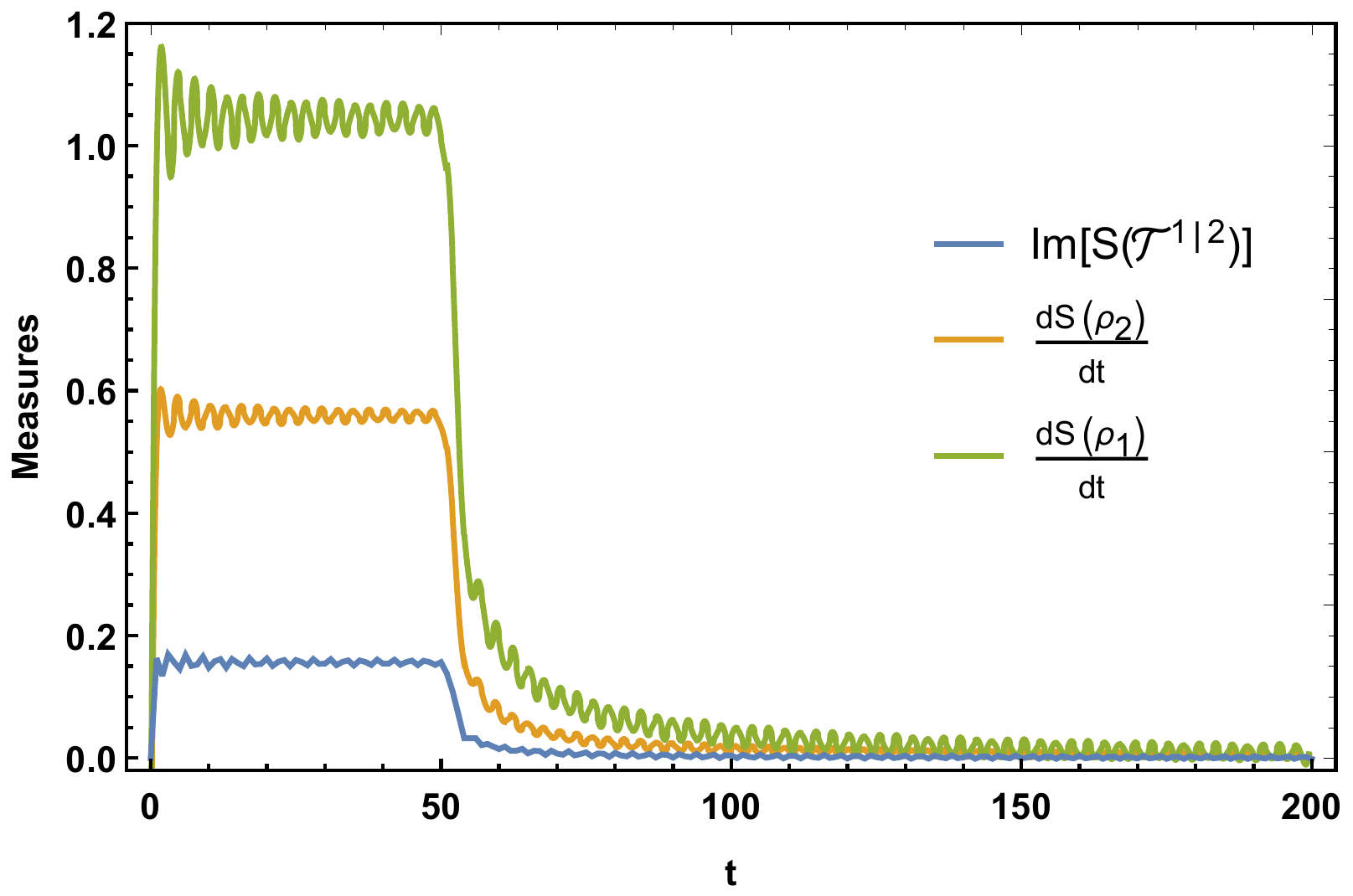}
\includegraphics[scale=0.3]{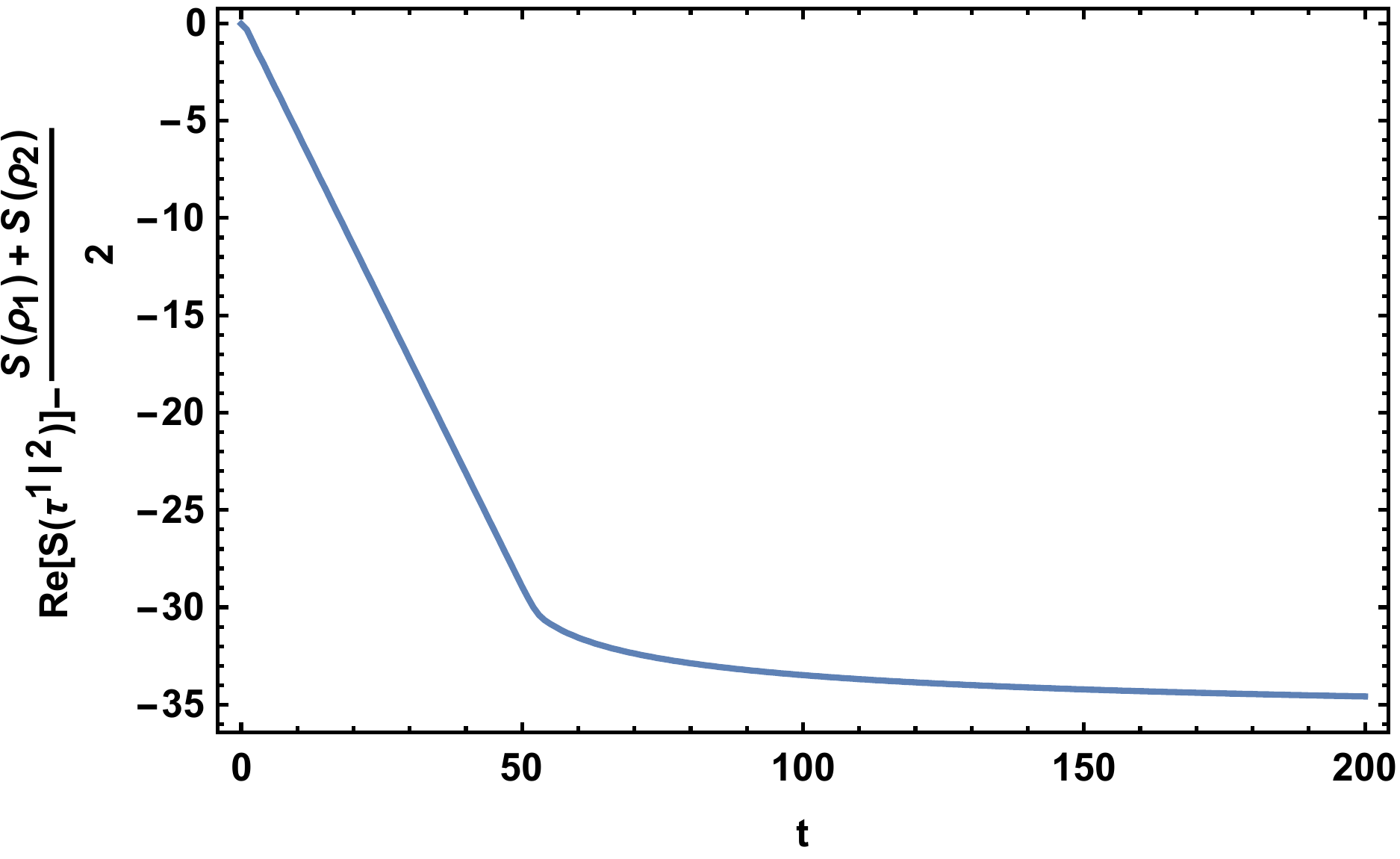}
\includegraphics[scale=0.3]{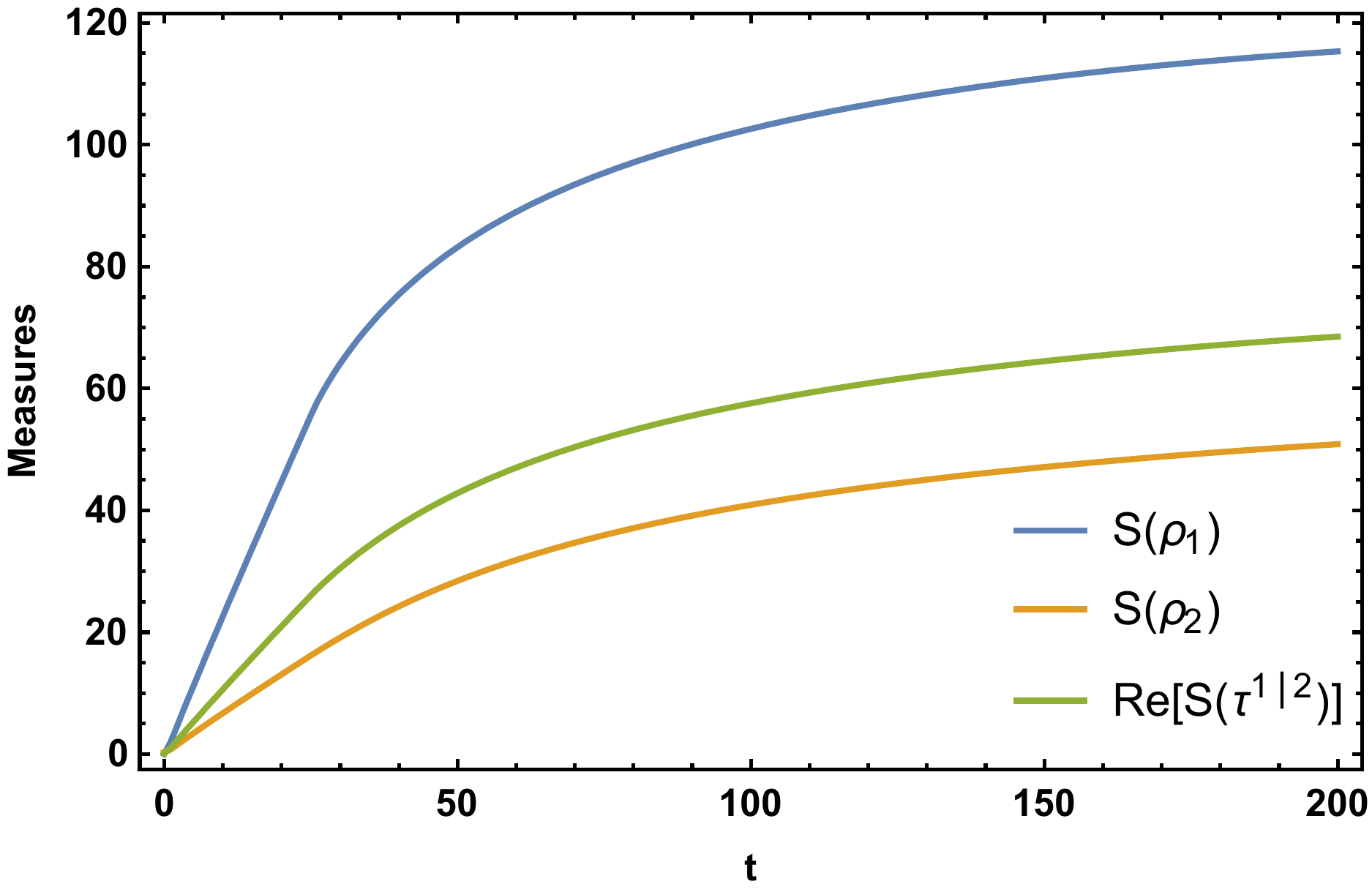}
\includegraphics[scale=0.3]{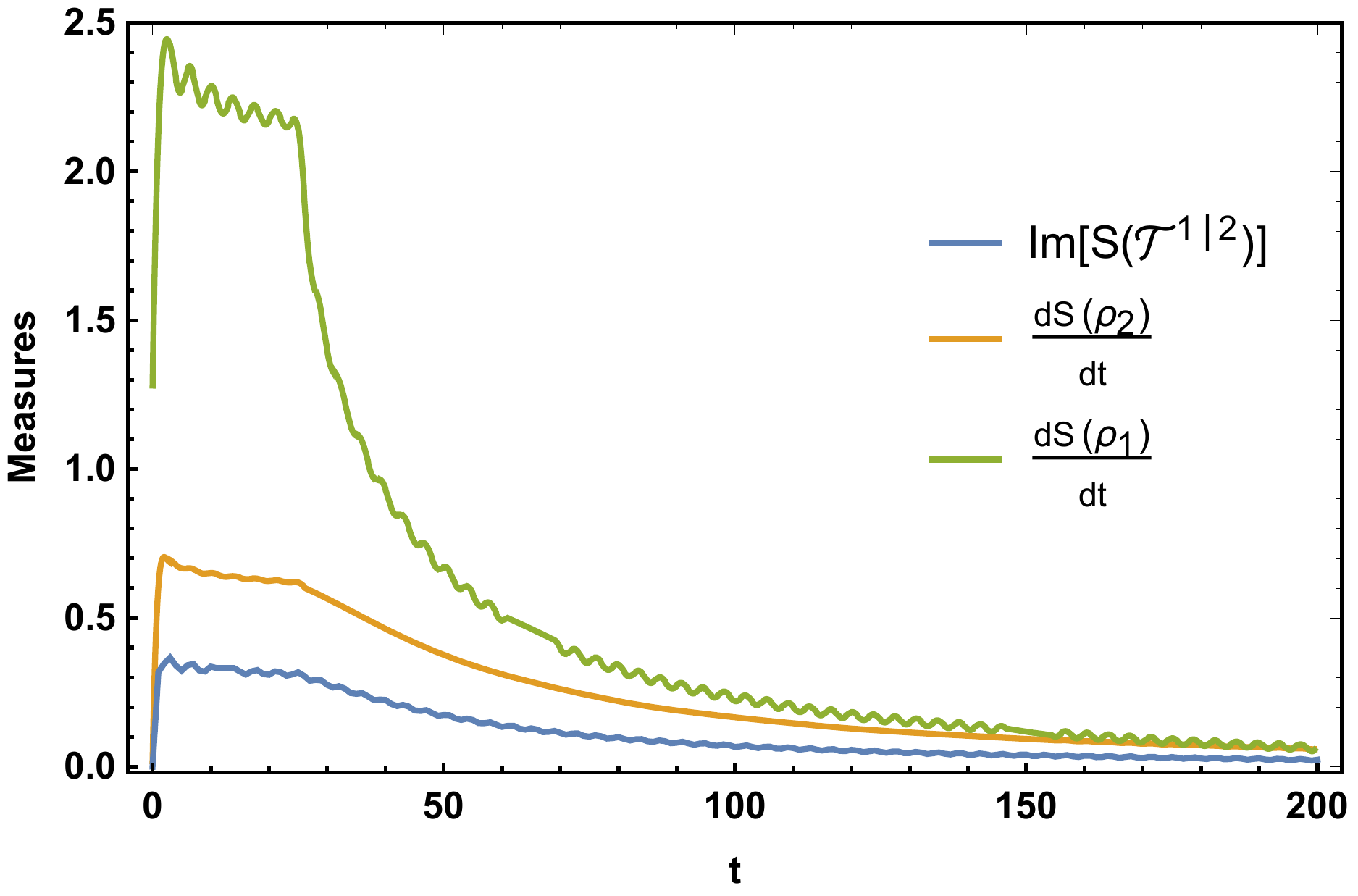}
\includegraphics[scale=0.3]{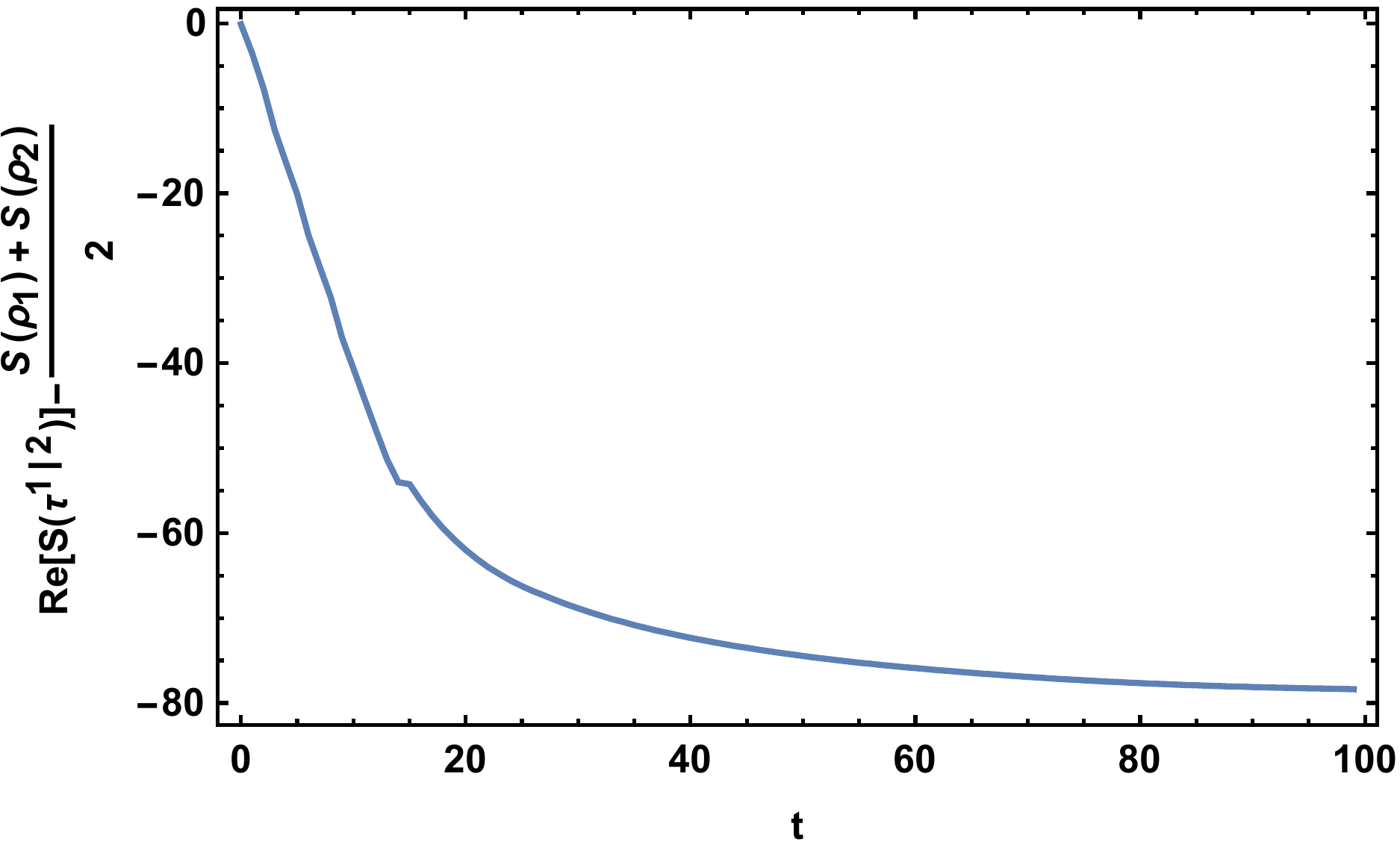}
\includegraphics[scale=0.35]{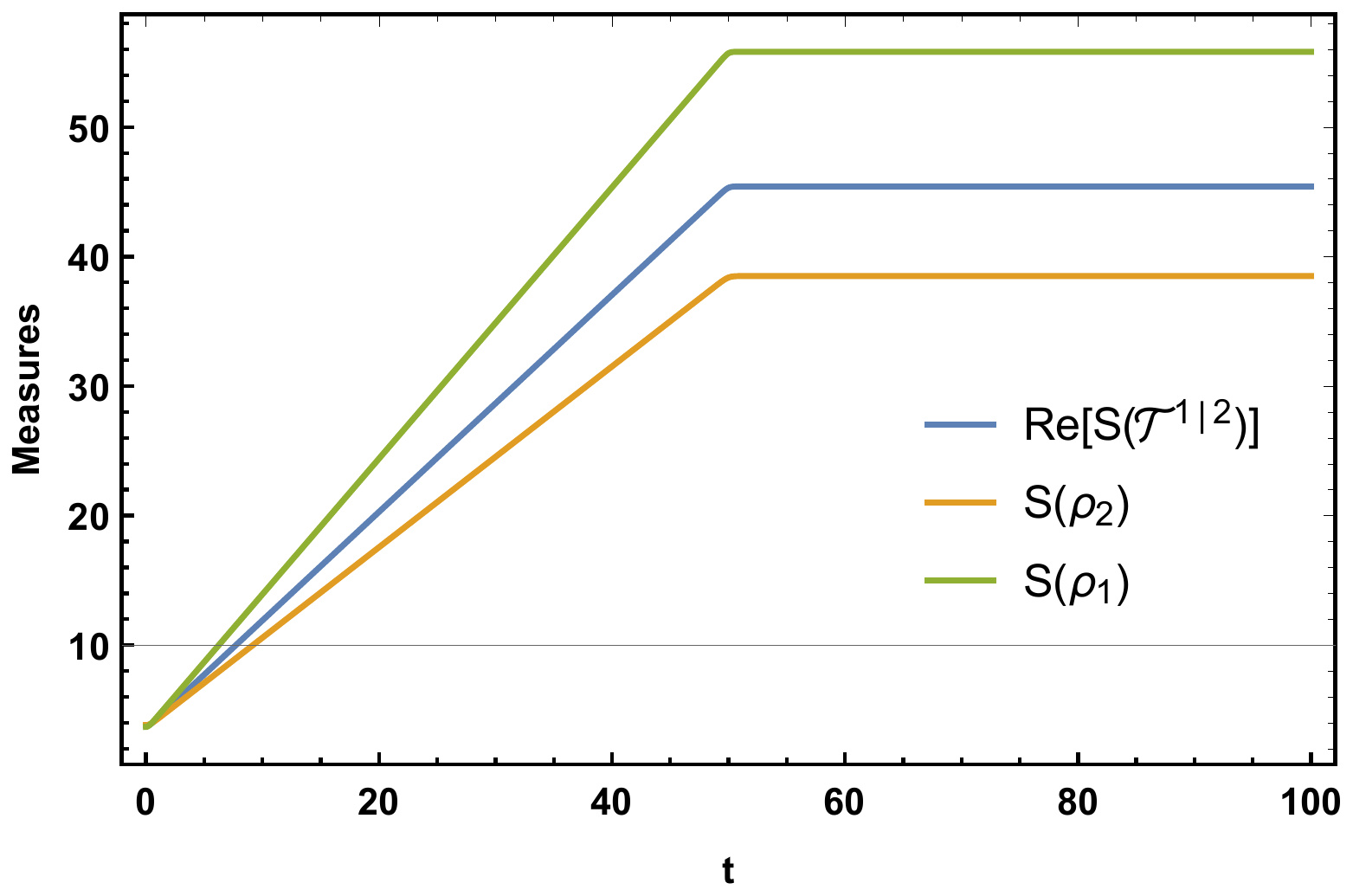}
\includegraphics[scale=0.35]{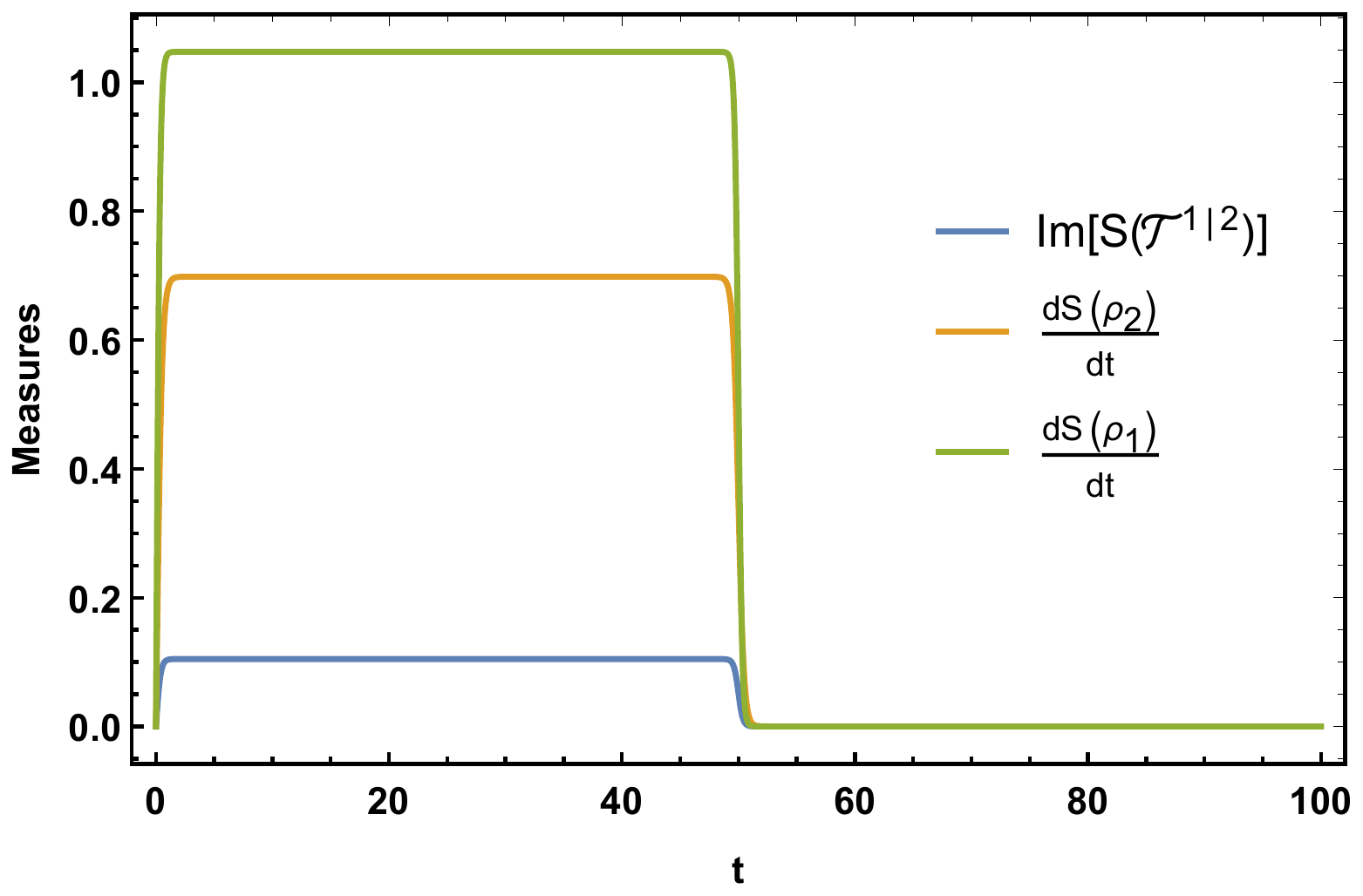}
\includegraphics[scale=0.35]{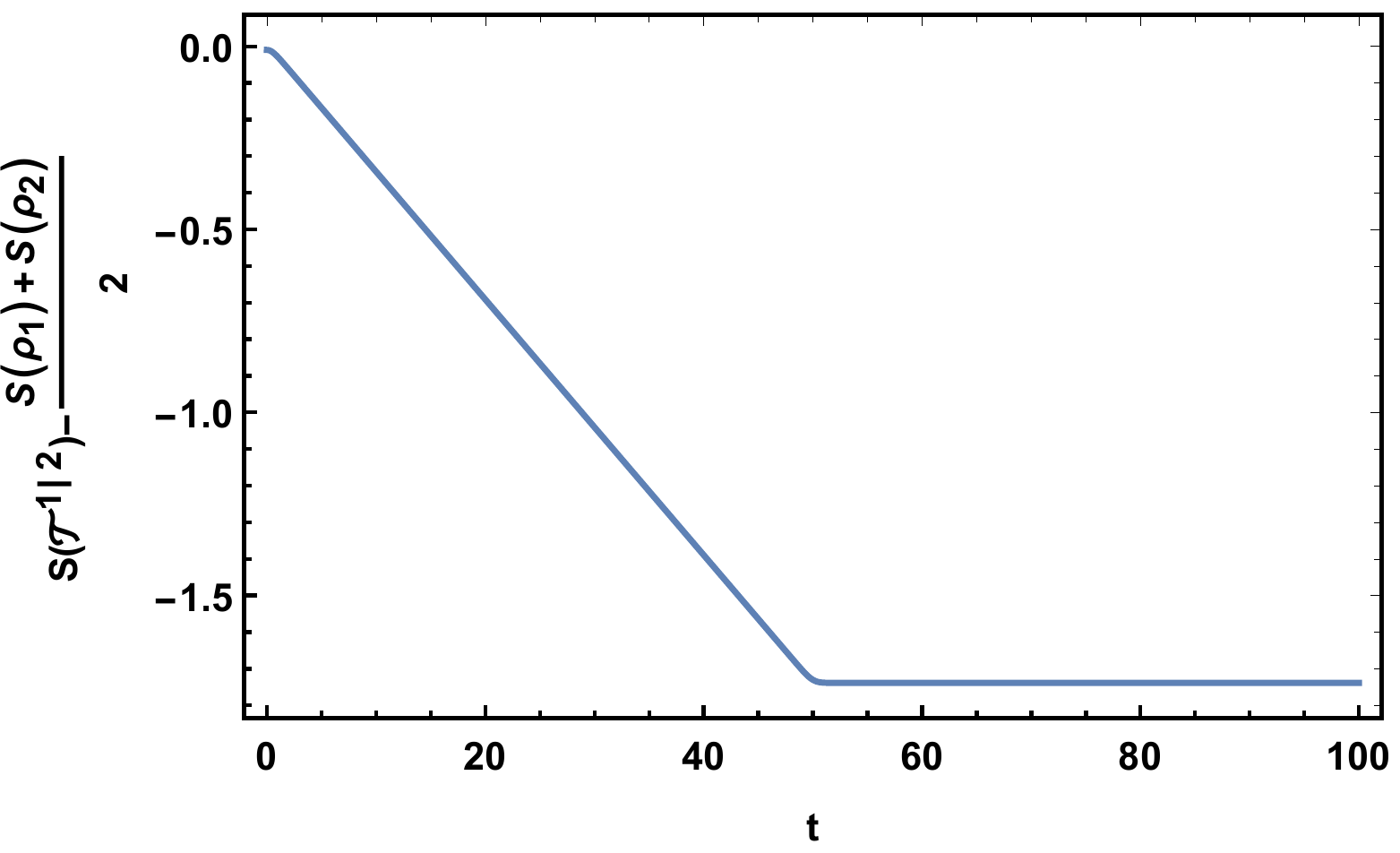}
\caption{Left: Real part of PE and entanglement entropies. Middle: Imaginary part of PE compared with the time derivative of $S(\rho_1)$ and $S(\rho_2)$. Right: $\Delta S$.
The first line corresponds to $z=z_1=z_2=1$ and the socond line corresponds to $z=z_1=z_2=2$. We have set $L_A=100$ on an infinite lattice and $m_1=1$,  $m_2=2$ and  $m=10^{-5}$. In the third line We present the free fermion analytic results where we set $\epsilon=10^{-5}$, $\alpha_1=\/2$ and $\alpha_2=3/4$.}
\label{fig:caseA}
\end{center}
\end{figure*}

In this section we utilize the correlator method to study time evolution of  pseudo entropy followed by quantum quenches in free theories. We consider the pre-quech state to be a Gaussian state, namely the vacuum of a massive theory with a given dynamical exponent and the post-quench state to be again a Gaussian state, the massless vacuum of the theory with the same dynamical exponent.

To find the corresponding correlators, we need to evaluate
\be\label{eq:QO0}
\langle\psi_1(t)|O|\psi_2(t)\rangle=\langle \psi_1(0)|e^{iHt}Oe^{-iHt}|\psi_2(0)\rangle
\ee
for $O=X,P,R$.
This expression contains three type of creation and annihilation operators corresponding to $H_1$ and $H_2$ as well as $H$, where the last evolves the states. 
We expand all the operators in terms of the creation and annihilation operators of $H_2$ as well as the states in terms of its energy modes.

We recall that we have already worked out the desired transformation in \eqref{eq:bog1}
\begin{align}
\begin{split}
a_{i,k}&=\alpha_{i2,k}\, a_{2,k} + \beta_{i2,k}\, a^{\dagger}_{2,k}
\\
a^{\dagger}_{i,k}&=\beta_{i2,k}\, a_{2,k} + \alpha_{i2,k}\, a^{\dagger}_{2,k}
\end{split}
\end{align}
where similar to \eqref{eq:ab}
\begin{align}
\begin{split}
\alpha_{i2,k}&=\frac{1}{2}\left(\sqrt{\frac{\omega_{i,k}}{\omega_{2,k}}}+\sqrt{\frac{\omega_{2,k}}{\omega_{i,k}}}\right)
\\
\beta_{i2,k}&=\frac{1}{2}\left(\sqrt{\frac{\omega_{i,k}}{\omega_{2,k}}}-\sqrt{\frac{\omega_{2,k}}{\omega_{i,k}}}\right).
\end{split}
\end{align}
In these expressions, $a_{i,k}$ stands for the annihilation operator of $\ket{\psi_1}$ or the one appearing in $H$.  
The relation between these two initial states is given by $|0 \rangle_1=\sum_{n=0}^\infty c_{2n}\,| 2n\rangle_{2}
$ 
where similar to \eqref{eq:scalarcn} we have
\be
c_{2n}=\left(-\frac{\beta_{12}}{\alpha_{12}}\right)^n\sqrt{\frac{(2n-1)!!}{2n!!}}\;c_0
\ee
So in this language \eqref{eq:QO0} reduces to
\be\label{eq:QO}
\langle\psi_1(t)|O|\psi_2(t)\rangle=\sum_{n=0}^{\infty} c_{2n} \langle 2n|e^{iHt}Oe^{-iHt}|0\rangle.
\ee
It is not hard to show that that correlators in the momentum space are given by
\begin{align}
\begin{split}
&X(k)=\frac{1}{\omega^2\left(\omega_1+\omega_2\right)}
\bigg[
\omega^2\cos^2(\omega t)
\\ & \;\;\;\;\;
+\omega_1\omega_2\sin^2(\omega t)
+\frac{i}{2}\omega(\omega_2-\omega_1)\sin(2\omega t)
\bigg]
\\
&P(k)=\frac{1}{\left(\omega_1+\omega_2\right)}
\bigg[
\omega \sin^2(\omega t)
\\ & \;\;\;\;\;
+\omega_1\omega_2\cos^2(\omega t)
-\frac{i}{2}\omega(\omega_2-\omega_1)\sin(2\omega t)
\bigg]
\\
&R(k)=\frac{1}{2\omega\left(\omega_1+\omega_2\right)}
\bigg[
\left(\omega_1\omega_2-\omega^2\right)\sin(2\omega t)
\\ & \;\;\;\;\;\;\;\;\;\;
+i\omega(\omega_2-\omega_1)\cos(2\omega t)
\bigg]
\end{split}
\end{align}
where $\omega$ corresponds to the post-quench Hamiltonian and on the right-hand side we have dropped the momentum index for simplicity. It is clear from these expressions that there is non-trivial imaginary contribution in the correlators which leads to non-trivial imaginary spectrum for the transition matrix.  

One can easily check that the above expressions in the $|\psi_2(0)\rangle\to|\psi_1(0)\rangle$ limit reduce to the standard quantum quench which gives the (real) spectrum of the reduced density matrix.

We have shown our numerical results for a finite interval, versus the analytical results for free fermion model for finite interval in the figure \ref{fig:caseA}. In this figure we have considered quenching from a gapped Hamiltonian to a gapeless one to compare our results with CFT results. The left column corresponds to the real part of pseudo entropy together with the corresponding entanglement entropies of both states, the middle column shows the imaginary part of pseudo entropy together with the \textit{time derivative} of the corresponding entanglement entropies, and the right column shows $\Delta S_{12}$ which is negative as expected in all cases.

\subsubsection*{On the imaginary part}
The evolution of the real part of pseudo entropy is very similar to that of entanglement entropies. On the other hand, we numerically observe that the evolution of the imaginary part is very similar to the time derivative of entanglement entropies. Let's take a look at our analytic expressions for the free fermion model to understand this behavior. From the expressions of the pseudo entropy and the corresponding entanglement entropies given in \eqref{eq:freefermionPRE} and
\eqref{eq:freefermionEE}, one can easily see that
$$2^{n}\frac{d^{n}S(\tau^{1|2})}{d\alpha_-^{n}}\Bigg|_{\alpha_-=0}=(-i)^n\frac{d^{n}S(\rho(\alpha_+))}{dt^{n}}$$
with this it is easy to see that
\begin{align}
\mathrm{Re}\left[S(\tau^{1|2})\right]&=\sum_{n=0}^{\infty}\frac{\alpha_-^{2n}}{(2n)!}\frac{d^{2n}S(\tau^{1|2})}{d\alpha_-^{2n}}\Bigg|_{\alpha_-=0}\;,
\\
\mathrm{Im}\left[S(\tau^{1|2})\right]&=\sum_{n=0}^{\infty}\frac{\alpha_-^{2n+1}}{(2n+1)!}\frac{d^{2n+1}S(\tau^{1|2})}{d\alpha_-^{2n+1}}\Bigg|_{\alpha_-=0}\;.
\end{align}
One can see that when $\alpha_-\ll \alpha_+$ (as it is in our figure \ref{fig:caseA}), the linear approximation in the above expansion works well and we have
\be
S(\tau^{1|2})=S(\rho(\alpha_+))-i \frac{\alpha_-}{2}\frac{dS(\rho(\alpha_+))}{dt}+\cdots\;,
\ee
where the linear approximation works well for the cases that $\alpha_1$ and $\alpha_2$ are very different even by an order of magnitude.

\subsection{The Gravity Dual for Quenched Transition Matrices}
\label{sec:quenchGraDual}

In Sec \ref{sec:quenchCFT}, we gave the Euclidean path integral (shown in Fig. \ref{fig:quenchPI}) corresponding to the transition matrix 
\begin{align}\label{eq:TM12}
    \tau^{1|2}(t) = \frac{e^{-itH} e^{-\alpha_1 H} |B\rangle \langle B|e^{-\alpha_2 H} e^{itH}}{\braket{B|e^{-(\alpha_1+\alpha_2) H}|B}} .
\end{align}
The setup is a boundary CFT (BCFT). If the CFT is holographic, we can use the AdS/BCFT correspondence \cite{Takayanagi:2011zk,Fujita:2011fp} to construct its gravity dual. We are going to consider this in this subsection. 

In one word, the resulting geometry is an eternal black-hole with inverse temperature $\beta = 2(\alpha_1 + \alpha_2)$. There is an end-of-the-world brane in the bulk whose location involves imaginary parts, reflecting the imaginary nature of the transition matrix (\ref{eq:TM12}). 

Readers can choose to skip this subsection, and this will not affect reading other parts.

In the following, we firstly review AdS/BCFT and the gravity dual of the density matrix for a conventional quenched state with inverse temperature $\beta = 4\alpha$:
\begin{align}\label{eq:DMquench}
    \rho(t) = \frac{e^{-itH} e^{-\alpha H} |B\rangle \langle B| e^{-\alpha H} e^{itH}}{\braket{B|e^{-2\alpha H}|B}}.
\end{align}
A recipe for constructing the gravity dual of a generaic 2D BCFT is given in \cite{Shimaji:2018czt,Caputa:2019avh}. However, we will not fully apply this recipe because the simplicity of the current setup allows us to make many shortcuts. The gravity dual for the global quench was firstly investigated in \cite{Hartman:2013qma} and further explored in \cite{Almheiri:2018ijj,Cooper:2018cmb,Antonini:2019qkt}. 

After that, we will replace the parameters properly to get the gravity dual for a transition matrix and discuss its properties.

\subsubsection*{The AdS/BCFT Correspondence}

\begin{figure}[t]
    \centering
    \includegraphics[width=8cm]{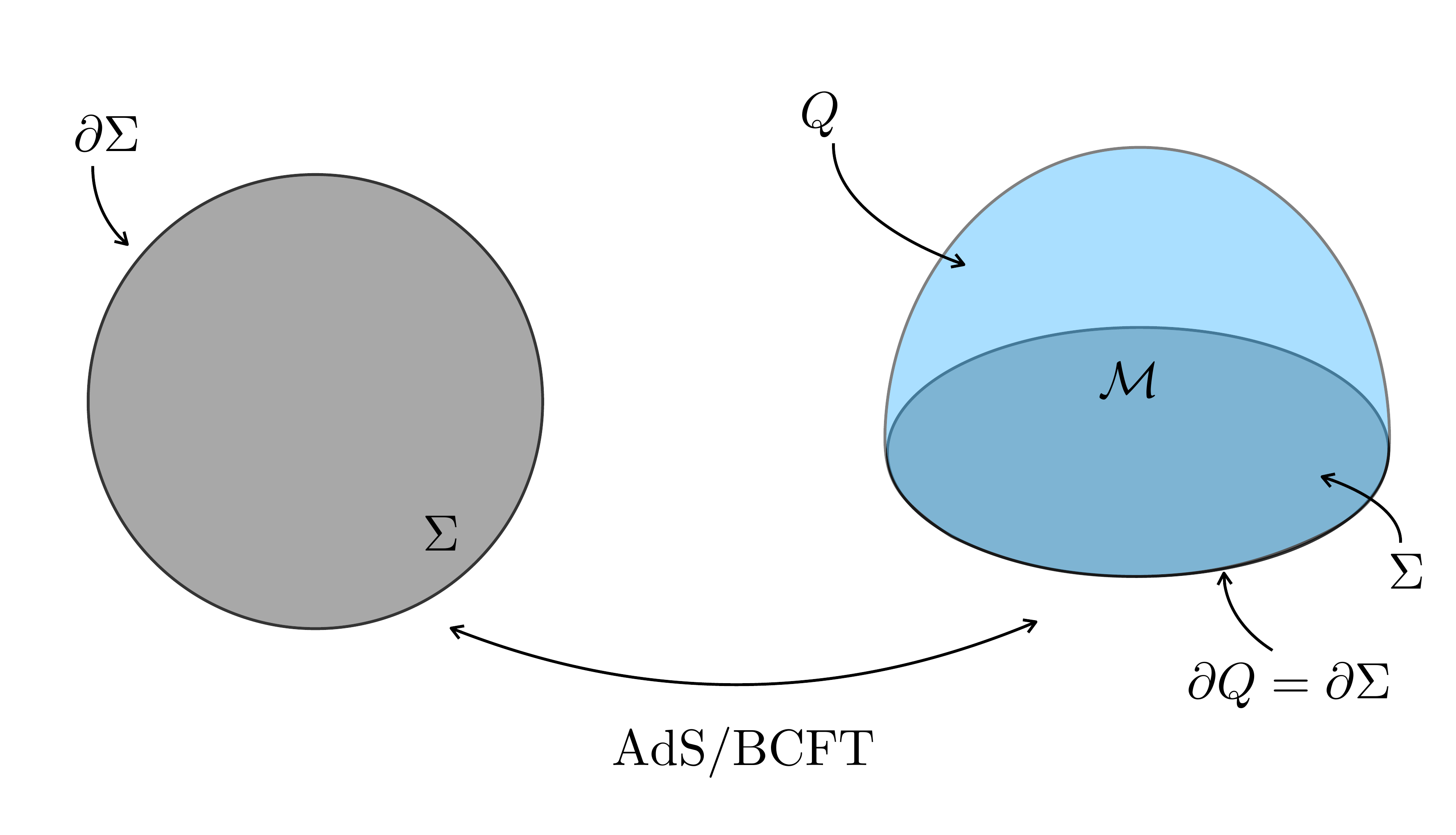}
    \caption{A sketch of the AdS/BCFT correspondence. The left half shows a 2D manifold $\Sigma$ (shown in grey) on which the BCFT is defined. Its gravity dual $\CM$ is shown in the right half. $\CM$ is a portion of a 3D asymptotically AdS spacetime with $\partial\CM = \Sigma\cup Q$. $Q$ (shown in blue) is an end-of-the-world brane with Neumann boundary condition (\ref{eq:Neumann}) imposed on it, and $\partial Q = \partial \Sigma$.}
    \label{fig:AdS/BCFT}
\end{figure}

Consider a BCFT defined on a manifold $\Sigma$. The idea of the AdS/BCFT correspondence is that, since $\Sigma$ has a boundary, its gravity dual $\CM$ should also contains an end-of-the-world brane $Q$ which satisfies $\partial Q = \partial \Sigma$ as a portion of $\partial \CM$. See Fig. \ref{fig:AdS/BCFT}. $\CM$ is determined by solving the Einstein equation in the bulk with Dirichlet boundary condition imposed on $\Sigma$ and Neumann boundary condition 
\begin{align}\label{eq:Neumann}K_{ab}-K h_{ab}+T h_{ab} = 0 .
\end{align}
imposed on $Q$. Here, $K_{ab}$, $h_{ab}$ and $T$ are the extrinsic curvature, the induced metric and the tension of $Q$, respectively. Note that the tension $T$ depends on the details of the BCFT's boundary condition.

\subsubsection*{Gravity Dual of a Global Quench}
\begin{figure}[H]
    \centering    \includegraphics[width=8cm]{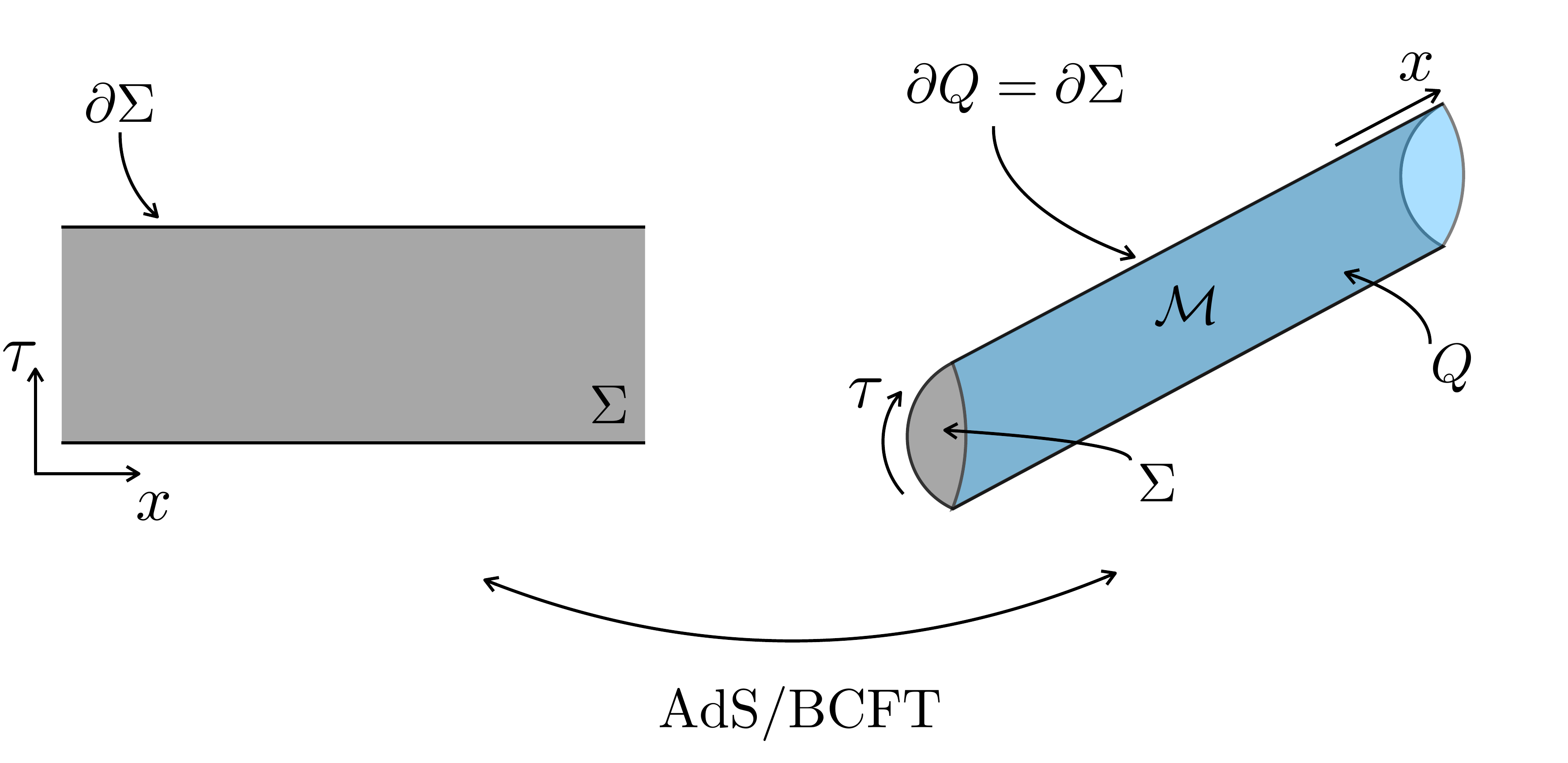}
    \caption{The AdS/CFT correspondence for a Euclidean strip. The asymptotic boundary $\Sigma$ and the end-of-the-world brane $Q$ are shown in grey and blue respectively.}
    \label{fig:DMquenchdual}
\end{figure}

\begin{figure}[H]
    \centering    \includegraphics[width=9cm]{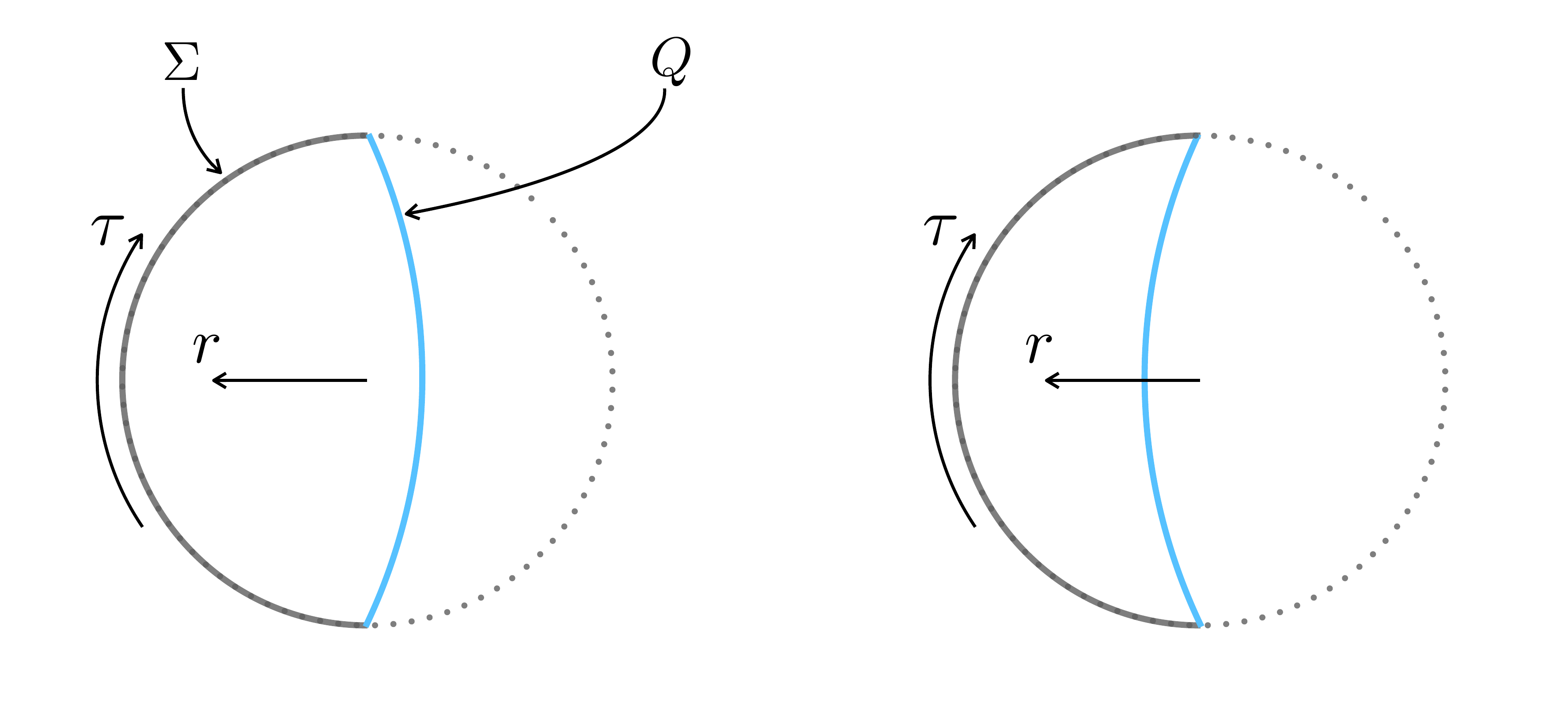}
    \caption{A $x={\rm const.}$ slice of the right panel of Fig. \ref{fig:DMquenchdual}. The solid grey (blue) line shows the asymptotic boundary $\Sigma$ (end-of-the-world brane $Q$), and the dotted grey line shows the asymptotic boundary of a regular Euclidean BTZ black hole. For positive (negative) tension $T$, the gravity dual looks like the left (right) panel where the larger (smaller) portion is taken.}
    \label{fig:EuclideanBTZ}
\end{figure}

The Euclidean path integral corresponding to the density matrix (\ref{eq:DMquench}) of a global quench is a strip with $-\alpha\leq \tau \leq \alpha$. After some messy but straightforward computation, one can figure out that its gravity dual is a Euclidean BTZ black hole with inverse temperature $\beta=4\alpha$
\begin{align}\label{eq:EuclideanBTZ}
    &ds^2 = \frac{dr^2}{r^2 - r_H^2} + r^2 dx^2 + \left(r^2 - r_H^2\right) d\tau^2 \\
    &r_H = \frac{2\pi}{\beta} = \frac{\pi}{2\alpha}
\end{align}
with an end-of-the-world brane $Q$ floating in the bulk. The location of $Q$ is given by
\begin{align}\label{eq:Qlocaltion}
    r_{\rm brane}(\tau) = \frac{r_H}{\sqrt{1-T^2}}\sqrt{1+T^2\tan^2\left(\frac{2\pi\tau}{\beta}\right)} .
\end{align}
for $T\neq0$, and 
\begin{align}\label{eq:zerobrane}
    \tau = \pm \alpha
\end{align}
for $T=0$.
The situation is sketched in Fig. \ref{fig:DMquenchdual}. Also, a spatial slice of the gravity dual is shown in Fig. \ref{fig:EuclideanBTZ}.

Note that, though $T$ and $-T$ give the same value of $r_{\rm brane}$ in Eq. (\ref{eq:Qlocaltion}), these two cases correspond to two different brane configuration. See Fig. \ref{fig:EuclideanBTZ}. For positive tension, the gravity dual is given by the larger portion. For negative tension, the gravity dual is given by the smaller portion.

The gravity dual of a global quench in Lorentzian signature can obtained by simply taking the analytic continuation $\tau\rightarrow it$ in Eq. (\ref{eq:EuclideanBTZ}) and (\ref{eq:Qlocaltion}). Note that however in the Lorentzian case, the $(t,x,r)$ coordinate only covers one AdS-Schwartzchild patch and is not suitable to probe the interior region of the black hole. We will shortly discuss how to probe the whole spacetime at the last part of this subsection. Also, the brane $Q$ intersects the AdS-Schwartzchild patch only when tension $T$ is negative. Therefore, we would like to restrict our discussion to negative tension cases when using the $(t,x,r)$ coordinate.

\subsubsection*{Gravity Dual of Transition Matrix \texorpdfstring{$\tau^{1|2}(t)$}{}}

Since we have already gotten the gravity dual for the density matrix of a globally quenched state, it is straightforward to obtain the grvaity dual for the transition matrix $\tau^{1|2}$. All one should do is to replace $\beta\rightarrow (\beta_1 + \beta_2)/2$ and $\tau \rightarrow it + (-\beta_1+\beta_2)/8$ in in Eq. (\ref{eq:EuclideanBTZ}) and (\ref{eq:Qlocaltion}). Note that $\beta_i = 4\alpha_i$. 

The metric is given by 
\begin{align}
    ds^2 = \frac{dr^2}{r^2 - \left(\frac{4\pi}{\beta_1+\beta_2}\right)^2} + r^2 dx^2 - \left(r^2 - \left(\frac{4\pi}{\beta_1+\beta_2}\right)^2\right) dt^2 . 
\end{align}
This is a BTZ black hole with an averaged inverse temperature $(\beta_1 + \beta_2)/2$. Besides, the brane $Q$ locates at 
\begin{align}
    &r_{\rm brane}(t)  \nonumber\\
    =& \frac{4\pi}{(\beta_1 + \beta_2)\sqrt{1-T^2}}\sqrt{1-T^2\tanh^2\left(\frac{4\pi t}{\beta_1 + \beta_2} -i \frac{\pi}{2}\frac{\beta_2 -\beta_1}{\beta_2 +\beta_1}\right)} .
\end{align}
for negative $T$. 

The key feature of the gravitational configuration for this transition matrix is that the location of the end-of-world brane is given by complex coordinate values, while the background metric itself is real-valued.

\subsubsection*{Complex Brane behind the Horizon}

To probe the region inside the wormhole, we need to extend our coordinate to, for example, the Kruskal coordinate. Let us use $t_-$ to denote the time of the patch we are focusing on and $t_+$ to denote the time of the other side of the wormhole. We can extend to the Kruskal coordinate by applying the diffeomorphism
\begin{align}
    \tilde{U} &= \pm \sqrt{\frac{r-r_H}{r+r_H}} ~ e^{r_H t_{\mp}} , \label{eq:Kruskal1}\\
    \tilde{V} &= \mp \sqrt{\frac{r-r_H}{r+r_H}} ~ e^{-r_H t_{\mp}} . \label{eq:Kruskal2}
\end{align}
The metric turns out to be 
\begin{align}
    ds^2 = \frac{-4 d\tilde{U} d\tilde{V} + (-1+\tilde{U}\tilde{V})^2 ~r_H^{~2} ~dx^2}{(1+\tilde{U}\tilde{V})^2}. 
\end{align}
Here, $\tilde{U}$ and $\tilde{V}$ are light cone coordinates. Let us also introduce a spacelike coordinate $\tilde{X}$ and a timelike coordinate $\tilde{T}$ for convenience: 
\begin{align}
    \tilde{U} = \tilde{T} - \tilde{X} , \\
    \tilde{V} = \tilde{T} + \tilde{X} .
\end{align}
By identifying $\beta\rightarrow (\beta_1 + \beta_2)/2$ and $\tau \rightarrow it_- + (-\beta_1+\beta_2)/8$, we can express the dual geometry of the transition matrix in the Kruskal coordinate. 

Let us consider a zero tension brane. Plugging (\ref{eq:zerobrane}) into (\ref{eq:Kruskal1}) and (\ref{eq:Kruskal2}) and erasing $r$, we get 
\begin{align}
    \tilde{X} = \frac{\tilde{T}}{i\tan\left(\frac{\pi \beta_1}{\beta_1+\beta_2}\right)}. 
\end{align}
This is the location of the zero tension brane in the gravity dual of the transition matrix. We can see that the coordinates should take complex values in this case. Note that when $\beta_1 = \beta_2$, the brane locates at $\tilde{X}=0$, and this gives the well-known result in the conventional global quench setup.

\section{Pseudo Entropy in Perturbed CFTs} \label{sec:pertpe}

In this section, we explore universal properties of pseudo entropy under perturbations of a given field theory. Especially we focus on perturbations of two dimensional CFT and would like to examine the difference between the pseudo entropy and the entanglement entropy.

\subsection{Toy Example: Two Qubit System}

Before we go on to studies of CFTs,  we would like to start with perturbations of two qubit states as a toy example. Consider the state with an angle $\theta_1$
\ba
|\psi_1\lb_{AB}=\cos\theta_1|00\lb_{AB}+\sin\theta_1|11\lb_{AB},
\ea
and simiarly $|\psi_2\lb_{AB}$ with another angle $\theta_2$.
We assume the range $0\leq \theta_1,\theta_2\leq \frac{\pi}{2}$.
The pseudo entropy for these states is found as
\ba
&& S(\tau^{\psi_1|\psi_2}_A)  \no
&& =-\left(\frac{\cos\theta_1\cos\theta_2}{\cos(\theta_1-\theta_2)}\right) \log\left(\frac{\cos\theta_1\cos\theta_2}{\cos(\theta_1-\theta_2)}\right)\no
&&-\left(\frac{\sin\theta_1\sin\theta_2}{\cos(\theta_1-\theta_2)}\right) \log\left(\frac{\sin\theta_1\sin\theta_2}{\cos(\theta_1-\theta_2)}\right).\no
\ea

We are interested in a small perturbation $\theta_2=\theta_1+\delta$. Then the interesting difference looks like
\ba
&& 2S(\tau^{\psi_1|\psi_2}_A) -\left(S(\rho^{\psi_1}_A)+S(\rho^{\psi_2}_A)\right) \no
&& \simeq f(\theta_1)\delta^2+O(\delta^3),  \label{difpetwo}
\ea
where we defined
\ba
f(\theta)=\frac{1}{2}+\frac{\cos 2\theta}{2}\cdot \log \tan^2\theta.
\label{fthea}
\ea
It is easy to see that this function takes both the positive and negative value. 
The positive values are localized around the maximally entangled points $\theta=\frac{\pi}{4}$.
On the other hand, when the state has small entanglement, the difference (\ref{difpetwo}) gets 
negative.

\subsection{First Law Like Relation}

Next we would like to examine a universal property of pseudo entropy under 
infinitesimal perturbations, which is analogous to the first law like relation in 
entanglement entropy \cite{Blanco:2013joa,Bhattacharya:2012mi}.
Consider two transition matrices: $\tau_0$ and its infinitesimally perturbed one $\tau$, whose difference is written as 
\ba
\delta\tau\equiv \tau-\tau_0.
\ea

Consider a generalization of relative entropy for the transition matrices, defined as
\ba
S(\tau|\tau_0)=\mbox{Tr}[\tau\log\tau]-\mbox{Tr}[\tau\log\tau_0].
\ea
If we expand $S(\tau|\tau_0)$ with respect to the infinitesimal $\delta\tau$, the linear term vanishes and the leading term is the quadratic one:
\ba
S(\tau|\tau_0)\!\simeq\! \int^\infty_0\! dt\!~ \mbox{Tr}\left[\frac{t}{(t+\tau_0)^2}\delta\tau
\frac{1}{t+\tau_0}\delta\tau\right].  \label{quadp}
\ea

Since the linear term vanishes, we obtain the first law like relation for pseudo entropy:
\ba
S(\tau)-S(\tau_0)\simeq \la H\lb_{\tau}-\la H\lb_{\tau_0}+O(\delta\tau^2),
\ea
where $H=-\log \tau_0$ is a `pseudo' modular Hamiltonian.

If we take into account the perturbation up to the quadratic order (\ref{quadp}), we get 
\ba
&& S(\tau)-S(\tau_0)\no
&&\simeq \la H\lb_{\tau}-\la H\lb_{\tau_0}-\int^\infty_0 dt\ \mbox{Tr}\left[\frac{t}{(t+\tau_0)^2}\delta\tau
\frac{1}{t+\tau_0}\delta\tau\right].  \no
\label{pertgh}
\ea

Consider two excited states $|\psi_1\lb$ and $|\psi_2\lb$ such that they are very closed to the vacuum $|0\lb$ in a given field theory. Then the difference between the pseudo entropy $\tau^{\psi_1|\psi_2}_A$ and the entanglement entropy of the vacuum
$S(\rho_A)$ is estimated from the first law as follows:
\ba
S(\tau^{\psi_1|\psi_2}_A)-S(\rho_A)\simeq \frac{\la \psi_2|H_A|\psi_1\lb}{\la \psi_2|\psi_1\lb}+O(\delta \tau^2), \no \label{perty}
\ea
where $H_A$ is the modular Hamiltonian given by $H_A=-\log\rho_A+h_0$, where the constant $h_0$ is chosen such that $\la 0|H_A|0\lb=0$. 

We can further consider the quadratic order term
by using (\ref{pertgh}). This shows the difference $S(\tau^{\psi_1|\psi_2}_A)+S(\tau^{\psi_2|\psi_1}_A)-S(\rho_{1A})-S(\rho_{2A})$ starts from quadratic order of the perturbations of the state, though the sign of the quadratic term depends on perturbations.

\subsection{Pseudo Entropy in Perturbed CFTs}

Now we would like to study the behavior of pseudo entropy when we perturb a two dimnensional 
CFT by a primary operator $O(x_1,x_2)$ with the conformal dimension $h$. The action is perturbed as follows
\ba
S=S_{CFT}+\int d^2x \lambda(x)O(x).  \label{pertl}
\ea
We take $|\psi_1\lb$ as the ground state of the original CFT and $|\psi_2\lb$ as that of the perturbed CFT. The reduced transition matrix $\tau_A$ and the inner product $\la \psi_2|\psi_1\lb$ 
can be described by choosing
\ba
\lambda(x)=\lambda\theta(x_1),
\ea
where $x_1$ is the coordinate of the Euclidean time and $\theta(x)=\frac{|x|}{2x}+\frac{1}{2}$ is the step function. 

The Euclidean two dimensional space $\mathbb{R}^2$ is described by the complex coordinate $(w,\bar{w})$, where $w=x_2+ix_1$. The subsystem $A$ is chosen to be the interval $0\leq x_2\leq L$ at the time $x_1=0$. The trace of products of the reduced transition matrix is given as follows 
via the standard replica method:
\ba
\mbox{Tr}[(\tau_A)^n]\!=\!\frac{\la e^{-\int d^2x \lambda(x) O(x)}\lb_{\Sigma_n}}
{\left(\la e^{-\int d^2x  \lambda(x) O(x)}\lb_{\Sigma_1}\right)^n},
\ea
where $\Sigma_n$ is the $n$-sheeted Riemann surface obtained by replicating the complex plane 
$\Sigma_1$ along the interval $A$.

\begin{figure}[H]
    \centering
    \includegraphics[width=7.5cm]{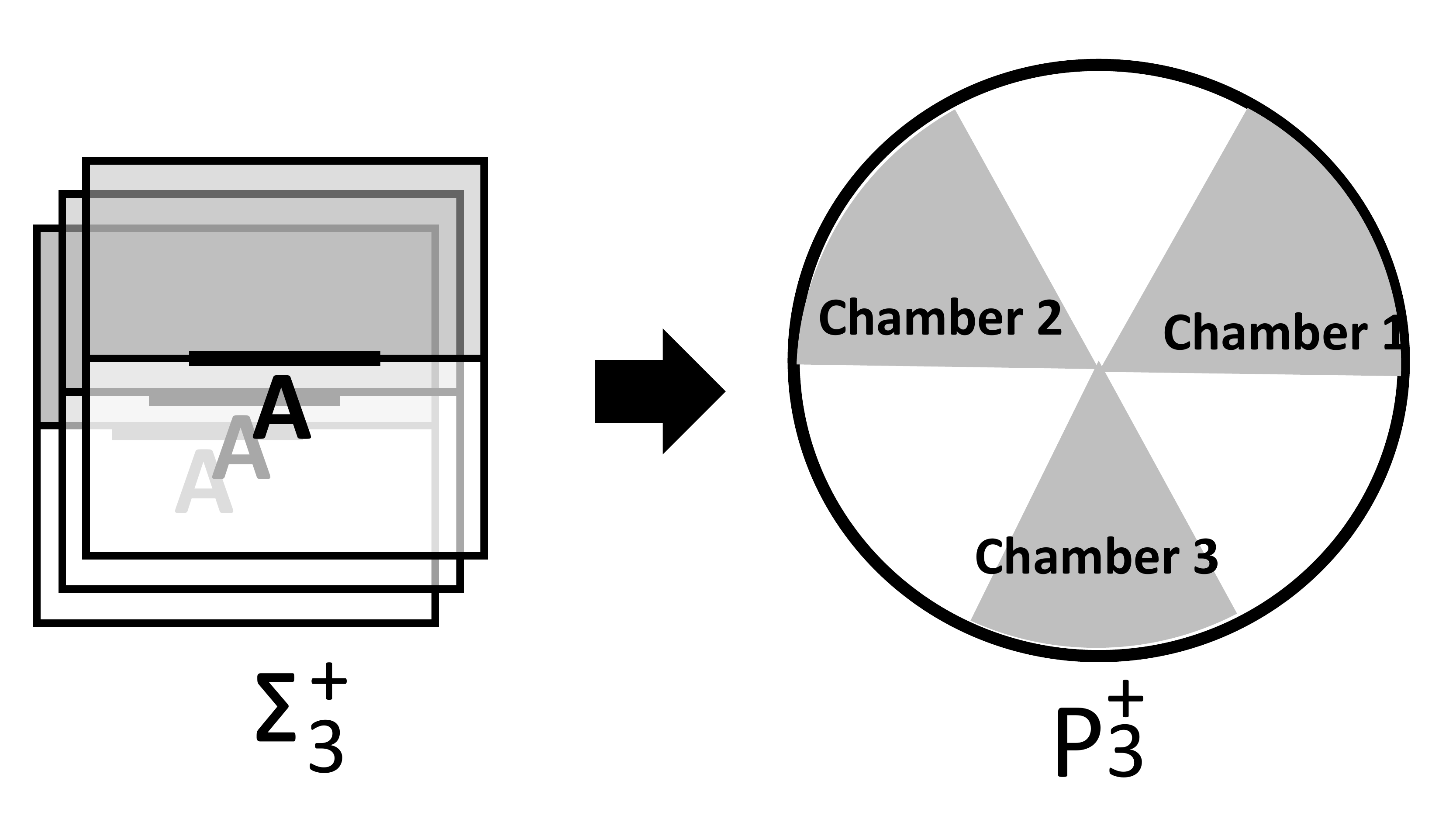}
    \caption{The conformal map from a n-replicated upper complex planes $\Sigma^+_n$ to a Pizza slice geometry $P^+_n$. We wrote $n=3$ explicitly in the above pictures. The colored regions in the left and right picture consist of $n=3$ disconnected regions, which are $\Sigma^+_n$ and 
    $P^+_n$, respectively. The $n$-sheeted surface with both colored and white region in the left 
    represents $\Sigma_n$.}
    \label{fig:replicam}
\end{figure}

We write the reduced transition matrix at $\lambda=0$, the vacuum of the original CFT, as  $\rho_A$ i.e. $\tau_A|_{\lambda=0}=\rho_A$. By considering the perturbation with respect to $\lambda$,
we can evaluate the the difference of $n$-th pseudo R\'{e}nyi entropy up to the quadratic order as follows

\ba
&& S^{(n)}(\tau_A)-S^{(n)}(\rho_A)=\frac{1}{1-n}\log 
\left[\frac{\mbox{Tr}[(\tau_A)^n]}{\mbox{Tr}[(\rho_A)^n]}\right]\no
&&\simeq \frac{\lambda^2}{2(1-n)}\Biggl[\int_{\Sigma^{+}_n} dw^2_1 dw^2_2 \la O(w_1)O(w_2)\lb_{\Sigma_n}\no
&&-n\int_{\Sigma^{+}_1} dw^2_1 dw^2_2 \la O(w_1)O(w_2)\lb_{\Sigma_1}\Biggr].
\ea
Here $\Sigma^{+}_n$ is the upper half of the surfaces, where the perturbation is localized,
as depicted in the left of Fig.\ref{fig:replicam}.

To calculate the two point functions on $\Sigma_n$, we employ the conformal map from 
$\Sigma_n$ into a single complex plane, as sketched in Fig.\ref{fig:replicam}, given by
\ba
z^n=\frac{w}{w-L}.  \label{mapn}
\ea
Since the two point function on a complex plane $\Sigma_1$ is given by the standard formula:
\ba
\la O(w_1)O(w_2)\lb_{\Sigma_1}=\frac{1}{|w_1-w_2|^{4h}},
\ea
the conformal invariance in two dimensional CFTs allows us to evaluate as follows:
\ba
&& \int_{\Sigma^{+}_n} d^2w_1 d^2w_2 \la O(w_1)O(w_2)\lb_{\Sigma_n}  \no
&&= \int_{P^{+}_n}d^2z_1 d^2z_2 \left|\frac{dz_1}{dw_1}\right|^{2(1-h)}
\left|\frac{dz_2}{dw_2}\right|^{2(1-h)} |z_1-z_2|^{-4h}\no
&&=\int_{P^{+}_n}d^2z_1 d^2z_2  \left|f(z_1)f(z_2)\right|^{2(1-h)}
|z_1-z_2|^{-4h},\no  \label{intzwq}
\ea
where we introduced the function
\ba
f(z)=\frac{nLz^{n-1}}{(z^n-1)^2}.
\ea
In the above, the region $P^{+}_n$ is the image of $\Sigma^+_{n}$ by the conformal 
transformation (\ref{mapn}), depicted in the right of Fig.\ref{fig:replicam}.
$P^{+}_n$ is explicitly given by
\ba
P^+_n=\{z=r e^{i\theta}|0\leq r<\infty,\ \theta\in Q_n\},
\ea
where
\ba
Q_n=\cup_{k=0}^{n-1}\left[\frac{2\pi k}{n},\frac{2\pi k}{n}+\frac{\pi}{n}\right].  \label{qset}
\ea
Below we call the $n$ disconnected regions in $Q_n$ (\ref{qset}) as $n$ chambers.

Since the integral (\ref{intzwq}) gets divergent when $z_1$ and $z_2$ get closer, we need 
a UV regularization. For this purpose, it is useful to rewrite the integral on $\Sigma_1$ 
in terms of $z$ coordinate via the map (\ref{mapn}) as follows
\ba
&&\int_{\Sigma^{+}_1} dw^2_1 dw^2_2 \la O(w_1)O(w_2)\lb_{\Sigma_1}\no
&&=\int_{P^{+}_1}d^2z_1 d^2z_2  
\left|f(z_1)f(z_2)\right|^{2(1-h)}
\left|\frac{g(z_1,z_2)}{z_1-z_2}\right|^{4h},\no
\ea
where we defined
\ba
g(z_1,z_2)=n\frac{(z_1-z_2)(z_1z_2)^{\frac{n-1}{2}}}{z^n_2-z^n_1},
\ea
and the single chamber region $P^+_1$ as
\ba
P^+_1=\{z=r e^{i\theta}|0\leq r<\infty,\ 0\leq \theta\leq \frac{\pi}{n}\}.
\ea
We would like to note the following useful properties:
\ba
\lim_{z_2\to z_1}g(z_1,z_2)=1,\ \   \mbox{and} \ \ \ |g(z_1,z_2)|\leq 1.\ \  \label{dwww}
\ea

In this way, the difference of pseudo entropy is represented as follows:
\ba
&& S^{(n)}(\tau_A)-S^{(n)}(\rho_A)\no
&& \simeq \frac{\lambda^2}{2(1-n)}\Biggl[\int_{P^{+}_n}d^2z_1 d^2z_2  |f(z_1)f(z_2)|^{2(1-h)}|z_1-z_2|^{-4h} \no
&&-n\int_{P^{+}_1}\!d^2z_1 d^2z_2\!   |f(z_1)f(z_2)|^{2(1-h)}\!
\left|\frac{g(z_1,z_2)}{z_1-z_2}\right|^{4h}\!\Biggr],
\ea 
If we set $g=1$ in the above integral, the contributions from the case where $z_1$ and $z_2$
cancels completely between the difference, which remains the negative contributions from the first 
integral (we assume $n> 1$) where $z_1$ and $z_2$ belong to different chambers. 
In addition we know the property 
$|g|< 1$. Thus we can conclude that the above difference is negative at the leading perturbation $O(\lambda^2)$: 
\ba
S^{(n)}(\tau_A)-S^{(n)}(\rho_A)< 0.
\ea
This shows the basic property of pseudo entropy in CFTs that it gets decreased under 
perturbations.

\subsection{Exactly Marginal Perturbation}

Finally, let us focus on the exactly marginal case i.e. $h=1$ to study more details.
The leading divergence of the integral of $z_1$ and $z_2$ is logarithmic and this occurs when $|z_1-z_2|$ vanishes. The difference of pseudo entropy is simplified as 
\ba
&& S^{(n)}(\tau_A)-S^{(n)}(\rho_A)\no
&& \simeq \frac{\lambda^2}{2(1-n)}\Biggl[\int_{P^{+}_n}d^2z_1 d^2z_2  \frac{1}{|z_1-z_2|^{4}}\no
&&-n\int_{P^{+}_1}d^2z_1 d^2z_2  \frac{1}{|z_1-z_2|^{4}}\Biggr].
\ea 

Since the UV divergences around $z_1=z_2$ are canceled out when both $z_1$ and $z_2$ are 
in the same chamber, we can focus on the region where
 $z_1$ and $z_2$ are in different chambers. 
Then the divergence arises in the two 
limits $z_1,z_2\to 0$ or $z_1,z_2\to \infty$, corresponding to the limits that the coordinate 
$w_1$ and $w_2$ both get closer to an end point of the interval $A$.

This consideration leads to the following estimation
\ba
&&  S^{(n)}(\tau_A)-S^{(n)}(\rho_A)\no
&&\simeq \!\frac{n\lambda^2}{2(1-n)}\!\!\int^{\infty}_0\!\! dr_1\! dr_2\! 
\int^{\frac{\pi}{n}}_0\! d\theta_1\! \sum^{n-1}_{k=1}\!\int^{\frac{2\pi k}{n}+\frac{\pi}{n}}_{\frac{2\pi k}{n}}
\!\!d\theta_2\frac{r_1r_2}{|r_1e^{i\theta_1}\!-\!r_2e^{i\theta_2}|^4}.\no  \label{intthephs}
\ea
The mentioned divergences arise around $r_1=r_2=0$ and  $r_1=r_2=\infty$. 

To estimate the integral in (\ref{intthephs})
we rewrite the integral using new variables as
$(r_1,r_2)=\rho (\cos\phi,\sin\phi)$, where $0\leq \rho<\infty$ and 
$0\leq\phi\leq\frac{\pi}{2}$. We regulate the divergence at $\rho=0$ 
by restricting $\delta<\rho$ and this cut off $\delta$ is related to the lattice spacing $\ep$ 
as $\delta\sim \frac{\ep}{L}$. Since the divergences from  $r_1=r_2=\infty$ is 
 equal to the one from $r_1=r_2=0$,  finally we can evaluate the logarithmic divergent 
 contribution as follows
\ba
 S^{(n)}(\tau_A)-S^{(n)}(\rho_A)\!
\simeq\! \frac{n\lambda^2}{1-n} c_n \log\frac{L}{\ep},\ \ \ \ 
\ea
where the coefficient $c_n$ is found as
\ba
c_n=\int^{\frac{\pi}{2}}_0\! d\phi\! \int^{\frac{\pi}{n}}_{0}d\theta_1 
\! \sum^{n-1}_{k=1}\!\int^{\frac{2\pi k}{n}+\frac{\pi}{n}}_{\frac{2\pi k}{n}}\!\f{d\theta_2}
{\left(1-\sin2\phi\cos(\theta_1-\theta_2)\right)^2}.\no
\ea

Explicit numerical analysis for $n=2,3,\ddd,10$ implies that $c_n$ is a monotonically increasing 
function of $n$. Thus we expect $\frac{c_n}{1-n}$ approaches to a negative $O(1)$ value $c'$, 
leading to the behavior of pseudo entropy at $n=1$:
\ba
 S(\tau_A)-S(\rho_A)\simeq\! -c'\lambda^2 \log\frac{L}{\ep}<0.  \label{difexac}
\ea
Note that the entanglement entropy of perturbed vacuum $|\psi_2\lb$ 
is identical to that of the unperturbed one $|\psi_1\lb$ 
as we consider exactly marginal perturbations. Therefore the above result (\ref{difexac}) coincides a half of 
the difference $2\Delta S_{12}=2S(\tau^{\psi_1|\psi_2}_A) -\left(S(\rho^{\psi_1}_A)+S(\rho^{\psi_2}_A)\right)$, 
which has been considered in this article time to time. 

\section{Holographic Pseudo Entropy for Perturbed CFTs} \label{sec:hol} 

In this section, we would like to study properties of pseudo entropy in CFTs by employing 
the holographic calculation given in \cite{Nakata:2021ubr}. The inner product $\la \psi_2|\psi_1\lb$
is dual to a partition function of gravity in a Euclidean time-dependent background, such that 
the initial half of gravitational path-integral is dual to the state $|\psi_1\lb$ in the dual CFT, while the latter half is dual to $|\psi_2\lb$. Similarly, by cutting along the middle time slice we can have a gravity dual of the transition matrix $\tau^{\psi_1|\psi_2}_A$. Therefore, via a straightforward extension of holographic entanglement entropy \cite{RT,HRT}, the holographic pseudo entropy in a $d$ dimensional CFT is computed from the area of minimal surface $\Gamma_A$ in $d+1$ dimensional asymptotically Euclidean AdS space as
\ba
S(\tau^{\psi_1|\psi_2}_A)=\frac{\mbox{Area}(\Gamma_A)}{4G_N},
\ea
where $G_N$ is the Newton constant of the $d+1$ dimensional gravity. $\Gamma_A$ ends on the boundary of $A$ on the AdS boundary and is homologous to $A$. Notice that $S(\tau^{\psi_1|\psi_2}_A)$ always takes real and positive values for a classical gravity dual with a real valued metric, which also shows 
$S(\tau^{\psi_1|\psi_2}_A)=S(\tau^{\psi_2|\psi_1}_A)$.

Below we will evaluate the difference $2\Delta S_{12}=2S(\tau^{\psi_1|\psi_2}_A) -\left(S(\rho^{\psi_1}_A)+S(\rho^{\psi_2}_A)\right)$ for two vacuum states $|\psi_1\lb$
and $|\psi_2\lb$ in two different quantum field theories. First we consider the case where 
the two field theories are both CFTs, related by an exactly marginal perturbation. In this case the difference always turns out to be negative.
Next we
study a class of example where two states are vacuum states of two different massive perturbations of 
a CFT. The result shows the difference can be both positive and negative depending on the detailed values of parameters.

\subsection{Holographic Pseudo Entropy in Gapless Phases via Janus Solutions}

An important class of gravity configurations dual to interfaces between CFTs related to each other via exactly marginal deformations is called Janus solutions \cite{Bak:2003jk,Freedman:2003ax,Clark:2004sb,Clark:2005te,DHoker:2006vfr,Bak:2007jm}.
A $(d+1)$-dimensional Janus solution takes the general form: 
\ba
ds^2=d\rho^2+e^{h(\rho)}ds^2_{AdS_d},  \label{xxe}
\ea
where $ds^2_{AdS(d)}$ is $d$ dimensional (Euclidean) AdS metric
\ba
ds^2_{AdS_d}=\frac{dy^2+\sum_{i=1}^{d-1}dx^i dx_i}{y^2}.
\ea
We are interested in the holographic pseudo entropy at the time slice of the dual CFT defined by $\rho=y=0$. We choose a subsystem $A$ on this $(d-1)$-dimensional time slice.

We assume the $\mathbb{Z}_2$ invariance $h(\rho)=h(-\rho)$ such that the minimal surface $\Gamma_A$ is localized on the time slice $\rho=0$.  
Also we assume both the future infinity and past infinity are dual to two different CFT vacua 
$|\psi_1\lb$ and $|\psi_2\lb$ with the same central charge $c$.
This forces us to set $h(\rho)\simeq 2|\rho|$ in the limit $\rho\to\pm\infty$. 
The coordinate $x$ describes the space direction of the dual Janus CFT. 
The limit $\rho=\infty$ (and $\rho=-\infty$) describes the upper (and lower) half plane of the interface CFT. The upper and lower CFT path-integral define the states $|\psi_1\lb$ and $|\psi_2\lb$ respectively.
In a Janus solution, we typically have a bulk scalar field which approaches two different values
in the two limits $\rho\to\pm\infty$. This means that the two states are related by an exactly marginal 
deformation.

For simplicity, we consider $d=2$ i.e. two dimensional CFTs \cite{Bak:2007jm} , below.
We choose the subsystem $A$ as an interval $-l/2\leq x \leq l/2$ at the location of Janus interface 
i.e. $\rho=y=0$ and calculate its holographic pseudo entropy. Due to the $\mathbb{Z}_2$ symmetry,  $\Gamma_A$ is the minimal surface situated on the $\rho=0$ slice. If we write the cutoff of $y$ as 
$\delta$, then the holographic pseudo entropy is estimated as follows:
\ba
S(\tau^{\psi_1|\psi_2}_A)=\frac{c}{3}e^{\frac{h(0)}{2}}\log\frac{l}{\delta},
\ea
where $c$ is the central charge of the CFT.

We are interested in whether this pseudo entropy is smaller than the original value of entanglement entropy in the CFT
\ba
S(\rho^{\psi_1}_{A})=S(\rho^{\psi_2}_A)=\frac{c}{3}\log\frac{l}{\ep},
\ea
where $\ep$ is the CFT UV cutoff.
We do not need to care about 
the difference between $\delta$ and $\ep$ as their difference is subleading for our purpose.
Below we would like to argue the difference $S(\tau^{\psi_1|\psi_2}_A)-S(\rho^{\psi_1}_{A})$
is negative, which is equivalent to the inequality
\ba
h(0)\leq 0.  \label{inrewdf}
\ea
We can generalize our coming argument to any higher  $d$ dimensions and we have a smaller value of holographic pseudo entropy when (\ref{inrewdf}) is satisfied.

To restrict our gravity configurations to solutions physically sensible solutions, we impose an Euclidean version of null energy condition
\ba
R_{\mu\nu}N^\mu N^\nu \geq 0,
\ea
where $N^\mu$ describes arbitrary null vector. In our Euclidean ansatz (\ref{xxe}),  we can choose as
\ba
N_{(1)}^\mu=(0,1,i),\ \ \ N_{(2)}^\mu =(1,e^{-h(\rho)}y i,0).\nonumber
\ea
The second one leads to the non-trivial constraint as follows
\ba
2e^{-h}-\de^2_\rho h\geq 0.  \label{ineqrf}
\ea

In the explicit example of 3D Janus solution given in \cite{Bak:2007jm}, the metric reads
\ba
e^{2h(\rho)}=\frac{1}{2}+\frac{\s{1-2\gamma^2}}{2}\cosh(2\rho),
\ea
where $\gamma$ parameterizes the Janus deformation.
This solution satisfies
\ba
2e^{-h}-\de^2_\rho h=2\gamma^2 e^{-h},
\ea
which is positive as expected.

Now, the $\mathbb{Z}_2$ symmetry and our assumptions on the asymptotic behavior are summarized as 
\ba
&& h'(0)=0, \ \ \ h(\rho)=h(-\rho),\no
&& h(\rho)\simeq  2\rho\ \ \ (\rho\to \infty). 
\ea 
Moreover, we would like to assume $h'(\rho)\geq 0$ as it holds in all known Janus solutions \cite{Bak:2003jk,Freedman:2003ax,Clark:2004sb,Clark:2005te,DHoker:2006vfr,Bak:2007jm}.
By multiplying $h'$ with (\ref{ineqrf}), we get
\ba
\frac{1}{2}\de_\rho[(\de_{\rho}h)^2]\leq \de_\rho[-2e^{-h}].
\ea
An integration of this from $\rho=0$ to $\rho=\infty$ leads to 
\ba
2\leq 2 e^{-h(0)}.  \label{eeefing}
\ea
This proves the expected inequality (\ref{inrewdf}).

In this way, under exactly marginal deformations, our holographic results show that
the difference 
$2\Delta S_{12}=2S(\tau^{\psi_1|\psi_2}_A) -\left(S(\rho^{\psi_1}_A)+S(\rho^{\psi_2}_A)\right)$
is negative.

\subsection{Holographic Pseudo Entropy in Gapped Phases}\label{sec:holPT}

Consider two states $|\psi_1\lb$ and  $|\psi_2\lb$ which are ground states of two different massive field theories.  In particular, we assume these field theories are given by two different massive deformations of a common gapless theory (i.e. CFT) such that these two states can be deformed into each other only passing through the gapless critical point. 
We analyzed such an example in a spin system in section \ref{sec:spinex}, where we found that the difference of pseudo entropy $2S(\tau^{\psi_1|\psi_2}_A) -\left(S(\rho^{\psi_1}_A)+S(\rho^{\psi_2}_A)\right)$ tends to be positive, though it is always negative when the two states are in the same phase.
Note also that such a setup typically occurs when we consider two ground states in two different topological phases, which are connected only though a gapless theory. Below we would like to study a holographic example of this type.

We expect the gravity duals of $|\psi_1\lb$ and $|\psi_2\lb$ to be massive deformations of an AdS.
Moreover, the gravity dual of the inner product $\la \psi_2|\psi_1\lb$ is given by a massive version of Janus solution. Since the two states are connected via a CFT, the gravity dual geometry near the time slice where $|\psi_1\lb$ and  $|\psi_2\lb$ are glued, gets close to the pure AdS geometry. 
By thinking this setup even intuitively, we can expect that the difference 
\ba
2\Delta S_{12}=2S(\tau^{\psi_1|\psi_2}_A) 
-\left(S(\rho^{\psi_1}_A)+S(\rho^{\psi_2}_A)\right)>0,  \no \label{pepad}
\ea
can be positive because the minimal surface area gets enhanced near the 
bulk time slice which looks like the pure AdS, compared with those in massive gravity duals.
Note that the assumption that the two states are connected with each other through a critical point guarantees that its gravity dual has such a time slice which enhances the minimal surface area.

To explicitly confirm the above expectation in an example, we 
consider the following Einstein-Scalar Theory in $d+1$ dimension:
\ba
I=-\frac{1}{16\pi G_N}\int d^{d+1}x
\s{g}\left(R-2\Lambda-g^{ab}\de_a\phi\de_b\phi-V(\phi)\right).\nonumber
\ea
The Einstein equation reads
\ba
&& R_{ab}-\frac{1}{2}Rg_{ab}+\Lambda g_{ab}
+\frac{1}{2}g_{ab}(\de\phi)^2-(\de_a\phi)(\de_b\phi)\no
&& +\frac{1}{2}V(\phi)g_{ab}=0. \nonumber
\ea
Consider a $d$ dimensional surface $\Sigma$ and write its unit normal vector as $n^a$.
 By extracting the component of Einstein equation in the normal direction to the surface $\Sigma$ 
we obtain
\ba
2R_{ab}n^an^b-R+2\Lambda+V(\phi)+(\de\phi)^2-2(n^a\de_a\phi)^2=0.\nonumber
\ea
This is written in terms of the intrinsic curvature $R^{(d)}$ and extrinsic curvature $K$ on $\Sigma$ as follows
\ba
&& -R^{(d)}+K_{ab}K^{ab}-K^2+2\Lambda+V(\phi)-(n^a\de_a\phi)^2\no
&& +\sum_{i=1}^{d}(t^{(i)a}\de_a\phi)^2=0, \nonumber
\ea
where $t^{(i)a}$ are the tangent unit vectors on $\Sigma$.

Now focus on a background with the scalar field excited in a $\mathbb{Z}_2$ invariant manner along the time slice
$\Sigma$. We take the $d+1$ dimensional coordinate to be $(\tau,x_1,x_2,\ddd,x_{d-1},z)$.  
Assuming the $\mathbb{Z}_2$ symmetric potential $V(\phi)=V(-\phi)$, 
we impose the boundary condition at the AdS boundary 
$z=\ep$: 
\ba
&& \phi(\tau)=\phi_*\ \ \ \mbox{when}\ \tau>0,\no
&& \phi(\tau)=-\phi_*\ \ \ \mbox{when}\ \tau<0.  \label{expscgtr}
\ea
We regard the states with the external field $\phi=\phi_*$ and $\phi=-\phi_*$ as the two 
different states 
$|\psi_1\lb$ and $|\psi_2\lb$, dual to ground states of two different massive theories defined by the two relevant deformations.

We expect both the metric and scalar field has the $\mathbb{Z}_2$ symmetry $\tau\to -\tau$ along the surface 
$\Sigma$ given by $\tau=0$. In this case, we have $K_{ab}=\phi=0$ on the $\mathbb{Z}_2$ symmetric time slice $\Sigma$. Thus, we obtain
\ba
R^{(d)}=2\Lambda+V(\phi)-(n^a\de_a\phi)^2.
\ea

In the special case of exactly marginal deformation, which corresponds to the genuine Janus solutions, we have $V(\phi)=0$. In this case, it is obvious that $R^{(d)}<2\Lambda$ and thus the AdS radius gets squeezed. This explains the difference is negative: 
\ba
2\Delta S_{12}=S(\tau^{\psi_1|\psi_2}_A) +S(\tau^{\psi_2|\psi_1}_A) 
-\left(S(\rho^{\psi_1}_A)+S(\rho^{\psi_2}_A)\right)<0,  \nonumber
\ea
which reproduces our result in the previous subsection i.e. (\ref{eeefing}).

Consider the case dual to two relevant deformations, which we are interested in. 
The mass $m$ of scalar  is related to the conformal dimension $\Delta$ of the operator as usual:
\ba
-m^2=\Delta(d-\Delta)>0.
\ea We assume the range 
$\frac{d}{2}<\Delta<d$. For generic $V$, it is not clear whether $R^{(d)}-2\Lambda$ is positive 
(the entropy increases) or negative 
(the entropy decreases). To make this analysis tractable, let us consider a small perturbation around the pure AdS$_{d+1}$
\ba
ds^2=\frac{dz^2+d\tau^2+\sum_{i=1}^{d-1}dx_i^2}{z^2}.
\ea
We simply set $V(\phi)=m^2\phi^2$ and treat the back reaction caused by the scalar field as a small perturbation. 

The solution of scalar field in the pure AdS$_{d+1}$ is expressed as 
\ba
&& \phi(\tau,x,z) \no
&&=\frac{\pi^{-d/2}\ep^{\Delta-d}\Gamma(\Delta)}{\Gamma(\Delta-d/2)}\!
\int\! d^{d-1}x'd\tau'\!\frac{z^\Delta\phi_0(\tau',x')}{(z^2+(\tau-\tau')^2+|x-x'|^2)^{\Delta}},
\nonumber
\ea
which satisfies
\ba
\lim_{\ep\to 0}\phi(\tau,x,z=\ep)=\phi_0(\tau,x).
\ea

In particular, when the boundary value $\phi_0(\tau,x)$ is a constant $\phi_*$ we simply have
\ba
\phi_*(\tau,x,z)=\left(\frac{z}{\ep}\right)^{d-\Delta}\phi_*.
\ea
In this static background, the scalar curvature is given by 
\ba
R^{(d)}_*=2\Lambda+m^2\left(\frac{z}{\ep}\right)^{2(d-\Delta)}\phi_*^2.
\ea
The minimal surface on $\Sigma$ with the above curvature gives the holographic entanglement entropy
$S(\rho^{\psi_1}_A)=S(\rho^{\psi_2}_A)$.

To calculate the holographic pseudo entropy $S(\tau^{\psi_1|\psi_2}_A)=S(\tau^{\psi_2|\psi_1}_A)$ perturbatively, we can consider the gravity dual given by the back reaction of the scalar with the boundary condition
\ba
\phi_0(\tau,x)=\mbox{sgn}(\tau)\cdot \phi_*. \nonumber
\ea
The solution of the scalar field under this boundary condition leads to
\ba
z\de_\tau\phi(\tau,x,z)|_{\tau=0}=\frac{2\Gamma(\Delta-d/2+1/2)}{\s{\pi}\Gamma(\Delta-d/2)}
\left(\frac{z}{\ep}\right)^{d-\Delta}\phi_*.  \nonumber
\ea
By noting the normal vector on $\Sigma$ is given by $n^\tau=z$ for the pure AdS,
on the surface $\Sigma$ at $\tau=0$, we find
\ba
R^{(d)}=2\Lambda-(z\de_\tau\phi)^2.
\ea
Thus the difference of the curvature on $\Sigma$ reads 
\ba
&& R^{(d)}-R^{(d)}_*  \no
&& =\left(\frac{z}{\ep}\right)^{2(d-\Delta)}\!\phi_*^2\!
\left[\Delta(d-\Delta)\!-\!\frac{4}{\pi}\left(\frac{\Gamma(\Delta-d/2+1/2)}{\Gamma(\Delta-d/2)}\right)^2\right].\no
\label{difget}
\ea
If this difference is positive, we expect
\ba
2\Delta S_{12}=S(\tau^{\psi_1|\psi_2}_A) +S(\tau^{\psi_2|\psi_1}_A) 
-\left(S(\rho^{\psi_1}_A)+S(\rho^{\psi_2}_A)\right)>0.  \no 
\label{qqqse}
\ea
This is because as the absolute value of the curvature gets larger, the area of 
the minimal surface for a fixed subsystem $A$ gets smaller. Note that in the current setup,
the metric on the time slice $\Sigma$ can be written in the following form, owing 
rotational and translational symmetry in the $x_i$ direction:
\ba
&& ds^2=\frac{F(\ti{z})d\ti{z}^2+\sum_{i=1}^{d-1}dx_i^2}{\ti{z}^2},\no
&& F(\ti{z})\equiv 1+A \phi_*^2 \ti{z}^{2(d-\Delta)}+O(\phi_*^4)
\ea
via a coordinate transformation $z\to \ti{z}$, where $A$ is an $O(1)$ constant. 
Since the value of the curvature $R^{(d)}$ is proportional to the value of $A$, the above simple 
relation between the curvature and area of minimal surfaces follows.

Indeed, from (\ref{difget}) 
we can find $R^{(d)}-R^{(d)}_*>0$ when $\Delta$ is close to $d/2$.
This provides a class of example where the pseudo entropy is enhanced such that (\ref{qqqse}) 
is satisfied.


\section{Discussions}
In this article, we studied fundamental properties of pseudo entropy 
in quantum field theories and spin systems via both numerical and analytical calculations.
First we numerically analyzed the two dimensional Lifshitz free scalar field theory.
This analysis demonstrates three basic properties of 
pseudo entropy in quantum many-body systems, namely, the area law behavior, the saturation behavior, and the non-positivity of difference between the pseudo entropy and averaged entanglement entropy in the same quantum phase. We also found an example where the strong subadditivity of pseudo entropy 
is violated. Next we numerically studied the XY spin model. In addition to the confirmation of the three 
properties, we found that the non-positivity of the difference can be violated when the initial and final state belong to different quantum phases. We also studied the time evolution of pseudo entropy after a global quantum quench, which shows that the imaginary part of pseudo entropy has an interesting characteristic behavior. Finally we explored analytical calculations for two dimensional CFTs and also for CFTs with gravity duals via the AdS/CFT. The conformal perturbation 
analysis again confirms the three basic properties. Our holographic analysis based on a scalar field perturbation in the gravity duals shows
that the difference (\ref{difpea}) can be positive when the initial state and final state are different relevant perturbations of an identical CFT vacuum. It is an interesting future problem to consider fully non-linear solutions in gravity duals and work out a general condition of positive difference. 
Our results imply the following simple intuitive explanation of the positivitiy of difference. 
When the initial and final state belong to different topological phases, we expect a gapless mode localized along an interface. Since this gapless mode, described by a lower dimensional CFT, enhances the pseudo entropy, the violation of the non-positivity of the difference can occur. It would be intriguing to consider its condensed matter implications.


\section*{Acknowledgements}
We are grateful to Kanato Goto, Reza Mohammadi-Mozaffar, Tatsuma Nishioka, Masahiro Nozaki, Shinsei Ryu and Yusuke Taki for useful discussions.
AM was supported by Alexander von Humboldt foundation via a postdoctoral fellowship during the early stages of this work.
TT is supported by the Simons Foundation through the ``It from Qubit'' collaboration,  Inamori Research Institute for Science and World Premier International Research Center Initiative (WPI Initiative) 
from the Japan Ministry of Education, Culture, Sports, Science and Technology (MEXT). 
TT is supported by JSPS Grant-in-Aid for Scientific Research 
(A) No.21H04469 and by JSPS Grant-in-Aid for Challenging Research (Exploratory) 18K18766.
KT was supported by the Simons Foundation through the ``It from Qubit'' collaboration and by JSPS Grant-in-Aid for Research Activity start-up 19K23441. KT is supported by JSPS Grant-in-Aid for Early-Career Scientists 21K13920. ZW is supported by the ANRI Fellowship and Grant-in-Aid for JSPS Fellows No. 20J23116. Some of the numerical calculations were carried out on Yukawa-21 at YITP in Kyoto University.

\appendix

\section{Path Integral Calculation of Gaussian Transition Matrix}\label{sec:app1}
In this appendix we introduce an alternative method for calculation of pseudo entropy for Gaussian states in quadratic scalar theories. The method we introduce here is based on direct calculation of the transition matrix from path integral formulation of the wave functionals. This method is a generalization of \cite{BKLS} which was introduced for calculation of entanglement entropy for Gaussian states in scalar theories. With this method we can calculate R\'{e}nyi pseudo entropies which agree with the results obtained by the correlator method and the operator method introduced in the main text.

We deal with the scalar field on $\mathbb{R}^d$ as a collection of coupled oscillators on a lattice of space points, labeled by capital Latin indices. The displacement at each point gives the value of the scalar field there. We consider two Gaussian states $\psi_{\alpha}$ $(\alpha=1,2)$,
\begin{equation}
\begin{split}
\braket{q_A|\psi_{\alpha}} = N^{(\alpha)} \exp \left[-\frac{1}{2}\sum_{A,B}q_A W_{AB}^{(\alpha)} q_B \right]
\end{split}  \label{wave function}
\end{equation}
where $q_A$ gives the displacement of the $A$-th oscillator, $W_{AB}^{(\alpha)}$ is a positive definite matrix 
and $N^{(\alpha)}$ is a normalization constant,
\begin{equation}
\begin{split}
N^{(\alpha)} = \left( \det \left( \frac{W_{AB}^{(\alpha)} }{\pi} \right) \right)^{1/4} .
\end{split}
\end{equation}
Now consider a region $\Omega$ in $\mathbb{R}^d$. 
The oscillators in this region will be specified by Greek letters, 
and those in the complement of $\Omega$, $\Omega^c$, will be specified by lowercase Latin letters.
We will use the following notation
\begin{alignat}{2}
W^{(\alpha)}_{AB} = \begin{pmatrix} 
W^{(\alpha)}_{ab} & W^{(\alpha)}_{a\beta}  \\
W^{(\alpha)}_{\alpha b} & W^{(\alpha)}_{\alpha \beta} 
\end{pmatrix} 
\equiv   \begin{pmatrix} 
A^{(\alpha)} & B^{(\alpha)}  \\
B^{(\alpha)T} & C^{(\alpha)} 
\end{pmatrix} .  
\end{alignat}
$W^{(\alpha)}_{AB}$ can be written as the correlation function of the momentum operators, 
\begin{equation}
\begin{split}
W^{(\alpha)}_{AB} = 2 \bra{\psi_{\alpha}} p_A p_B \ket{\psi_{\alpha}}  \label{normalization}
\end{split}
\end{equation}
where $p_A$ and $p_B$ are the momentum operators and obey the canonical commutation relations $[q_A, p_B]=i\delta_{AB}$.

We calculate the reduced transition matrix,
\begin{equation}
\begin{split}
\bra{q_a}\mathrm{Tr}_{\Omega^c} [\ket{\psi_{1}}\bra{\psi_{2}}] \ket{q_b} 
= \int dq_\alpha \braket{q_a, q_\alpha|\psi_{1}} \braket{\psi_{2}|q_b, q_\alpha}.
\end{split}  
\end{equation}
For simplicity, we use the vector notation $(q^{(1)})_a=q_a$ and $(q^{(2)})_a=q_a$.
We perform the Gaussian integral and obtain the matrix elements of the reduced transition matrix $\tau_\Omega^{1|2}=\mathrm{Tr}_{\Omega^c} \left[\frac{\ket{\psi_{1}}\bra{\psi_{2}}}{\braket{\psi_{2}|\psi_{1}}} \right]$, 
\begin{align}\label{scalar transition matrix}
\begin{split}
&
\bra{q^{(1)}}\tau_\Omega^{1|2} \ket{q^{(2)}} 
= \frac{N'}{\sqrt{\det \frac{\bar{C}}{\pi}}} 
\times
\\&\;\;\;\;\;\;\;\;\;\;\;\;\;\;
\exp \left[-\frac{1}{2}  (q^{(1)T}, q^{(2)T}) M
\begin{pmatrix}
q^{(1)}  \\
q^{(2)}
\end{pmatrix} \right]
\end{split}  
\end{align}
where
\begin{alignat}{2}
M = \begin{pmatrix} 
X^{(1)} & 2Y   \\
2Y^{T} & X^{(2)}  
\end{pmatrix} 
\end{alignat}
and 
\begin{equation}
\begin{split}
&X^{(\alpha)}=A^{(\alpha)} -\frac{1}{2} B^{(\alpha)} \bar{C}^{-1} B^{(\alpha)T}, \\
&Y=-\frac{1}{4} B^{(1)} \bar{C}^{-1} B^{(2)T} ,  \\
&\bar{C} = \frac{1}{2} \left(C^{(1)} +C^{(2)} \right) \\
&N'=\sqrt{\det \frac{\bar{W}}{\pi}}, \\
&\bar{W}=\frac{1}{2}\left(W^{(1)}+W^{(2)}\right).
\end{split}  \label{vol}
\end{equation}
From (\ref{scalar transition matrix}), we obtain
\begin{align}\label{trace n power}
\begin{split}
&\mathrm{Tr} \left(\tau_\Omega^{1|2} \right)^n =N'^{n} \left(\det \frac{\bar{C}}{\pi}\right)^{-n/2} \int dq^{(1)} \cdots dq^{(n)}
\times
\\ &
\;\;\;\;\;\;\;\;
\exp \left[-  (q^{(1)T},\cdots, q^{(n)T}) M_n
\begin{pmatrix}
q^{(1)}  \\
\vdots \\
q^{(n)}
\end{pmatrix} \right]
\end{split}  
\end{align}
where 
\begin{equation}
M_n =
\begin{pmatrix}
\bar{X} & Y & 0 & \cdots &0 &Y   \\
Y& \bar{X}&Y&\cdots &0&0 \\
0&Y&\bar{X}&\cdots &0&0 \\
\vdots &\vdots &\vdots & &  \vdots &\vdots \\
0&0&0&\cdots &\bar{X}&Y \\
Y&0&0&\cdots &Y&\bar{X} \\
\end{pmatrix} , \label{M_n}
\end{equation}
here
\begin{equation}
\begin{split}
\bar{X}=\frac{1}{2}\left(X^{(1)}+X^{(2)}\right).
\end{split} 
\end{equation}
We rewrite $M_n$ as
\begin{equation}
M_n = \dfrac{1}{2} \left(\bar{X}^{1/2} \otimes I_{n} \right) \tilde{M_n} \left(\bar{X}^{1/2} \otimes I_{n} \right), \label{M_n2}
\end{equation}
where $I_{n}$ is the $n\times n$ identity matrix and
\begin{equation}
\tilde{M_n} =
\begin{pmatrix}
2 & -Z & 0 & \cdots &0 &-Z^{T}   \\
-Z^{T}&2&-Z&\cdots &0&0 \\
0&-Z^{T}&2&\cdots &0&0 \\
\vdots &\vdots &\vdots & &  \vdots &\vdots \\
0&0&0&\cdots &2&-Z \\
-Z&0&0&\cdots &-Z^{T}&2 \\
\end{pmatrix} . \label{tilde M_n}
\end{equation}
Here,
\begin{equation}
Z=-2\bar{X}^{-1/2} Y \bar{X}^{-1/2}  . \label{matz}
\end{equation}
From (\ref{trace n power}) and (\ref{M_n2}), we obtain
\begin{align} \label{trace n power 2}
\begin{split}
\mathrm{Tr} \left(\tau_\Omega^{1|2} \right)^n 
&= \left(\det \left(\bar{A}-\bar{B}\bar{C}^{-1} \bar{B}^T \right) \right)^{n/2}
\times
\\& \;\;\;\;\;\;
(\det \bar{X})^{-n/2} \left(\det \frac{\tilde{M}_n}{2} \right)^{-1/2} .
\end{split} 
\end{align}
We can diagonalize $\tilde{M}_n$ by Fourier transformation and obtain
\begin{equation}
\begin{split}
&\det \frac{\tilde{M}_n}{2} = \prod_{k=0}^{n-1} \det \left[1-\left(\cos\frac{2\pi k}{n} \bar{Z}+i\sin\frac{2\pi k}{n} \tilde{Z}\right)\right]
\end{split}  \label{det Mn}
\end{equation}
where
\begin{equation}
\begin{split}
&\bar{Z}=\frac{1}{2}\left(Z+Z^T\right), \\
&\tilde{Z}=\frac{1}{2}\left(Z-Z^T\right).
\end{split}
\end{equation}
Finally, from (\ref{trace n power 2}) and (\ref{det Mn}), we obtain
\begin{equation}
\begin{split}
&S^{(n)}(\tau_\Omega^{\psi_{1}|\psi_{2}}) =\frac{1}{1-n} \ln \mathrm{Tr}\left[ (\tau_\Omega^{\psi_{1}|\psi_{2}})^n \right] \\
&=\frac{1}{1-n} \left[\frac{n}{2} \ln \det \left(\bar{A}-\bar{B}\bar{C}^{-1} \bar{B}^T \right) -\frac{n}{2} \ln \det \bar{X} \right. \\
&\left. -\frac{1}{2} \sum_{k=0}^{n-1} \ln \det \left[1-\left(\cos\frac{2\pi k}{n} \bar{Z}+i\sin\frac{2\pi k}{n} \tilde{Z}\right)\right]  \right].
\end{split}
\end{equation}
Using the explicit expression for $W$ as
\begin{equation}
W^{(\alpha)}_{rs}=\frac{1}{N^d} \sum_{k} \omega_{\alpha,k} e^{2\pi i k(r-s)/N},
\end{equation}
one can see that this method numerically leads to the same results as the correlator and the operator methods R\'{e}nyi pseudo entropy of Gaussian states on a finite lattice.


\end{document}